\documentclass[reqno,11pt]{amsart}
\usepackage[utf8]{inputenc}
\usepackage{enumitem}
\usepackage{esint}
\usepackage{graphicx}
\usepackage{amscd}
\usepackage{slashed}
\usepackage{amssymb}
\usepackage{mathtools} 
\usepackage{pstricks-add}
\usepackage{epsfig}
\usepackage{pst-grad} 
\usepackage{pst-plot} 
\usepackage[mathscr]{eucal}
\numberwithin{equation}{section}
\allowdisplaybreaks[1]

\title[Entangled Quantum States of Causal Fermion Systems]{Entangled Quantum States of Causal Fermion Systems and Unitary Group Integrals}

\author[F.\ Finster]{Felix Finster}
\address{Fakult\"at f\"ur Mathematik \\ Universit\"at Regensburg \\ D-93040 Regensburg \\ Germany}
\email{finster@ur.de}

\author[N.\ Kamran]{Niky Kamran}
\address{Department of Mathematics and Statistics \\ McGill University \\ Montr{\'e}al \\ Canada}
\email{nkamran@math.mcgill.ca}

\author[M.\ Reintjes]{Moritz Reintjes \\ \\ July 2023} 
\address{Department of Mathematics \\ City University of Hong Kong \\ SAR Hong Kong}
\email{moritzreintjes@gmail.com}

\newtheorem{Def}{Definition}[section]
\newtheorem{Thm}[Def]{Theorem}
\newtheorem{Prp}[Def]{Proposition}
\newtheorem{Lemma}[Def]{Lemma}
\newtheorem{Remark}[Def]{Remark}
\newtheorem{Corollary}[Def]{Corollary}
\newtheorem{Example}[Def]{Example}

\newcommand{\Thanks}{\vspace*{.5em} \noindent \thanks}
\newcommand{\beq}{\begin{equation}}
\newcommand{\eeq}{\end{equation}}
\newcommand{\Proof}{\begin{proof}}
\newcommand{\QED}{\end{proof} \noindent}
\newcommand{\QEDrem}{\ \hfill $\Diamond$}

\newcommand{\la}{\langle}
\newcommand{\ra}{\rangle}
\newcommand{\bra}{\mathopen{<}}
\newcommand{\ket}{\mathclose{>}}
\newcommand{\Sl}{\mathopen{\prec}}
\newcommand{\Sr}{\mathclose{\succ}}

\newcommand{\C}{\mathbb{C}}
\newcommand{\R}{\mathbb{R}}
\newcommand{\1}{\mbox{\rm 1 \hspace{-1.05 em} 1}}

\newcommand{\N}{\mathbb{N}}
\newcommand{\Pdd}{\mbox{$\partial$ \hspace{-1.2 em} $/$}}
\newcommand{\slsh}{\mbox{ \hspace{-1.13 em} $/$}}

\renewcommand{\H}{\mathscr{H}}
\newcommand{\h}{\mathfrak{h}}
\newcommand{\U}{{\rm{U}}}
\newcommand{\SU}{{\rm{SU}}}

\newcommand{\G}{\mathscr{G}}

\newcommand{\bep}{\begin{pmatrix}}
\newcommand{\enp}{\end{pmatrix}}

\renewcommand{\O}{\mathscr{O}}
\renewcommand{\Re}{\text{Re}}
\renewcommand{\Im}{\text{Im}}
\newcommand{\F}{{\mathscr{F}}}

\newcommand{\D}{{\mathscr{D}}}

\renewcommand{\O}{{\mathscr{O}}}
\renewcommand{\L}{{\mathcal{L}}}
\newcommand{\Sact}{{\mathcal{S}}}
\newcommand{\s}{{\mathfrak{s}}}

\newcommand{\Lin}{\text{\rm{L}}}

\newcommand{\T}{{\mathscr{T}}}

\newcommand{\Fock}{{\mathcal{F}}}
\newcommand{\fermi}{{\mathrm{{f}}}}
\newcommand{\bose}{{\mathrm{{b}}}}
\newcommand{\lab}{{\mathrm{{lab}}}}
\newcommand{\he}{{\mathrm{{he}}}}
\renewcommand{\le}{{\mathrm{{le}}}}
\renewcommand{\sp}{{\mathrm{{sp}}}}

\newcommand{\scrM}{\myscr M}

\newcommand{\Gtest}{\Gamma^\text{\rm{{test}}}}
\newcommand{\Gdiff}{\Gamma^\text{\rm{{diff}}}}
\newcommand{\Gfermi}{\Gamma^\text{\rm{{f}}}}
\newcommand{\Jfermi}{\mathfrak{J}^\text{\rm{{f}}}}
\newcommand{\Jdiff}{\mathfrak{J}^\text{\rm{{diff}}}}
\newcommand{\Ctest}{C^\text{\rm{{test}}}}
\newcommand{\J}{\mathfrak{J}}
\newcommand{\Jin}{\mathfrak{J}^\text{\rm{{in}}}}
\newcommand{\Jlin}{\mathfrak{J}^\text{\rm{{lin}}}}
\newcommand{\Jtest}{\mathfrak{J}^\text{\rm{{test}}}}

\newcommand{\Jvary}{\mathfrak{J}^\text{\rm{{vary}}}}
\newcommand{\Glin}{\Gamma^\text{\rm{{lin}}}}

\newcommand{\dyn}{{\text{\rm{dyn}}}}
\newcommand{\scrU}{{\mathscr{U}}}
\newcommand{\scrA}{{\mathscr{A}}}
\renewcommand{\div}{{\rm{div}}\,}
\newcommand{\reg}{{\text{\rm{reg}}}}
\newcommand{\bu}{{\mathbf{u}}}
\newcommand{\bv}{\mathbf{v}}

\newcommand{\A}{\myscr A}

\newcommand{\hol}{\text{\rm{hol}}}
\newcommand{\ah}{\text{\rm{ah}}}
\newcommand{\symm}{\text{s}}

\DeclareFontFamily{OT1}{rsfso}{}
\DeclareFontShape{OT1}{rsfso}{m}{n}{ <-7> rsfso5 <7-10> rsfso7 <10-> rsfso10}{}
\DeclareMathAlphabet{\myscr}{OT1}{rsfso}{m}{n}

\DeclareMathOperator{\re}{Re}
\DeclareMathOperator{\im}{Im}

\DeclareMathOperator{\Tr}{Tr}
\DeclareMathOperator{\tr}{tr}

\DeclareMathOperator{\diag}{diag}

\DeclareMathOperator{\supp}{supp}

\DeclareMathOperator{\sign}{sign}
\renewcommand{\u}{\mathfrak{u}}
\renewcommand{\v}{\mathfrak{v}}

\newcommand{\bitem}{\begin{itemize}[leftmargin=2em]}
\newcommand{\eitem}{\end{itemize}}

\newcommand{\scrt}{T}

\renewcommand{\sc}{\text{\rm{sc}}}
\newcommand{\x}{\mathbf{x}}

\begin{document}

\maketitle

\begin{abstract}
This paper is dedicated to a detailed analysis and computation of quantum states of
causal fermion systems.
The mathematical core is to analyze integrals over the unitary group asymptotically
for a large dimension of the group, for various integrands with a specific scaling behavior
in this dimension. It is shown that, in a well-defined limiting case, the localized refined pre-state is positive
and allows for the description of general entangled states.
\end{abstract}

\tableofcontents

\section{Introduction and Overview of Main Results} \label{secintro}
The theory of {\em{causal fermion systems}} is a recent approach to fundamental physics
(see the basics in Section~\ref{secprelim}, the reviews~\cite{dice2014, nrstg, review}, the textbooks~\cite{cfs, intro}
or the website~\cite{cfsweblink}).
In this approach, spacetime and all objects therein are described by a measure~$\rho$
on a set~$\F$ of linear operators on a Hilbert space~$(\H, \la .|. \ra_\H)$. 
The physical equations are formulated by means of the so-called {\em{causal action principle}},
a nonlinear variational principle where an action~$\Sact$ is minimized under variations of the measure~$\rho$.
The present paper is a major step in an ongoing research program aimed at giving a conceptually clear and mathematically sound {\em{derivation of quantum field theory}} within the theory of causal fermion systems.
In simple terms, the goal of this program is to show that, in a well-defined limiting case,
the dynamics of a causal fermion system can be described as in quantum field theory by a unitary time evolution
on bosonic and fermionic Fock spaces.
This program was initiated in~\cite{qft}, where quantum electrodynamics was obtained from the
causal action principle under certain simplifying assumptions and approximations, which were
stated but not yet justified in a fully satisfying way.
The first step in the detailed treatment based on the mathematical structures of causal fermion systems
is~\cite{fockbosonic}, where a connection between causal variational principles
and a dynamics on bosonic Fock spaces was established.
More recently, in~\cite{fockfermionic} it was shown that a causal fermion system in a Minkowski-type spacetime
gives rise to a distinguished quantum state. Moreover, various modifications and refinements of this
construction were given. The aim of the present paper is to analyze the different definitions in detail
and to identify the construction which is most suitable for the description of entanglement.
To this end, we need to delve deeper into the question of how to compute these quantum states.

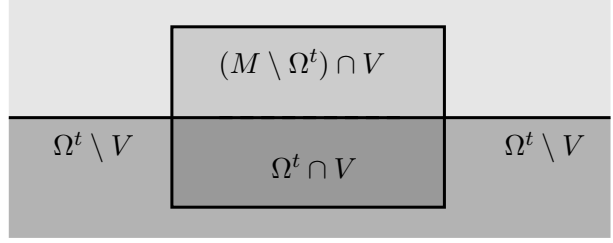
\begin{figure}[tb]
\psscalebox{1.0 1.0} 
{
\begin{pspicture}(0,26.0)(8.02,29.2)
\definecolor{colour0}{rgb}{0.7019608,0.7019608,0.7019608}
\definecolor{colour1}{rgb}{0.9019608,0.9019608,0.9019608}
\definecolor{colour2}{rgb}{0.6,0.6,0.6}
\definecolor{colour3}{rgb}{0.8,0.8,0.8}
\psframe[linecolor=colour0, linewidth=0.04, fillstyle=solid,fillcolor=colour0, dimen=outer](8.01,27.6)(0.01,26.0)
\psframe[linecolor=colour1, linewidth=0.04, fillstyle=solid,fillcolor=colour1, dimen=outer](8.01,29.2)(0.01,27.6)
\pspolygon[linecolor=colour2, linewidth=0.02, fillstyle=solid,fillcolor=colour2](2.1814287,27.612143)(5.8,27.58609)(5.802857,27.296106)(5.8,27.121767)(5.792857,26.738081)(5.78,26.435)(2.1942856,26.429794)(2.1771429,27.119688)
\pspolygon[linecolor=colour3, linewidth=0.02, fillstyle=solid,fillcolor=colour3](2.1642857,28.785)(5.7785716,28.807858)(5.8085713,28.576166)(5.79,28.285154)(5.792857,27.950018)(5.7957144,27.609285)(2.1757143,27.615)(2.1842856,28.035082)(2.1671429,28.43066)
\psline[linecolor=black, linewidth=0.04](0.01,27.6)(8.01,27.6)
\psline[linecolor=black, linewidth=0.04, linestyle=dashed, dash=0.17638889cm 0.10583334cm](2.81,27.6)(5.21,27.6)
\psframe[linecolor=black, linewidth=0.04, dimen=outer](5.821429,28.828571)(2.1585715,26.38857)
\rput[bl](0.6,27){$\Omega^t \setminus V$}
\rput[bl](6.6,27){$\Omega^t \setminus V$}
\rput[bl](3.5,26.8){$\Omega^t \cap V$}
\rput[bl](2.8,28.1){$(M \setminus \Omega^t) \cap V$}
\end{pspicture}
}
\caption{Decomposition of the vacuum spacetime in the localized state.}
\label{figlocstate}
\end{figure}

Our main objective is to show that general entangled states can indeed be described with the help of the
{\em{localized refined partition function}}~$Z^t_V$, which can be written as
\beq
 Z^t_V\big( \alpha, \beta, \tilde{\rho} \big) := 
 \fint_\G d\mu_\G \big(\scrU_< \big)  \fint_\G d\mu_\G \big( \scrU_> \big) \:
 e^{\alpha N \T^t_V \big(\tilde{\rho}, T_{\scrU_<, \scrU_>}  \rho \big)} \:. \label{Zreflocfinal} 
\eeq
Before giving an overview of our results, we make a few comments on the structure of~$Z^t_V$.
In the above formula, $\tilde{\rho}$ is the measure describing the interacting spacetime, whereas~$\rho$ describes
the Minkowski vacuum. The partition function is a double integral over
a compact group~$\G \simeq \U(N)$ of unitary operators on the Hilbert space~$\H$
(here~$\mu_\G$ is the Haar measure on this group, and by $\fint$ we always denote a
normalized integral).
The integrand involves the exponential of the functional~$\T^t_V$ which relates
the interacting and vacuum measures at time~$t$ within a bounded spatial region described by
a subregion~$V$ of the vacuum spacetime (see Figure~\ref{figlocstate}, where the support of
the measure~$M:= \supp \rho$ is the vacuum spacetime,
and~$\Omega^t$ is the region in the past of time~$t$;
for details see Section~\ref{secprelim} below).
The functional~$\T^t_V$ has the mathematical structure of a {\em{nonlinear surface layer integral}}
(see~\eqref{Tgen} and the preliminaries in Section~\ref{secosinonlin}).
The {\em{localized refined pre-state}}~$\omega^t_V$ can be written symbolically as
\begin{align}
\omega^t_V \big( \cdots \big) &:= \frac{1}{Z^t_V\big( \alpha, \beta, \tilde{\rho} \big)} \fint_\G d\mu_\G \big(\scrU_< \big)  \fint_\G d\mu_\G \big( \scrU_> \big) \:
e^{\alpha N \T^t_V \big(\tilde{\rho}, T_{\scrU_<, \scrU_>}  \rho \big)} \,\big( \cdots \big) \:, \label{omegareflocfinal}
\end{align}
where the dots on the left stand for an operator in the observable algebra~$\A$ formed of the
linearized fields in the vacuum spacetime.
The dots on the right, on the other hand, stand for suitable surface layer integrals which again involve the
linearized fields in the vacuum.
The structure of~\eqref{omegareflocfinal} has some resemblance with the path integral formulation
of quantum theory, where the $n$-point functions are obtained by integrating an exponential of the classical action
over field configurations, taking the fields as insertions.
In this formalism, even the density operator has been constructed in~\cite{page-density}.
However, the similarity to the path integral formalism
does not seem to extend beyond a formal analogy, because in~\eqref{omegareflocfinal}
we integrate over unitary operators, not over field configurations. Moreover, the surface layer integral~$\T^t_V$
in the exponent is not the classical action (or the causal action), but instead a surface layer integral which
can be understood as a device for ``comparing'' the measures~$\tilde{\rho}$ and~$\rho$.
In this ``comparison'' it is important that the unitary operators~$\scrU_<$ and~$\scrU_>$ come up in
a nonlinear way.

In general terms, the localized refined pre-state describes the interacting measure~$\tilde{\rho}$
using the familiar objects of the vacuum spacetime (linear bosonic and fermionic fields).
The measure~$\tilde{\rho}$ should be thought of as having a very complicated structure,
both on microscopic and macroscopic scales.
More specifically, all the objects of the interacting physical system are encoded in the {\em{physical wave functions}},
being a family of spinorial wave functions in spacetime.
The collective behavior of all these wave functions gives rise to the usual spacetime
structures (causality, metric, particles, fields, etc.).
Due to the mutual interaction of all the wave functions, this collective behavior can be intricate,
including {\em{long-range correlations}} and {\em{dephasing effects}} between sub-families of waves
propagating in different spatial directions.
The role of the integration over the unitary group is to detect all these phenomena
as encoded in the interacting measure~$\tilde{\rho}$ and to
quantify them in the familiar language of quantum field theory.
The interplay of these effects  gives rise to entanglement,
as will be worked out and made precise in this paper (see the discussion at the end of Section~\ref{secentangle}).

In order to get into a well-defined limiting case in which the above integrals over the unitary group
can be computed, we consider the following asymptotics:
\bitem
\item[(i)] The causal fermion systems involve an ultraviolet cutoff on a length scale~$\varepsilon$
(which can be thought of as the Planck length). We consider the asymptotics~$\varepsilon \searrow 0$,
while keeping the length scales of macroscopic physics fixed.
\item[(ii)] For any given~$\varepsilon>0$, we also consider the asymptotics~$N \rightarrow \infty$
where the dimension of the Hilbert space~$\H$ tends to infinity.
Since the parameter~$N$ also appears in the exponent of~$Z^t_V$ and~$\omega^t_V$,
this also amounts to a suitable rescaling of the nonlinear surface layer integral.
Since~$\H$ describes the system also outside the spacetime region~$V$,
the limit~$N \rightarrow \infty$ can be thought of as the infinite volume limit when the
size of the whole system tends to infinity, while keeping the bounded spacetime region~$V$ fixed.
\eitem
In physical terms, the quantum state tells us about the outcome of
measurements performed at time~$t$ in the spacetime region~$V$.
With this in mind, the above asymptotics reflect the physical facts
that the Planck length is much smaller than all other relevant length scales (point~(i))
and that a physical measurement takes place in a spatial region which is typically
much smaller than the size of the universe (point~(ii)).
This asymptotics will be worked out in the main sections of our paper
using Gaussian integrals and saddle points techniques
(Sections~\ref{secintunit}--\ref{secrefinecompute}).

We now outline our main results. We first prove that, in the above limiting case, the
localized refined pre-state is positive, i.e.\
\beq \label{state}
\omega^t_V : \A \rightarrow \C \qquad \text{with} \qquad
\omega^t_V(A^* A) \geq 0 \quad \text{for all~$A \in \A$}
\eeq
(see Theorem~\ref{thmpositive} in Section~\ref{secpositive}).
Thus it is a {\em{quantum state}} as used in the algebraic formulation of quantum field theory.
Next, we show that our quantum state
allows for the general description of {\em{entanglement}} (Section~\ref{secentangle}).
We even get a concrete prescription
for how to encode a given entangled Fock state in the measure~$\tilde{\rho}$
(see Section~\ref{secentangle}).
We remark that the usual notion of a quantum state described by a density operator acting on a Hilbert space
is obtained from the algebraic formulation~\eqref{state} by
constructing representations of the field algebra (for details see~\cite[Section~4.5]{fockfermionic}).
Then the quantum state~$\omega^t$ is obtained from the density operator by taking the expectation value, i.e.\
\[ \omega^t(A) = \tr_\Fock(\sigma^t A) \qquad \text{for all~$A \in \A$} \:. \]
This density operator gives all the familiar structures of quantum field theory.
In particular, the density operator gives rise to notions of {\em{entropy}} like the von Neumann entropy~$S = - \tr_\F \big( \sigma^t \, \log \sigma^t \big)$, the relative entropy and the entanglement entropy of a spatial subregion.
We remark that the only but very important point which at this stage
is still missing compared to standard quantum field theory is the {\em{dynamics}} of the quantum state.
This will be worked out in detail in the upcoming paper~\cite{fockdynamics}.

The mathematical core of our analysis
is the computation of specific integrals over the unitary group, asymptotically 
when the dimension of the group gets large. Similar integrals have been studied
in the context of {\em{random matrix theory}} and {\em{lattice gauge theories}}
(see for example the standard textbooks~\cite{mehta, rothe}).
Despite similarities to problems studied in this context
(like the Harish-Chandra integral~\cite{mehta, mcswiggen}, the Itzykson-Zuber model~\cite{itzykson-zuber-planar},
the Gaussian asymptotics of group integrals in~\cite{weingarten, collinsb, ipsen-kieburg} and
diagrammatic approaches~\cite{creutz}), these results do not immediately apply to our problem.
The main difference is that, in our case, the integrand has an explicit dependence on the dimension of
the group, changing the asymptotics as this dimension tends to infinity.
We take advantage of the fact that our integrand only depends on the matrix entries on a fixed subspace.
This makes it possible to carry out the integral over all other matrix entries, giving the reduction formula
in Theorem~\ref{thmgenint} (for related results see~\cite{neretin,friedman-mello,fyodorov, akemann-kieburg}).
From this formula one can deduce the Gaussian asymptotics (see Proposition~\ref{prpgauss})
as well as the leading asymptotics if the integrand is a product depending on the matrix entries
of two orthogonal subspaces
(see Propositions~\ref{prpfact} and~\ref{prpfact2}).
We then apply these results to specific integrands involving exponentials of the matrix elements
(see Proposition~\ref{prpsaddlelin} and Theorem~\ref{thmsaddle}).

We finally comment on the physical picture behind our constructions.
In the usual description of quantum theory, the expectation value of a measurement is
given by the quantum state applied to the corresponding observables.
With this in mind, the construction of a quantum state of a causal fermion system
can be understood as the preparation of a measurement device which can be used
for performing certain measurements,
where the notion of a ``measurement device'' is meant more generally not for an
experimental apparatus, but rather
for a mathematical procedure for extracting information from the causal fermion system.
In this analogy, the freedom in the construction of the state is not problematic or surprising; it
can be understood similar to the fact that different experimental setups can be used to measure
the same physical quantity. From this perspective, we need to address the question
which construction captures the physical essence of quantum fields,
including entanglement. Answering this question also shows that the theory of causal fermion systems
is indeed capable of describing the effects of quantum field theory.

The paper is organized as follows. After giving the necessary preliminaries
on causal fermion systems, quantum states and Fock spaces (Section~\ref{secprelim}),
in Section~\ref{secqintro} we begin with qualitative considerations which explain why
it is preferable to consider the
localized refined pre-state (Sections~\ref{secgeneral}--\ref{seclocalize}). We also specify the precise mathematical setup (Section~\ref{seclocdetail}).
In Section~\ref{secintunit} we review the general methods which will be used later in this paper for
the computation of the group integrals. In Section~\ref{secmodel} we consider
group integrals of exponentials with a specific scaling in the dimension~$N$.
These integrals have a similar structure as the refined partition function~\eqref{Zreflocfinal}.
They can be used for the detailed analysis of the localized refined pre-state~\eqref{omegareflocfinal},
as is worked out in Section~\ref{secrefinecompute}.
In Section~\ref{secentangle} it is shown how entangled states can be described.
In Appendix~\ref{appA} an alternative method is given for computing group integrals.
This method is superseded by the stronger and more suitable methods in Section~\ref{secgauss}.
We explain it nevertheless, because it gives an alternative way of understanding
why group integrals simplify when evaluating them asymptotically for large~$N$.

\section{Preliminaries} \label{secprelim}
\subsection{Basics on Causal Fermion Systems}
This section provides the necessary background on causal fermion systems.
Here we introduce all the basic structures needed later on.
We keep the presentation brief and refer for details to corresponding articles and textbooks.

\subsubsection{Causal Fermion Systems and the Causal Action Principle}
We begin with the abstract definitions.
\begin{Def} \label{defcfs} (causal fermion systems) {\em{ 
Given a separable complex Hilbert space~$\H$ with scalar product~$\la .|. \ra_\H$
and a parameter~$n \in \N$ (the {\em{``spin dimension''}}), we let~$\F \subset \Lin(\H)$ be the set of all
symmetric\footnote{Here by a symmetric operator~$A$ we mean that~$\la A u | v \ra_\H =
\la u | A v \ra_\H$ for all~$u,v \in \H$. For bounded operators as considered here,
the notions ``symmetric'' and ``selfadjoint'' coincide.} 
 operators on~$\H$ of finite rank, which (counting multiplicities) have
at most~$n$ positive and at most~$n$ negative eigenvalues. On~$\F$ we are given
a positive measure~$\rho$ (defined on a $\sigma$-algebra of subsets of~$\F$).
We refer to~$(\H, \F, \rho)$ as a {\em{causal fermion system}}.
}}
\end{Def} \noindent
A causal fermion system describes a spacetime together
with all structures and objects therein.
In order to single out the physically admissible
causal fermion systems, one must formulate physical equations. To this end, we impose that
the measure~$\rho$ should be a minimizer of the causal action principle,
which we now introduce. For any~$x, y \in \F$, the product~$x y$ is an operator of rank at most~$2n$. 
However, in general it is no longer a symmetric operator because~$(xy)^* = yx$,
and this is different from~$xy$ unless~$x$ and~$y$ commute.
As a consequence, the eigenvalues of the operator~$xy$ are in general complex.
We denote these eigenvalues counting algebraic multiplicities
by~$\lambda^{xy}_1, \ldots, \lambda^{xy}_{2n} \in \C$
(more specifically,
denoting the rank of~$xy$ by~$k \leq 2n$, we choose~$\lambda^{xy}_1, \ldots, \lambda^{xy}_{k}$ as all
the non-zero eigenvalues and set~$\lambda^{xy}_{k+1}, \ldots, \lambda^{xy}_{2n}=0$).
We introduce the Lagrangian and the causal action by
\begin{align}
\text{\em{Lagrangian:}} && \L(x,y) &= \frac{1}{4n} \sum_{i,j=1}^{2n} \Big( \big|\lambda^{xy}_i \big|
- \big|\lambda^{xy}_j \big| \Big)^2 \label{Lagrange} \\
\text{\em{causal action:}} && \Sact(\rho) &= \iint_{\F \times \F} \L(x,y)\: d\rho(x)\, d\rho(y) \:. \label{Sdef}
\end{align}
The {\em{causal action principle}} is to minimize~$\Sact$ by varying the measure~$\rho$
under the following constraints,
\begin{align}
\text{\em{volume constraint:}} && \rho(\F) = \text{const} \quad\;\; & \label{volconstraint} \\
\text{\em{trace constraint:}} && \int_\F \tr(x)\: d\rho(x) = \text{const}& \label{trconstraint} \\
\text{\em{boundedness constraint:}} && \iint_{\F \times \F} 
|xy|^2
\: d\rho(x)\, d\rho(y) &\leq C \:, \label{Tdef}
\end{align}
where~$C$ is a given parameter, $\tr$ denotes the trace of a linear operator on~$\H$, and
the absolute value of~$xy$ is the so-called spectral weight,
\beq \label{sw}
|xy| := \sum_{j=1}^{2n} \big|\lambda^{xy}_j \big| \:.
\eeq
This variational principle is mathematically well-posed if~$\H$ is finite-dimensional.
For the existence theory and the analysis of general properties of minimizing measures
we refer to~\cite{continuum, lagrange} and~\cite[Chapter~12]{intro}.
In the existence theory one varies in the class of regular Borel measures
(with respect to the topology on~$\Lin(\H)$ induced by the operator norm),
and the minimizing measure is again in this class. With this in mind, here we always assume that~$\rho$
is a {\em{regular Borel measure}}.

\subsubsection{Spacetime and Physical Wave Functions}
Let~$\rho$ be a {\em{minimizing}} measure. {\em{Spacetime}}
is defined as the support of this measure,
\[ 
M := \supp \rho \:. \]
Thus the spacetime points are symmetric linear operators on~$\H$.
On~$M$ we consider the topology induced by~$\F$ (generated by the operator norm
on~$\Lin(\H)$). Moreover, the measure~$\rho|_M$ restricted to~$M$ gives a volume
measure on spacetime. This makes spacetime into a {\em{topological measure space}}.

The operators in~$M$ contain a lot of information which, if interpreted correctly,
gives rise to spacetime structures like causal and metric structures, spinors
and interacting fields (for details see~\cite[Chapter~1]{cfs}).
Here we restrict attention to those structures needed in this paper.
We begin with a basic notion of causality:

\begin{Def} (causal structure) \label{def2}
{\em{ For any~$x, y \in \F$, the product~$x y$ is an operator
of rank at most~$2n$. We denote its non-trivial eigenvalues (counting algebraic multiplicities)
by~$\lambda^{xy}_1, \ldots, \lambda^{xy}_{2n}$.
The points~$x$ and~$y$ are
called {\em{spacelike}} separated if all the~$\lambda^{xy}_j$ have the same absolute value.
They are said to be {\em{timelike}} separated if the~$\lambda^{xy}_j$ are all real and do not all 
have the same absolute value.
In all other cases (i.e.\ if the~$\lambda^{xy}_j$ are not all real and do not all 
have the same absolute value),
the points~$x$ and~$y$ are said to be {\em{lightlike}} separated. }}
\end{Def} \noindent
Restricting the causal structure of~$\F$ to~$M$, we get causal relations in spacetime.

Next, for every~$x \in \F$ we define the {\em{spin space}}~$S_x$ by~$S_x = x(\H)$;
it is a subspace of~$\H$ of dimension at most~$2n$.
It is endowed with the {\em{spin inner product}} $\Sl .|. \Sr_x$ defined by
\[ 
\Sl u | v \Sr_x = -\la u | x v \ra_\H \qquad \text{(for all $u,v \in S_x$)}\:. \]
A {\em{wave function}}~$\psi$ is defined as a function
which to every~$x \in M$ associates a vector of the corresponding spin space,
\[ 
\psi \::\: M \rightarrow \H \qquad \text{with} \qquad \psi(x) \in S_xM \quad \text{for all~$x \in M$}\:. \]
A wave function~$\psi$ is said to be {\em{continuous}} at~$x$ if
for every~$\varepsilon>0$ there is~$\delta>0$ such that
\beq \label{wavecontinuous}
\big\| \sqrt{|y|} \,\psi(y) -  \sqrt{|x|}\, \psi(x) \big\|_\H < \varepsilon
\qquad \text{for all~$y \in M$ with~$\|y-x\| \leq \delta$}
\eeq
(where~$|x|$ is the absolute value of the symmetric operator~$x$ on~$\H$, and~$\sqrt{|x|}$
is the square root thereof).
Likewise, $\psi$ is said to be continuous on~$M$ if it is continuous at every~$x \in M$.
We denote the set of continuous wave functions by~$C^0(M, SM)$
(where~$SM := \cup_{x \in M} S_xM$ generalizes the
spinor bundle; for details see~\cite[Section~3]{topology}).

It is an important observation that every vector~$u \in \H$ of the Hilbert space gives rise to a unique
wave function. To obtain this wave function, denoted by~$\psi^u$, we simply project the vector~$u$
to the corresponding spin spaces,
\[ 
\psi^u \::\: M \rightarrow \H\:,\qquad \psi^u(x) = \pi_x u \in S_xM \:. \]
We refer to~$\psi^u$ as the {\em{physical wave function}} of~$u \in \H$.
A direct computation shows that the physical wave functions are continuous
(in the sense~\eqref{wavecontinuous}). Associating to every vector~$u \in \H$
the corresponding physical wave function gives rise to the {\em{wave evaluation operator}}
\beq \label{weo}
\Psi \::\: \H \rightarrow C^0(M, SM)\:, \qquad u \mapsto \psi^u \:.
\eeq
Evaluating at a fixed spacetime point~$x \in M$, we obtain a corresponding mapping
\[ \Psi(x) \::\: \H \rightarrow S_xM\:, \qquad u \mapsto \psi^u(x)\:. \]
Every~$x \in M$ can be written as (for the derivation see~\cite[Lemma~1.1.3]{cfs})
\beq
x = - \Psi(x)^* \,\Psi(x) \label{Fid} \:.
\eeq
In words, every spacetime point operator is the {\em{local correlation operator}} of the wave evaluation operator
at this point (for details see~\cite[\S1.1.4 and Section~1.2]{cfs}).

\subsubsection{The Kernel of the Fermionic Projector} \label{secPxy}
For computations, it is most convenient to work with the {\em{kernel of the fermionic projector}}~$P(x,y)$
which can be defined in terms of the wave evaluation operator~\eqref{weo} by
\beq \label{Pxydef}
P(x,y) := -\Psi(x)\, \Psi(y)^* \::\: S_y \rightarrow S_x \:.
\eeq
It can be regarded as the basic object of the theory because the Lagrangian as well as the
constraints can be computed from it. This is based on the observation that the non-trivial eigenvalues
of the operator product~$xy$ coincide with the eigenvalues of the {\em{closed chain}}~$A_{xy}$ defined by
\[ A_{xy} := P(x,y)\, P(y,x) : S_x \rightarrow S_x \]
(for details see for example~\cite[\S1.1.3]{cfs}).
First variations of the Lagrangian can be written as (for details see~\cite[\S1.4.1]{cfs})
\beq \label{delLdef}
\delta \L(x,y) = 2 \re \Tr_{S_y} \!\big( Q(y,x)\, \delta P(x,y) \big) \:,
\eeq
where~$Q(x,y)$ is a symmetric kernel, i.e.\
\beq \label{Qxydef}
Q(x,y) \::\: S_y \rightarrow S_x \qquad \text{and} \qquad Q(x,y)^* = Q(y,x) \:.
\eeq
We will use this variational formula in Section~\ref{seclowenergy}.

\subsubsection{Causal Fermion Systems in Minkowski Space} \label{seccfsmink}
The prime examples of causal fermion systems are constructed from Dirac wave functions
in Minkowski space. For these systems, the causal action principle has been studied in detail
in~\cite{cfs}.
In particular, it is known that in a specific limiting case, referred to as the {\em{continuum limit}},
the measure describing the Minkowski vacuum is indeed a minimizer of the
causal action principle. Moreover, the continuum limit analysis developed in~\cite{cfs}
associates a specific class
of critical measures with systems of Dirac particles in Minkowski space which interact via classical bosonic fields
which satisfy classical field equations.

We now recall a few basics on causal fermion systems in Minkowski space, which will be needed
in the later computations in Section~\ref{secrefinecompute}.
More detailed introductions can be found in~\cite[Section~1.2]{cfs} or~\cite[Chapter~5]{intro}.
In order to describe the Minkowski vacuum, one considers the Hilbert space~$(\H, (.|.)_m)$
of solutions of the Dirac equation~$(i \Pdd - m) \psi=0$ of mass~$m$ with the scalar product
\[ ( \phi | \psi )_m := \int_{\R^3} \Sl \phi \,|\, \gamma^0 \psi
\Sr(t, \vec{x}) \:d^3x 
\]
(where~$\Sl .|. \Sr$ is the spin inner product of signature~$(2,2)$; this scalar product
does not depend on time due to current conservation).
Next, we choose~$(\H, \la .|. \ra_\H)$ as the subspace of~$\H$ of all solutions of negative
energy (i.e.\ negative frequency) with the induced scalar product.
For the construction of a corresponding causal fermion system, one needs to introduce
an {\em{ultraviolet regularization}} on a microscopic length scale~$\varepsilon$, which can
be thought of as the Planck length. To this end, one introduces a so-called {\em{regularization operator}}~$({\mathfrak{R}}_\varepsilon)$ as an operator which maps~$\H$ to the continuous wave functions,
\[ 
{\mathfrak{R}}_\varepsilon \::\: \H \rightarrow C^0(\scrM, S\scrM) \:. \]
Next, for any~$x \in \scrM$ one introduces the {\em{local correlation operator}}~$F^\varepsilon$ by
\[ F^\varepsilon(x) := - {\mathfrak{R}}_\varepsilon(x)^* {\mathfrak{R}}_\varepsilon(x) \::\: \H \rightarrow \H \:. \]
Taking into account that the inner product on the Dirac spinors at~$x$ has signature~$(2,2)$,
the local correlation operator~$F^\varepsilon(x)$ is a symmetric operator on~$\H$
of rank at most four, which has at most two positive and at most two negative eigenvalues.
We thus obtain a mapping
\[ 
F^\varepsilon \::\: \scrM \rightarrow \F \:, \]
where~$\F \subset \Lin(\H)$ is the set of all
symmetric operators on~$\H$ of finite rank, which (counting multiplicities) have
at most two positive and at most two negative eigenvalues.
Finally, we introduce the measure~$\rho^\varepsilon$ on~$\F$ as the push-forward
of the volume measure on~$\scrM$
\[ 
\rho^\varepsilon := F^\varepsilon_* \mu \:. \]
We thus obtain a causal fermion system of spin dimension two.

For the purposes of this paper, it suffices to consider the simplest regularization,
the so-called $i \varepsilon$-regularization (for details see for example~\cite[\S2.4.1]{cfs}).
Its effect is seen most easily in the kernel of the fermionic projector~\eqref{Pxydef},
which can be computed to be the integral over the lower mass shell with a convergence-generating
factor~$e^{\varepsilon k^0}$, i.e.\
\[ P^\varepsilon(x,y) = \int \frac{d^4k}{(2 \pi)^4}\:(\slashed{k}+m)\: \delta(k^2-m^2)\: \Theta(-k^0)\: 
e^{\varepsilon k^0}\: e^{-ik(x-y)} \:. \]
For any~$\varepsilon>0$, the kernel~$P^\varepsilon(x,y)$ is a smooth function.
However, in the limit~$\varepsilon>0$, this kernel converges to a distribution which has singularities
on the light cone (i.e.\ if~$x$ and~$y$ have lightlike separation).
In a naive computation, these singularities give rise to divergent contributions to the causal action.
The continuum limit analysis gives a systematic method for studying the causal action principle
asymptotically for small~$\varepsilon$. In this paper, we shall not need the details.
Instead, we can make do with the scaling behavior in~$\varepsilon$ as captured by the
following simple formalism, which was introduced in~\cite{action}. The singularity of a composite expression
in the kernel of the fermionic projector (like the Lagrangian or variational derivatives thereof, the
spectral weight~$|xy|$, etc.) is described by the {\em{degree on the light cone}}.
The scaling behavior of a term of degree~$L$ on the light cone is given by
\beq \label{cont0}
\sim \frac{1}{(\varepsilon \,|\vec{\xi}|)^L} \qquad \text{if~$|\vec{\xi}| \gg \varepsilon$ and~$
\big| |t|-|\vec{\xi}| \big| \lesssim \varepsilon$}\:,
\eeq
where~$\xi := y-x$ with components~$\xi = (t, \vec{\xi})$ (thus~$t$ is the time distance between~$x$ and~$y$).
When integrating over spacetime, the $t$-integration can be carried out across the light cone,
compensating one factor of~$\varepsilon$. Thus the term~\eqref{cont0}
can be written as a distribution supported on the light cone of the form
\beq \label{cont1}
\simeq \frac{1}{\varepsilon^{L-1}\, |t|^L}\: \delta\big( |t| - |\vec{\xi}| \big)
\simeq \frac{1}{(\varepsilon\, |t|)^{L-1}}\: \delta\big( \xi^2\big) \:,
\eeq
where~$\simeq$ means ``up to a smooth prefactor''
(here~$\xi^2$ is again the Minkowski inner product).
The general {\em{formalism of the continuum limit}} (see~\cite[Chapter~4]{pfp} or~\cite[Section~2.4]{cfs})
gives a systematic method of computing the smooth prefactors in~\eqref{cont1} including
precise error terms.

\subsubsection{Connection to the Setting of Causal Variational Principles} \label{seccfscvp}
For the analysis of the causal action principle it is most convenient to get into the
simpler setting of {\em{causal variational principles}}. In this setting, $\F$ is a (possibly non-compact)
smooth manifold of dimension~$m \geq 1$ and~$\rho$ a positive Borel measure on~$\F$.
Moreover, we are given a non-negative function~$\L : \F \times \F \rightarrow \R^+_0$
(the {\em{Lagrangian}}) with the following properties:
\bitem
\item[(i)] $\L$ is symmetric: $\L(x,y) = \L(y,x)$ for all~$x,y \in \F$.\label{Cond1}
\item[(ii)] $\L$ is lower semi-continuous, i.e.\ for all \label{Cond2}
sequences~$x_n \rightarrow x$ and~$y_{n'} \rightarrow y$,
\[ \L(x,y) \leq \liminf_{n,n' \rightarrow \infty} \L(x_n, y_{n'})\:. \]
\eitem
The {\em{causal variational principle}} is to minimize the action
\beq \label{Sact} 
\Sact (\rho) = \int_\F d\rho(x) \int_\F d\rho(y)\: \L(x,y) 
\eeq
under variations of the measure~$\rho$, keeping the total volume~$\rho(\F)$ fixed
({\em{volume constraint}}).
If the total volume~$\rho(\F)$ is finite, one minimizes~\eqref{Sact}
over all regular Borel measures with the same total volume.
If the total volume~$\rho(\F)$ is infinite, however, it is not obvious how to implement the volume constraint,
making it necessary to proceed as follows.
We need the following additional assumptions:
\bitem
\item[(iii)] The measure~$\rho$ is {\em{locally finite}}
(meaning that any~$x \in \F$ has an open neighborhood~$U$ with~$\rho(U)< \infty$).\label{Cond3}
\item[(iv)] The function~$\L(x,.)$ is $\rho$-integrable for all~$x \in \F$, giving
a lower semi-continuous and bounded function on~$\F$. \label{Cond4}
\eitem
Given a regular Borel measure~$\rho$ on~$\F$, we then vary over all
regular Borel measures~$\tilde{\rho}$ with
\[ 
\big| \tilde{\rho} - \rho \big|(\F) < \infty \qquad \text{and} \qquad
\big( \tilde{\rho} - \rho \big) (\F) = 0 \]
(where~$|.|$ denotes the total variation of a measure).
These variations of the causal action are well-defined.
The existence theory for minimizers is developed in~\cite{noncompact}.

There are several ways to get from the causal action principle to causal variational principles,
as we now recall. If the Hilbert space~$\H$ is {\em{finite-dimensional}} and the total volume~$\rho(\F)$
is finite, one can proceed as follows:
As a consequence of the trace constraint~\eqref{trconstraint}, for any minimizing measure~$\rho$
the local trace is constant in spacetime, i.e.\
there is a real constant~$c \neq 0$ such that (see~\cite[Theorem~1.3]{lagrange} or~\cite[Proposition~1.4.1]{cfs})
\[ 
\tr x = c \qquad \text{for all~$x \in M$} \:. \]
Restricting attention to operators with fixed trace, the trace constraint~\eqref{trconstraint}
is equivalent to the volume constraint~\eqref{volconstraint} and may be disregarded.
The boundedness constraint, on the other hand, can be treated with a Lagrange multiplier.
More precisely, in~\cite[Theorem~1.3]{lagrange} it is shown that for every minimizing measure~$\rho$, 
there is a Lagrange multiplier~$\kappa>0$ such that~$\rho$ is a critical point of the causal action
with the Lagrangian replaced by
\[ 
\L_\kappa(x,y) := \L(x,y) + \kappa\, |xy|^2 \:, \]
leaving out the boundedness constraint.
Having treated the constraints, the difference to causal variational principles is that
in the setting of causal fermion systems, the set of operators~$\F \subset \Lin(\H)$ does not necessarily have
the structure of a manifold.
In order to give this set a manifold structure,
we need to assume that a given minimizing measure~$\rho$ (for the Lagrangian~$\L_\kappa$)
is {\em{regular}} in the sense that all operators in its support
have exactly~$n$ positive and exactly~$n$ negative eigenvalues.
This leads us to introduce the set~$\F^\reg$ 
as the set of all operators~$F$ on~$\H$ with the following properties:
\begin{itemize}[leftmargin=2em]
\item[(i)] $F$ is symmetric, has finite rank and (counting multiplicities) has
exactly~$n$ positive and~$n$ negative eigenvalues. \\[-0.8em]
\item[(ii)] The trace is constant, i.e
\[ 
\tr(F) = c>0 \:. \]
\end{itemize}
The set~$\F^\reg$ has a smooth manifold structure
(see the concept of a flag manifold in~\cite{helgason} or the detailed construction
in~\cite[Section~3]{gaugefix}). In this way, the causal action principle becomes an
example of a causal variational principle.

This finite-dimensional setting has the drawback that the total volume~$\rho(\F)$ of spacetime
is finite, which is not suitable for describing asymptotically flat spacetimes or spacetimes of
infinite lifetime like Minkowski space. Therefore, it is important to also consider
the {\em{infinite-dimensional setting}} where~$\dim \H=\infty$ and consequently also~$\rho(\F) = \infty$ 
(see~\cite[Exercise~1.3]{cfs}). In this case, the set~$\F^\reg$
has the structure of an infinite-dimensional Banach manifold (for details see~\cite{banach}).
Here we shall not enter the subtleties of this infinite-dimensional analysis.
Instead, we get by with the following simple method: Given a minimizing measure~$\rho$,
we choose~$\F^\reg$ as a finite-dimensional manifold which contains~$M:=\supp \rho$.
We then restrict attention to variations of~$\rho$ in the class of regular Borel measures on~$\F^\reg$.
In this way, we again get into the setting of causal variational principles.
We refer to this method by saying that we {\em{restrict attention to locally compact variations}}.
Keeping in mind that the dimension of~$\F^\reg$ can be chosen arbitrarily large, this
method seems like a sensible technical simplification.

For ease in notation, in what follows we will omit the superscript ``$\reg$.'' Thus~$\F$ stands
for a smooth (in general non-compact) manifold which contains the support~$M$ of the minimizing
measure~$\rho$.

\subsubsection{The Euler-Lagrange Equations and Jet Spaces} \label{secEL}
A minimizer of a causal variational principle
satisfies the following {\em{Euler-Lagrange (EL) equations}}.
For a suitable value of the parameter~$\s>0$,
the lower semi-continuous function~$\ell : \F \rightarrow \R_0^+$ defined by
\[ 
\ell(x) := \int_M \L(x,y)\: d\rho(y) - \s \]
is minimal and vanishes on spacetime~$M:= \supp \rho$,
\beq \label{EL}
\ell|_M \equiv \inf_\F \ell = 0 \:.
\eeq
The parameter~$\s$ can be understood as the Lagrange parameter
corresponding to the volume constraint. For the derivation and further details we refer to~\cite[Section~2]{jet}.

The EL equations~\eqref{EL} are nonlocal in the sense that
they make a statement on the function~$\ell$ even for points~$x \in \F$ which
are far away from the spacetime~$M$.
It turns out that for the applications we have in mind, it is preferable to
evaluate the EL equations only locally in a neighborhood of~$M$.
This leads to the {\em{restricted EL equations}} introduced in~\cite[Section~4]{jet}.
Here we give a slightly less general version of these equations which
is sufficient for our purposes. In order to explain how the restricted EL equations come about,
we begin with the simplified situation in which the function~$\ell$ is smooth.
In this case, the minimality of~$\ell$ implies that the derivative of~$\ell$
vanishes on~$M$, i.e.\
\beq \label{ELrestricted}
\ell|_M \equiv 0 \qquad \text{and} \qquad D \ell|_M \equiv 0
\eeq
(where~$D \ell(p) : T_p \F \rightarrow \R$ is the derivative).
In order to combine these two equations in a compact form,
it is convenient to consider a pair~$\u := (a, \bu)$
consisting of a real-valued function~$a$ on~$M$ and a vector field~$\bu$
on~$T\F$ along~$M$, and to denote the combination of 
multiplication and directional derivative by
\beq \label{Djet}
\nabla_{\u} \ell(x) := a(x)\, \ell(x) + \big(D_\bu \ell \big)(x) \:.
\eeq
Then the equations~\eqref{ELrestricted} imply that~$\nabla_{\u} \ell(x)$
vanishes for all~$x \in M$.
The pair~$\u=(a,\bu)$ is referred to as a {\em{jet}}.

In the general lower-continuous setting, one must be careful because
the directional derivative~$D_\bu \ell$ in~\eqref{Djet} need not exist.
Our method for dealing with this problem is to restrict attention to vector fields
for which the directional derivative is well-defined.
Moreover, we must specify the regularity assumptions on~$a$ and~$u$.
To begin with, we always assume that~$a$ and~$\bu$ are {\em{smooth}} in the sense that they
have a smooth extension to the manifold~$\F$ (for more details see~\cite[Section~2.2]{fockbosonic}).
Thus the jet~$\u$ should be
an element of the jet space
\[ \J_\rho := \big\{ \u = (a,\bu) \text{ with } a \in C^\infty(M, \R) \text{ and } \bu \in \Gamma(M, T\F) \big\} \:, \]
where~$C^\infty(M, \R)$ and~$\Gamma(M,T\F)$ denote the space of real-valued functions and vector fields
on~$M$, respectively, which admit a smooth extension to~$\F$.

Clearly, the fact that a jet~$\u$ is smooth does not imply that the functions~$\ell$
or~$\L$ are differentiable in the direction of~$\u$. This must be ensured by additional
conditions which are satisfied by suitable subspaces of~$\J_\rho$
which we now introduce.
First, we let~$\Gdiff_\rho$ be those vector fields for which the
directional derivative of the function~$\ell$ exists,
\[ \Gdiff_\rho = \big\{ \bu \in C^\infty(M, T\F) \;\big|\; \text{$D_{\bu} \ell(x)$ exists for all~$x \in M$} \big\} \:. \]
This gives rise to the jet space
\[ \Jdiff_\rho := C^\infty(M, \R) \oplus \Gdiff_\rho \;\subset\; \J_\rho \:. \]
For the jets in~$\Jdiff_\rho$, the combination of multiplication and directional derivative
in~\eqref{Djet} is well-defined. 
We choose a linear subspace~$\Jtest_\rho \subset \Jdiff_\rho$ with the property
that its scalar and vector components are both vector spaces,
\[ \Jtest_\rho = \Ctest(M, \R) \oplus \Gtest_\rho \;\subseteq\; \Jdiff_\rho \:, \]
and the scalar component is nowhere trivial in the sense that
\beq \label{Cnontriv}
\text{for all~$x \in M$ there is~$a \in \Ctest(M, \R)$ with~$a(x) \neq 0$}\:.
\eeq
Moreover, compactly supported jets are always denoted by a subscript zero, like for example
\[ 
\Jtest_{\rho,0} := \{ \u \in \Jtest_\rho \:|\: \text{$\u$ has compact support} \} \:. \]
Then the {\em{restricted EL equations}} read (for details cf.~\cite[(eq.~(4.10)]{jet})
\beq \label{ELtest}
\nabla_{\u} \ell|_M = 0 \qquad \text{for all~$\u \in \Jtest_\rho$}\:.
\eeq
Before going on, we point out that the restricted EL equations~\eqref{ELtest}
do not hold only for minimizers, but also for critical points of
the causal action. With this in mind, all the methods and results of this paper 
do not apply only to
minimizers, but more generally to critical points of the causal variational principle.
For brevity, we also refer to a measure which satisfies the restricted EL equations~\eqref{ELtest}
as a {\em{critical measure}}.

\subsubsection{Minkowski-Type Spacetimes} \label{secminktype}
For technical simplicity, in this paper we restrict attention to spacetimes which are smooth and
have the same topology as Minkowski space. This will make it possible to work with
the usual coordinates~$(t, \vec{x})$ and corresponding foliations by hypersurfaces~$t = \text{const}$.
Let~$(\H, \F, \rho)$ be a causal fermion system, which may be thought of as describing either the vacuum
or the interacting physical system. We assume that~$\rho$ is a critical point of the
causal action principle. Moreover, we assume that
the corresponding spacetime~$M:= \supp \rho$ is diffeomorphic to a four-dimensional spacetime
with trivial topology, i.e.\
\[ M \simeq \scrM := \R^4 \:. \]
Next, we assume that, using this identification, the measure~$\rho$ is absolutely continuous with respect
to the Lebesgue measure with a smooth weight function, i.e.\
\beq \label{rhoh}
d\rho = h(x)\: d^4x \qquad \text{with} \quad h \in C^\infty(\scrM, \R^+) \:.
\eeq
We also assume that~$h$ is bounded from above and below, i.e.\ there should be a constant~$C>1$ with
\[ 
\frac{1}{C} \leq h(x) \leq C \qquad \text{for all~$x \in \scrM$} \:. \]

We also denote the coordinate~$x^0$ as time function~$\scrt$,
\beq \label{scrtdef}
\scrt \::\: M \rightarrow \R\:,\qquad (t, \x) \mapsto t \:.
\eeq
For any~$t \in \R$, we let~$\Omega^t$ be the past of~$t$,
\beq \label{OmegaT}
\Omega^t := \{ x \in \scrM \:|\: \scrt(x) \leq t \} \:.
\eeq

\subsubsection{Surface Layer Integrals for Jets} \label{secosi}
Surface layer integrals were first introduced in~\cite{noether}
as double integrals of the general form
\beq \label{osi}
\int_\Omega \bigg( \int_{M \setminus \Omega} (\cdots)\: \L_\kappa(x,y)\: d\rho(y) \bigg)\, d\rho(x) \:,
\eeq
where $(\cdots)$ stands for a suitable
differential operator formed of jets.
A surface layer integral generalizes the concept of a surface integral over~$\partial \Omega$
to the setting of causal fermion systems.
The connection can be understood most easily in the
case when~$\L_\kappa(x,y)$ vanishes
unless~$x$ and~$y$ are close together. In this case, we only get a contribution to~\eqref{osi}
if both~$x$ and~$y$ are close to the boundary of~$\Omega$.
A more detailed explanation of the idea of a surface layer integral is given in~\cite[Section~2.3]{noether}.

In the present paper, we always choose the set~$\Omega$ according to~\eqref{OmegaT} as the past of a time~$t$.
We now recall those surface layer integrals for jets which will be of relevance in this paper.
\begin{Def} \label{defosi} We define the following surface layer integrals,
\begin{align}
\gamma^t_\rho \::\: \J_{\rho, \sc} &\rightarrow \R \qquad \text{(conserved one-form)} \notag \\
\gamma^t_\rho(\v) &= \int_{\Omega} d\rho(x) \int_{M \setminus \Omega} d\rho(y)\:
\big( \nabla_{1,\v} - \nabla_{2,\v} \big) \L(x,y) \label{gamma} \\
\sigma^t_\rho \::\: \J_{\rho, \sc} \times \J_{\rho, \sc} &\rightarrow \R  \qquad \text{(symplectic form)} \notag \\
\sigma^t_\rho(\u, \v) &= \int_{\Omega} d\rho(x) \int_{M \setminus \Omega} d\rho(y)\:
\big( \nabla_{1,\u} \nabla_{2,\v} - \nabla_{2,\u} \nabla_{1,\v} \big) \L(x,y) \label{sigma} \\
(.,.)^t_\rho \::\: \J_{\rho, \sc} \times \J_{\rho, \sc} &\rightarrow \R \qquad \text{(surface layer inner product)} \notag \\
(\u, \v)^t_\rho &= \int_{\Omega} d\rho(x) \int_{M \setminus \Omega} d\rho(y)\:
\big( \nabla_{1,\u} \nabla_{1,\v} - \nabla_{2,\u} \nabla_{2,\v} \big) \L(x,y) \:. \label{sprod}
\end{align}
\end{Def} \noindent
Here~$\J_{\rho, \sc}$ denotes the jets in~$\Jvary_\rho$ with spatially compact support
(for details see~\cite[Section~5.3]{linhyp}),
where~$\Jvary_\rho$ is a suitably chosen subspace of~$\Jtest_\rho$
(for details see~\cite[Section~3.2]{linhyp}).

\subsubsection{The Linearized Field Equations} \label{seclinfield}
In simple terms, the linearized field equations
describe variations of the measure~$\rho$ which preserve the EL equations.
More precisely, we consider variations where we multiply~$\rho$ by a non-negative
function and take the push-forward with respect to a mapping from~$M$ to~$\F$.
Thus we consider families of measures~$(\tilde{\rho}_\tau)_{\tau \in (-\delta, \delta)}$ 
of the form
\[ 
\tilde{\rho}_\tau = (F_\tau)_* \big( f_\tau \, \rho \big) \:, \]
where the~$f_\tau$ and~$F_\tau$ are smooth,
\[ f_\tau \in C^\infty\big(M, \R^+ \big) \qquad \text{and} \qquad
F_\tau \in C^\infty\big(M, \F \big) \:, \]
depend smoothly on the parameter~$\tau$
and have the properties~$f_0(x)=1$ and~$F_0(x) = x$ for all~$x \in M$
(moreover, the star denotes the push-forward measure, which is defined
for a subset~$\Omega \subset \F$ by~$((F_\tau)_*\mu)(\Omega)
= \mu ( F_\tau^{-1} (\Omega))$; see for example~\cite[Section~3.6]{bogachev}).
We assume that the measures~$(\tilde{\rho}_\tau)_{\tau \in (-\delta, \delta)}$ satisfy
the EL equations~\eqref{EL} for all~$\tau$.
Then the infinitesimal generator of the variation denoted by
\[ 
\v(x) := \frac{d}{d\tau} \big( f_\tau(x), F_\tau(x) \big) \Big|_{\tau=0} \:, \]
satisfies the {\em{linearized field equations}}
\[ 
0 = \la \u, \Delta \v \ra(x) := 
\nabla_\u \bigg( \int_M \big( \nabla_{1, \v} + \nabla_{2, \v} \big) \L_\kappa(x,y)\: d\rho(y) - \nabla_\v \,\s \bigg) \:, \]
which hold for all~$\u \in \Jtest_\rho$ and all~$x \in M$
(for details see~\cite[Section~3.3]{perturb}).
We denote the vector space of all solutions of the linearized field equations by~$\Jlin_\rho$.

The linearized field equations harmonize with the structure of surface layer integrals
in the sense that linearized solutions give rise to the following {\em{conservation laws}} or
almost conserved surface layer integrals (for details see~\cite{noether, jet, osi} or~\cite[Chapter~9]{intro}):
\bitem
\item[{\rm{(a)}}] The conserved one-form is time independent if~$\v$ is a linearized solution with vanishing scalar
component.
\item[{\rm{(b)}}] The symplectic form is independent of~$t$ for all linearized solutions~$\u$ and~$\v$.
\item[{\rm{(c)}}] The surface layer inner product is conserved up to quadratic corrections to the linearized field equations.
\eitem
We remark that the quadratic corrections will not be of relevance in this paper because
we shall consider the linearized fields only in the non-interacting vacuum spacetime.

\subsubsection{Inner Solutions, Arranging Jets without Scalar Components} \label{seczeroscalar}
We now briefly recall the definition of inner solutions as introduced in~\cite[Section~3]{fockbosonic}.

\begin{Def} \label{definner} An {\bf{inner solution}} is a jet~$\v$ of the form
\[ \v = (\div \bv, \bv) \qquad \text{with} \qquad \bv \in \Gamma(M, TM) \:, \]
where the divergence is taken with respect to the measure in~\eqref{rhoh},
\[ \div \bv := \frac{1}{h}\: \partial_j \big( h\, \bv^j \big) \:. \]
\end{Def}
Under suitable regularity and decay assumptions, an inner solution solves the
linearized field equations (for details see~\cite[Section~3.1]{fockbosonic}).
We denote these inner solutions by~$\Jin_\rho$. In~\cite[Proposition~3.6]{fockbosonic}
it is shown that inner solutions can be used for testing. With this in mind, we always assume that
\[ 
\Jin_\rho \subset \Jtest_\rho \:. \]

Inner solutions can be regarded as infinitesimal generators of transformations of~$M$
which leave the measure~$\rho$ unchanged. Therefore, inner solutions do not change the
causal fermion system, but merely describe symmetry transformations of the measure.
With this in mind, we can modify solutions of the linearized field equations by adding
inner solutions. For conceptual clarity, it is preferable that the diffeomorphism
generated by an inner solution does not change the global time function~$T$ in~\eqref{scrtdef}.
Therefore, we consider vector fields which are tangential to the
hypersurfaces~$N_t := T^{-1}(t)$. As shown in~\cite[Lemma~2.7]{pmt}, the divergence of
such a vector field can be arranged to be any given function~$a \in C^\infty(M, \R)$.
Thus, by adding the corresponding inner solutions we can achieve that all linearized solutions
have no scalar components. Therefore, in what follows we may restrict attention to linearized
solutions~$\v \in \Jlin$ with vanishing scalar component. We also write these jets as
\[ \v = (0, \bv) \qquad \text{with} \qquad \bv \in \Glin_\rho \:. \]

\subsubsection{The Dynamical Wave Equation and the Extended Hilbert Space} \label{secdynwave}
The restricted EL equations~\eqref{ELtest} can be expressed in terms of the physical
wave equations. This gives rise to the {\em{dynamical wave equation}} as introduced in~\cite{dirac}.
The solutions of this wave equation form a Hilbert space, the {\em{extended Hilbert space}}.
We now recall a few concepts and results from~\cite{dirac}.
Our starting point is the formula~\eqref{Fid} which expresses the spacetime point operator
as a local correlation operator. Varying the wave evaluation operator gives a vector field~$\bu$ on~$\F$
along~$M$,
\beq \label{ufermi}
\bu(x) = -\delta \Psi(x)^*\, \Psi(x) - \Psi(x)^*\, \delta \Psi(x) \:.
\eeq
In order to make mathematical sense of this formula in agreement with the concept of restricting
attention to locally compact variations, we choose a finite-dimensional subspace~$\H^\fermi \subset \H$, i.e.\
\beq \label{Hfermidef}
f^\fermi := \dim \H^\fermi  < \infty
\eeq
and impose the following assumptions on~$\delta \Psi$
(similar variations were first considered in~\cite[Section~7]{perturb}):
\begin{itemize}[leftmargin=2em]
\item[\rm{(a)}] 
The variation is trivial on the orthogonal complement of~$\H^\fermi$,
\[ \delta \Psi |_{(\H^\fermi)^\perp} = 0 \:. \]
\item[\rm{(b)}] The variations of all physical wave functions are continuous and compactly supported, i.e.
\[ \delta \Psi : \H \rightarrow C^0_0(M, SM) \:. \]
\end{itemize}
Before going on, we point out that the choice of~$\H^\fermi$ is {\em{not canonical}}.
Ultimately, one would like to exhaust~$\H$ by a sequence of finite-dimensional subspaces~$\H^\fermi_1 \subset
\H^\fermi_2 \subset \cdots$ and take the limit. Here we shall not enter this analysis, but 
for technical simplicity we rather choose~$\H^\fermi$ as a finite-dimensional subspace of sufficiently large dimension.

We choose~$\Gfermi_{\rho,0}$ as a space of vector fields of the form~\eqref{ufermi}.
For convenience, we identify the vector field with the first variation~$\delta \Psi$ and
write~$\delta \Psi \in \Gfermi_{\rho,0}$ (this representation of~$\bu$ in terms of~$\delta \Psi$
may not be unique, but this is of no relevance for what follows).
Choosing trivial scalar components, we obtain a corresponding space of jets~$\Jfermi_{\rho, 0}$,
referred to as the {\em{fermionic jets}}. We always assume that the fermionic jets are
admissible for testing, i.e.\
\[ \J^\fermi_{\rho,0} := \{0\} \oplus \Gfermi_{\rho,0} \;\subset\; \Jtest_{\rho,0} \:. \]
Moreover, in analogy to the condition~\eqref{Cnontriv} for the scalar components of the test jets,
we assume that the variation can have arbitrary values at any spacetime point, i.e.\
\[ 
\text{for all~$x \in M, \chi \in S_x$ and~$\phi \in \H^\fermi$ there is~$\delta \Psi \in \Gfermi_{\rho,0}$ with~$\delta \Psi(x)\, \phi = \chi$}\:. \]

Evaluating first variations of the causal action in direction of fermionic jets, one
obtains the {\em{EL equation for the physical wave functions}}
\beq \label{ELQ}
\int_M Q(x,y)\, \Psi^\fermi(y) \:d\rho(y) = \mathfrak{r}\, \Psi^\fermi(x) \qquad \text{for all~$x \in M$}\:,
\eeq
where~$\mathfrak{r} \in \R$ is the Lagrange parameter of the trace constraint,
$Q(x,y)$ is the kernel~\eqref{Qxydef} (with~$\L$ replaced by the $\kappa$-Lagrangian), 
and~$\Psi^\fermi:=\Psi|_{\H^\fermi}$ denotes the restriction of the wave evaluation operator
to the finite-dimensional subspace~$\H^\fermi$.
More details on this formulation of the EL equations
and many computations can be found in~\cite[Sections~1.4 and~2.6 as well as Chapters~3-5]{cfs}.
One way of looking at these equations is to regard~\eqref{ELQ} for given~$Q(x,y)$
as a linear equation describing
the dynamics of the physical wave functions. In~\cite{dirac}, this wave equation
was extended to more general solutions which form the so-called 
{\em{extended Hilbert space}}~$(\H^{\fermi,t}_\rho, \la .|. \ra^t_\rho)$.
The dynamics in~$\H^{\fermi,t}_\rho$ is described by the so-called {\em{dynamical wave equation}}
\[ 
Q^\dyn \psi = 0 \:, \]
where~$Q^\dyn$ is an integral operator with a symmetric integral kernel
\[ Q^\dyn(x,y) \,:\, S_y \rightarrow S_x \:. \]
The scalar product~$\la .|. \ra^t_\rho$ at time~$t$ has the form
\begin{align*} 
\la \psi | \phi \ra^t_\rho = -2i \,\bigg( \int_{\Omega^t} \!d\rho(x) \int_{M \setminus \Omega^t} \!\!\!\!\!\!\!d\rho(y) 
&- \int_{M \setminus \Omega^t} \!\!\!\!\!\!\!d\rho(x) \int_{\Omega^t} \!d\rho(y) \bigg)\\
&\times
\Sl \psi(x) \:|\: Q^\dyn(x,y)\, \phi(y) \Sr_x \:.
\end{align*}
This scalar product is conserved, i.e.\ time independent. With this in mind, we can drop to upper index~$t$
and denote the extended Hilbert space simply by~$(\H^{\fermi}_\rho, \la .|. \ra_\rho)$.

\subsubsection{A Conserved Nonlinear Surface Layer Integral} \label{secosinonlin}
\begin{figure}[tb]
\psscalebox{1.0 1.0} 
{
\begin{pspicture}(0,26.0)(9.62,29.2)
\definecolor{colour0}{rgb}{0.7019608,0.7019608,0.7019608}
\definecolor{colour1}{rgb}{0.9019608,0.9019608,0.9019608}
\psframe[linecolor=colour0, linewidth=0.04, fillstyle=solid,fillcolor=colour0, dimen=outer](9.61,27.6)(6.01,26.0)
\psframe[linecolor=colour1, linewidth=0.04, fillstyle=solid,fillcolor=colour1, dimen=outer](9.61,29.2)(6.01,27.6)
\psframe[linecolor=colour0, linewidth=0.04, fillstyle=solid,fillcolor=colour0, dimen=outer](3.61,27.6)(0.01,26.0)
\psframe[linecolor=colour1, linewidth=0.04, fillstyle=solid,fillcolor=colour1, dimen=outer](3.61,29.2)(0.01,27.6)
\psline[linecolor=black, linewidth=0.04](0.01,27.6)(3.61,27.6)
\psline[linecolor=black, linewidth=0.04](6.01,27.6)(9.61,27.6)
\psline[linecolor=black, linewidth=0.04, arrowsize=0.05291667cm 2.0,arrowlength=1.4,arrowinset=0.0]{<->}(3.41,28.6)(6.41,26.6)
\psline[linecolor=black, linewidth=0.04, arrowsize=0.05291667cm 2.0,arrowlength=1.4,arrowinset=0.0]{<->}(2.81,26.8)(6.81,28.4)
\rput[bl](7.3,28){$M \setminus \Omega^t$}
\rput[bl](7.7,26.8){$\Omega^t$}
\rput[bl](1.3,28){$\tilde{M} \setminus \tilde{\Omega}^t$}
\rput[bl](1.7,26.8){$\tilde{\Omega}^t$}
\rput[bl](2.55,26.7){$x$}
\rput[bl](3.15,28.5){$x$}
\rput[bl](6.5,26.5){$y$}
\rput[bl](6.9,28.3){$y$}
\end{pspicture}
}
\caption{The nonlinear surface layer integral.}
\label{figosinl}
\end{figure}
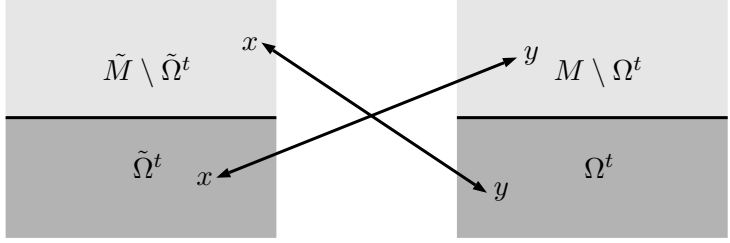

In~\cite[Section~4]{fockbosonic} a nonlinear surface layer integral~$\gamma^t(\rho, \tilde{\rho})$
was introduced which can be used for comparing the measure~$\tilde{\rho}$ describing the interacting system
with the vacuum measure~$\rho$. As the starting point, we let~$\rho$ and~$\tilde{\rho}$ be
two critical measures  on~$\F$, which describe the
vacuum and the interacting system, respectively. In order to relate the interacting
spacetime~$\tilde{M} := \supp \tilde{\rho}$ with the vacuum spacetime, we choose a mapping
\[ \Phi : \tilde{M} \rightarrow M \:, \]
which we assume to be measurable (in the sense that~$\Phi^{-1}(U)$ is $\tilde{\rho}$-measurable
for every $\rho$-measurable set~$U \subset M$).
Choosing a foliation~$(N_t)_{t \in \R}$ of~$M$, the past sets~$\Omega^t \subset M$ 
give rise to corresponding past sets~$\tilde{\Omega}^t := \Phi^{-1}(\Omega_t) \subset \tilde{M}$.
Then the nonlinear surface layer integral at time~$t$ is defined by
(see~\cite[Definition~4.1]{fockbosonic} and Figure~\ref{figosinl})
\beq \label{osinl}
\gamma^t(\tilde{\rho}, \rho) =
\int_{\tilde{\Omega}^t} d\tilde{\rho}(x) \int_{M \setminus \Omega^t} d\rho(y)\: \L(x,y)
- \int_{\tilde{M} \setminus \tilde{\Omega}^t}  d\tilde{\rho}(x) \int_{\Omega^t} d\rho(y) \: \L(x,y) \:.
\eeq
In order to ensure that the nonlinear surface layer integral is well-defined and finite, we
need to make the following assumption.
\begin{Def} \label{defsla}
The measures~$\rho$ and~$\tilde{\rho}$ are {\bf{surface layer admissible}} if for all~$t \in \R$,
both integrals in~\eqref{osinl} are finite,
\[ \int_{\tilde{\Omega}^t} d\tilde{\rho}(x) \int_{M \setminus \Omega^t} d\rho(y)\: \L(x,y) < \infty\:,\qquad
 \int_{\tilde{M} \setminus \tilde{\Omega}^t}  d\tilde{\rho}(x) \int_{\Omega^t} d\rho(y)\: \L(x,y) < \infty \:. \]
\end{Def}

By a suitable choice of the mapping~$\Phi : \tilde{M} \rightarrow M$
one can arrange that the nonlinear surface layer integral is conserved.
In order to construct~$\Phi$, we introduce a measure~$\nu$ on~$M$ and
a measure~$\tilde{\nu}$ on~$\tilde{M}:=F(M)$ 
(the so-called {\em{correlation measures}}) by
\begin{align*}
d\nu(x) &:= \bigg( \int_{\tilde{M}} \L(x,y)\: d\tilde{\rho}(y) \bigg) \: d\rho(x) \qquad \text{and} \qquad
d\tilde{\nu}(x) := \bigg( \int_M \L(x,y)\: d\rho(y) \bigg) \: d\tilde{\rho}(x) \:.
\end{align*}
As shown in~\cite[Appendix~A]{fockbosonic},
the conservation law is related to properties of the correlation measures:
\begin{Prp} The surface layer integral~\eqref{osinl} vanishes for every compact~$\Omega \subset M$
if and only if~$\nu = \Phi_* \tilde{\nu}$.
\end{Prp} \noindent
As worked out in detail again in~\cite[Appendix~A]{fockbosonic}, the existence of such a
diffeomorphism~$\Phi$ follows from a general result in~\cite{greene-shiohama}.

\subsection{Notions of Quantum States of Causal Fermion Systems} \label{secstate}
We now recall the definition of the quantum state and various refinements
of the construction as given in~\cite{fockfermionic}. Our presentation is very brief
and is limited to listing all the objects and definitions.
We will come back to these notions in more detail in Section~\ref{secqintro},
where we will discuss them with regard to entanglement and explain
how the quantum states can be computed.

\subsubsection{The Field Algebra of the Vacuum Spacetime} \label{secalgebra}
We let~$(\H, \F, \rho)$ be the causal fermion system describing the vacuum.
We now introduce the algebra of observables, denoted by~$\A$.
It is the $*$-algebra generated by the bosonic and fermionic field operators
defined as follows.
The {\em{fermionic creation and annihilation operators}} are denoted by
\[ \Psi^\dagger(\phi) \quad \text{and} \quad \Psi(\overline{\phi}) \qquad \text{for} \qquad
\phi \in \H^\fermi_{\rho, \sc} \:. \]
They satisfy the canonical anti-commutation relations
\[ 
\big\{ \Psi(\overline{\phi}), \Psi^\dagger(\phi') \big\} = \la  \phi | \phi' \ra^t_\rho \:, \]
and all other operators anti-commute,
\[ \big\{ \Psi(\overline{\phi}), \Psi(\overline{\phi'}) \big\} = 0 = \big\{ \Psi^\dagger(\phi), \Psi^\dagger(\phi') \big\} \:. \]

The definition of the {\em{bosonic field operators}} involves
the choice of a {\em{complex structure}}. We always work with the canonical
complex structure~$J$ is induced on~$\Glin_{\rho, \sc}$ by the surface layer integrals~$(.,.)^t_\rho$
and~$\sigma^t_\rho$ (see~\eqref{sprod} and~\eqref{sigma}; in the considered vacuum spacetime
these surface layer integrals are both time independent).
The construction, which is carried out in detail in~\cite[Section~6.3]{fockbosonic}, is summarized
as follows. We assume that the surface layer integral~$(.,.)^t_\rho$ restricted to~$\Glin_{\rho, \sc}
\times \Glin_{\rho, \sc}$ is positive semi-definite. Dividing out the null space and forming the completion,
we obtain a real Hilbert space denoted by~$(\h^\R, (.,.)^t_\rho)$.
For the construction of~$J$, one assumes that~$\sigma^t_\rho$ is bounded relative to the scalar product~$(.,.)^t_\rho$. Then we can represent~$\sigma^t_\rho$ as
\[ 
\sigma^t_\rho(u, v) = (u,\, \mathscr{T}\, v)^t_\rho \:, \]
where~$\mathscr{T}$ is a uniquely determined bounded operator on the Hilbert space~$\h^\R$.
Assuming that~$\mathscr{T}$ is invertible, we set
\[ 
J := -(-\mathscr{T}^2)^{-\frac{1}{2}}\: \mathscr{T} \:. \]
Next, we complexify the Hilbert space~$\h^\R$ and denote
its complexification by~$\h^\C$. On this complexification, the operator~$J$ has eigenvalues~$i$
and~$-i$. The corresponding eigenspaces are referred to as the {\em{holomorphic}} and
{\em{anti-holomorphic subspaces}}, respectively.
We write the decomposition into holomorphic and anti-holomorphic components as
\[ 
\bv = \bv^\hol + \bv^\ah \:. \]
We also complexify the symplectic form to a {\em{sesquilinear}}
form on~$\h^\C$ (i.e.\ anti-linear in its first and linear in its second argument).
On the holomorphic jets we introduce a scalar product~$(.|.)^t_\rho$ by
\[ (.|.)^t_\rho := \sigma^t_\rho( \,.\,, J \,.\, ) \::\: \Gamma^\hol_\rho \times \Gamma^\hol_\rho \rightarrow \C \:. \]
Taking the completion gives a Hilbert space, which we denote by~$(\h, (.|.)^t_\rho)$.
This scalar product has the useful property that
\[ 
\im ( u| v)^t_\rho = \im \sigma^t_\rho(u, J v) = \re \sigma^t_\rho(u, v) \:. \]
The {\em{bosonic creation and annihilation operators}} are denoted by
\[ a^\dagger(z) \quad \text{and} \quad a(\overline{z}) \qquad \text{with} \qquad
z \in \h \]
(the overline in~$a(\overline{z})$ serves as a reminder that this operator is anti-linear).
They satisfy the canonical commutation relations
\beq \label{CCR}
\big[ a(\overline{z}), a^\dagger(z') \big] = ( z | z' )^t_\rho \:,
\eeq
and all other operators commute,
\[ \big[ a(\overline{z}), a(\overline{z'}) \big] =  0 = \big[ a^\dagger(z), a^\dagger(z') \big]\:. \]

We define~$\A$ as the unital $*$-algebra generated by the above field operators.

\subsubsection{The Partition Function} \label{secZ}
Our starting point consists of two causal fermion systems~$(\H, \F, \rho)$ (describing the vacuum)
and~$(\tilde{\H}, \tilde{\F}, \tilde{\rho})$ (describing the interacting system).
We assume that both measures~$\rho$ and~$\tilde{\rho}$ are critical points of the causal action.
Moreover, we assume that both spacetimes~$M$ and~$\tilde{M}$ are of Minkowski type (see Section~\ref{secminktype}) with given time functions~$T$ and~$\tilde{T}$.

The idea for constructing a quantum state is ``compare''
these two systems at a given fixed time and to try to describe the interacting system in terms of linearized fields and
wave functions in the vacuum spacetime. This ``comparison'' can also be understood as a ``measurement''
performed in the interacting spacetime using objects from the vacuum spacetime as ``measurement devices.''
In more technical terms, we want to work with the nonlinear surface layer integral
(see Section~\ref{secosinonlin}) and variations thereof. 
When working with causal fermion systems, however,
there is the complication that the two causal fermion systems
are defined on two different Hilbert spaces~$\H$ and~$\tilde{\H}$.
Therefore, in order to make sense of the nonlinear surface layer integral, we need to identify the
Hilbert spaces~$\H$ and~$\tilde{\H}$ by a unitary transformation denoted by~$V$,
\[ 
V : \H \rightarrow \tilde{\H} \qquad \text{unitary} \:. \]
Then operators in~$\tilde{F}$ can be identified with operators in~$\F$ by the unitary transformation,
\[ 
\F = V^{-1}\, \tilde{\F}\, V \:. \]
An important point to keep in mind is that this identification is not canonical, but that it leaves the freedom
to transform the operator~$V$ according to
\[ 
V \rightarrow V \scrU \qquad \text{with} \qquad \scrU \in \Lin(\H) \text{ unitary} \:. \]
For ease of notation, in what follows we always identify~$\H$ and~$\tilde{\H}$ via~$V$,
making it possible to always work in the Hilbert space~$\H$. Then the non-uniqueness of the identification
still shows up in the unitary transformation of the vacuum measure
\beq \label{rhotrans}
\rho \mapsto \scrU \rho \:,
\eeq
where~$\scrU \rho$ is defined by
\beq \label{Urhodef}
(\scrU \rho)(\Omega) := \rho \big( \scrU^{-1} \,\Omega\, \scrU \big) 
\qquad \text{for} \qquad \Omega \subset \F \:.
\eeq

The method for treating the unitary freedom~\eqref{rhotrans} is to integrate over
the transformation~$\scrU$. This leads to the definition of the partition function,
obtained by integrating the exponential of the nonlinear surface layer integral.

\begin{Def} \label{defZ}
The {\bf{partition function}} is defined by
\beq \label{Zdef}
Z^t\big( \beta, \tilde{\rho} \big) = \fint_\G \exp \Big( \beta \,\gamma^t \big(\tilde{\rho}, \scrU \rho \big) \Big) \: d\mu_\G(\scrU) \:,
\eeq
where~$\G$ is chosen as a compact Lie subgroup of the unitary group,
\[ \G \subset \U(\H) \quad \text{compact} \:. \]
\end{Def} \noindent
Here~$\gamma^t(\tilde{\rho}, \scrU \rho \big))$ is the nonlinear surface layer integral~\eqref{osinl}
for the unitarily transformed measure~\eqref{Urhodef}, i.e.\
\beq \label{osinlU}
\begin{split}
\gamma^t(\tilde{\rho}, \scrU \rho) &=
\int_{\tilde{\Omega}^t} d\tilde{\rho}(x) \int_{M \setminus \Omega^t} d\rho(y)\: 
\L\big(x, \scrU y \scrU^{-1} \big) \\
&\quad\: - \int_{\tilde{M} \setminus \tilde{\Omega}^t}  d\tilde{\rho}(x) \int_{\Omega^t} d\rho(x) \: 
\L\big(x, \scrU y \scrU^{-1} \big) \:.
\end{split}
\eeq
It is one of the main goals of the present paper to compute integrals over the unitary
group as in~\eqref{Zdef}.

We finally comment on the choice of the group~$\G$. The most natural choice is
the whole unitary group~$\G = \U(\H)$. But clearly, this choice is possible only if the
Hilbert space~$\H$ is finite-dimensional. For this reason, in~\cite{fockfermionic}
we chose~$\G = \U( \H^\fermi)$, where~$\H^\fermi$ is the finite-dimensional subspace
of~$\H$ already introduced in the context of the dynamical wave equation~\eqref{Hfermidef}.
In this paper, it will be preferable to choose~$\H$ as a finite-dimensional Hilbert space
and~$\G=\U(\H)$ as the whole unitary group; this will be discussed and explained in detail
in Section~\ref{seclocdetail}.

\subsubsection{The Quantum State of a Causal Fermion System} \label{secstateglobal}
The general idea behind the construction of the quantum state is to probe the
interacting spacetime with the objects of the non-interacting spacetime.
More specifically, it is a linear mapping from the algebra~$\A$
(the field algebra of the vacuum spacetime; see Section~\ref{secalgebra}) to the complex
numbers which is positive, i.e.\
\[ \omega : \A \rightarrow \C \qquad \text{with} \qquad
\omega(A^* A) \geq 0 \quad \text{for all~$A \in \A$}\:. \]
It has the general structure
\beq \label{omegatdef}
\omega^t(A) = 
\frac{1}{Z^t\big( \beta, \tilde{\rho} \big)}
\int_\G \big( \cdots \big) \:e^{\beta\, \gamma^t(\tilde{\rho}, \scrU \rho)} \: d\mu_\G(\scrU) \:,
\eeq
where~$(\cdots)$ stands for insertions formed of surface layer integrals which
depend on~$A$. In more detail,
\begin{align}
&\omega^t\Big( a^\dagger(z'_1) \cdots a^\dagger(z'_p)\:\Psi^\dagger(\phi'_1) \cdots \Psi^\dagger(\phi'_{r'}) \;
a(\overline{z_1}) \cdots a(\overline{z_q}) \: \Psi(\overline{\phi_1}) \cdots \Psi(\overline{\phi_r}) \Big) \notag \\
&:= \frac{1}{Z^t\big( \beta, \tilde{\rho} \big)} \: \delta_{r' r}\:\frac{1}{r!} \sum_{\sigma, \sigma' \in S_{r}}
(-1)^{\sign(\sigma)+\sign(\sigma')} \notag \\
&\quad\;\: \times
\int_\G e^{\beta\, \gamma^t(\tilde{\rho}, \scrU \rho)}\;
\la \phi_{\sigma(1)} \,|\, \pi^{\fermi, t}\, \phi'_{\sigma'(1)} \ra^t_\rho
\cdots \la \phi_{\sigma(r)} \,|\, \pi^{\fermi, t} \, \phi'_{\sigma'(r)} \ra^t_\rho  \notag \\
&\qquad\quad \times 
D_{z'_1} \gamma^t(\tilde{\rho}, \scrU \rho) \cdots D_{z'_p} \gamma^t(\tilde{\rho}, \scrU \rho)\;
D_{\overline{z}_1} \gamma^t(\tilde{\rho}, \scrU \rho) \cdots D_{\overline{z}_q} \gamma^t(\tilde{\rho}, \scrU \rho) \; d\mu_\G(\scrU) \:, \label{insertions}
\end{align}
where the bosonic insertions are derivatives of the nonlinear surface layer integral,
and~$\pi^{f,t}$ is a projection operator which, in non-technical terms, tells us how the
physical wave functions of the interacting spacetime look like in the non-interacting spacetime
(for details see~\cite[Section~4.3]{fockfermionic} and the discussion in Remark~\ref{remcompare}).

\subsubsection{The Refined Pre-State} \label{secrefined}
The {\em{refined pre-state}} was introduced in~\cite[Section~5.3]{fockfermionic}
with the intention of getting more detailed phase information on the interacting measure.
We now briefly recall the definition. In Section~\ref{secnoentangle}, we will come back to the
refined pre-state and explain why it is needed in order to account for the physical
effect of entanglement. The basic idea is to work, instead of~$\scrU$, with two unitary transformations~$\scrU_<$ and~$\scrU_>$,
and to replace the unitarily transformed measure~$\scrU \rho$ in~\eqref{Urhodef} by
the measure~$T_{\scrU_<, \scrU_>} \rho$ defined by
\[ \big( T_{\scrU_<, \scrU_>} \rho \big)(\Omega) := \rho \big( \scrU_>^{-1} \,\Omega\, \scrU_< \big) \]
(note that this measure may be supported on non-symmetric operators).
For the refined partition function~$Z^t_{\text{\rm{ref}}}$ and the
refined pre-state~$\omega^t_{\text{\rm{ref}}}$ one modifies~\eqref{Zdef} and~\eqref{omegatdef}
by working with a double integral over the unitary group~$\G$, i.e.\ symbolically
\begin{align}
Z^t_{\text{\rm{ref}}}\big( \beta, \tilde{\rho} \big) &:= \fint_\G d\mu_\G \big(\scrU_< \big)  \fint_\G d\mu_\G \big( \scrU_> \big) \: e^{\beta \,\gamma^t \big(\tilde{\rho}, T_{\scrU_<, \scrU_>}  \rho \big)} \label{Zrefined} \\
\omega^t_{\text{\rm{ref}}} \big( \cdots \big) &:= \frac{1}{Z^t\big( \beta, \tilde{\rho} \big)} 
\fint_\G d\mu_\G \big(\scrU_< \big)  \fint_\G d\mu_\G \big( \cdots \big) \;e^{\beta \,\gamma^t \big(\tilde{\rho}, T_{\scrU_<, \scrU_>}  \rho \big)}  \:. \label{omegarefined}
\end{align}
Similar to~\eqref{insertions}, the insertions are formed of suitable surface layer integral.
We postpone the details to the thorough construction and discussion in Section~\ref{secinsert}.

We finally remark that by a {\em{pre-state}} we mean a linear functional from~$\A$ to~$\C$,
which does not necessarily need to be positive. Therefore, before applying the refined pre-state
in physical applications, we need to show positivity (see Section~\ref{secpositive}).

\subsubsection{The Localized Refined Pre-State} \label{seclocalized}
The quantum state defined in Section~\ref{secstateglobal} was defined globally in space.
For most applications, however, it seems more suitable to define a state describing only a {\em{bounded
spatial region}}, which can be thought of as our laboratory or the subsystem of our present universe
accessible to measurements. This picture will also be very helpful for the constructions in the present paper,
because it will make it possible to take the limit when the dimension of the group~$\G$ tends to
infinity, while leaving our laboratory unchanged (for more details see Section~\ref{seclocalize} below).
In order to implement this picture, we choose a subset~$V \subset M$
of the vacuum spacetime such that its intersection with the surface layer contains the spatial region of interest
(see Figure~\ref{figlocstate}).
Correspondingly, the set~$\tilde{V} := \Phi^{-1}(V) \subset \tilde{M}$ is the region of the
interacting spacetime region which is probed by the laboratory.
We define the {\em{localized nonlinear surface layer integral}}~$\gamma^{t,V}$ by
restricting all the integrals in the nonlinear surface layer integral~\eqref{osinlU} to~$V$
respectively~$\tilde{V}$,
\beq \label{osinlU2}
\gamma^t_V(\tilde{\rho}, \rho) =
\int_{\tilde{\Omega}^t \cap \tilde{V}} d\tilde{\rho}(x) \int_{V \setminus \Omega^t} d\rho(y)\: 
\L(x,y) - \int_{\tilde{V} \setminus \tilde{\Omega}^t}  d\tilde{\rho}(x) \int_{\Omega^t \cap V} d\rho(y) \: 
\L(x,y) \:,
\eeq
and similarly for the unitarily transformed measure as used in the refined state,
\begin{align*} 
\gamma^t_V(\tilde{\rho}, T_{\scrU_<, \scrU_>}  \rho) &=
\int_{\tilde{\Omega}^t \cap \tilde{V}} d\tilde{\rho}(x) \int_{V \setminus \Omega^t} d\rho(y)\: 
\L\big( x, \scrU_> \,y\, \scrU_<^{-1} \big) \\
&\quad\: - \int_{\tilde{V} \setminus \tilde{\Omega}^t}  d\tilde{\rho}(y) \int_{\Omega^t \cap V} d\rho(x) \: 
\L\big( x, \scrU_> \,y\, \scrU_<^{-1} \big) \:.
\end{align*}
Next, we replace all surface layer integrals in~\eqref{Zrefined}
and~\eqref{omegarefined} by the corresponding localized surface layer integrals.
After doing so, however, one must take into account that the unitary group~$\G$
typically includes the unitary transformations describing the symmetries of Minkowski space
like time translations and spatial translations.
In order to remove the freedom of performing translations in Minkowski
space, it is convenient to insert the factor~$e^{\alpha \T^t_V}(\tilde{\rho}, \scrU \rho)$
 into the integrand of the partition function and the state, where~$\T^t_V(\tilde{\rho}, \scrU \rho)$ involves
two spacetime integrals,
\begin{align}
&\T^t_V( \tilde{\rho}, T_{\scrU_<, \scrU_>} \rho) \notag \\
&:= \bigg( \int_{\tilde{\Omega}^t \cap \tilde{V}}\!\! d\tilde{\rho}(x) \int_{\Omega^t \cap V} \!\!d\rho(y)
+ \int_{(\tilde{M} \setminus \tilde{\Omega^t}) \cap \tilde{V}} \!\!\!\!\!d\tilde{\rho}(x)
\int_{(M \setminus \Omega^t) \cap V} \!\!\!\!\!d\rho(y) \bigg) \big| x\, \scrU_> \,y\, \scrU_<^{-1} \big|^2 \:.
\label{Tgen}
\end{align}
(here the absolute value is again the spectral weight~\eqref{sw}). Thus the
{\em{refined localized partition function}}~$Z_V$ and the {\em{refined localized state}}~$\omega^t_V$ are defined
similar to~\eqref{Zrefined} and~\eqref{omegarefined} by
\begin{align}
 Z^t_V\big( \alpha, \beta, \tilde{\rho} \big) &:= 
 \fint_\G d\mu_\G \big(\scrU_< \big)  \fint_\G d\mu_\G \big( \scrU_> \big) \:
 e^{\alpha \T^t_V \big(\tilde{\rho}, T_{\scrU_<, \scrU_>}  \rho \big) + \beta \gamma^t_V \big(\tilde{\rho}, T_{\scrU_<, \scrU_>}  \rho \big)} \label{Zrefloc} \\
\omega^t_V \big( \cdots \big)
&:= \frac{1}{Z^t_V\big( \alpha, \beta, \tilde{\rho} \big)} \fint_\G d\mu_\G \big(\scrU_< \big)  \fint_\G d\mu_\G \big( \scrU_> \big) \notag \\
&\qquad\qquad\qquad\qquad \times \:
e^{\alpha \T^t_V \big(\tilde{\rho}, T_{\scrU_<, \scrU_>}  \rho \big) + \beta \gamma^t_V \big(\tilde{\rho}, T_{\scrU_<, \scrU_>}  \rho \big)} \: \big( \cdots \big) \:, \label{omegarefloc}
\end{align}
where we again abbreviated the insertions by~$(\cdots)$. 
The point is that, in contrast to~$\gamma^t_V$, the functional~$\T^t_V$ is {\em{not}}
a surface layer integral (due to the plus sign in~\eqref{Tgen}).
The general picture is that, choosing~$\alpha$
sufficiently large (and~$V$ as a set of finite measure), the main contribution to the group integral is obtained when
the functional~$\T^t_V$ is maximal. This maximum should be attained when
the sets~$\Omega^t \cap V$ and~$(M \setminus \Omega^t) \cap V$
coincide corresponding sets of the interacting spacetime. In this way, the freedom in
performing translations in Minkowski is fixed.
This general picture will be confirmed and worked out in detail in the present paper.

\subsection{Basics on Fock Spaces and Entanglement} \label{secfermifock}
In this section, we recall the basics on Fock spaces and explain the phenomenon of entanglement.

For the construction of the {\em{bosonic Fock space}}, we take {\em{symmetrized}} products of the
Hilbert space~$(\h, (.|.)^t_\rho)$ of complexified linearized solutions as
introduced in Section~\ref{secalgebra}.
We let~$\h^n = \h \otimes \cdots \otimes \h$ be the $k$-fold tensor product, endowed with
the natural scalar product
\[ 
( \psi_1 \otimes \cdots \otimes \psi_k \,|\, \phi_1 \otimes \cdots \otimes \phi_k )^t_\rho
:= ( \psi_1 | \phi_1 )^t_\rho \: \cdots\: ( \psi_k | \phi_k )^t_\rho \:. \]
We denote total symmetrization by an index~$\symm$, i.e.\
\[ 
\big( \psi_1 \otimes \cdots \otimes \psi_k \big)_\symm := \frac{1}{k!} \sum_{\sigma \in S_k}
\psi_{\sigma(1)} \otimes \cdots \otimes \psi_{\sigma(k)} \:, \]
where~$S_k$ denotes the group of all permutations.
These totally symmetrized products, referred to as {\em{product states}} or {\em{factorizable states}},
do in general not form a vector space. But they generate a linear subspace denoted by
\[ \Fock^\bose_{\rho, k} := \overline{(\h^k)_\symm } \:\subset\: \h^k \:. \]
The {\em{bosonic Fock space}}~$(\Fock^\bose_\rho, \la .|.\ra_{\Fock^\bose_\rho})$ is the direct sum of the $n$-particle spaces,
\[ \Fock^\bose_\rho = \bigoplus_{k=0}^\infty \Fock^\bose_{\rho, k} \:. \]

Likewise, the {\em{fermionic Fock space}} is constructed by taking totally {\em{anti-symmetric}} products
of the one-particle Hilbert space~$(\H^\fermi_\rho, \la .|. \ra_\rho)$.
We let~$(\H^\fermi_\rho)^k = \H^\fermi_\rho \otimes \cdots \otimes \H^\fermi_\rho$ be the $k$-fold tensor product, endowed with the natural scalar product
\[ 
\la \psi_1 \otimes \cdots \otimes \psi_k \,|\, \phi_1 \otimes \cdots \otimes \phi_k \ra_{\Fock^\fermi_\rho}
:= \la \psi_1 | \phi_1 \ra_\rho\: \cdots\: \la \psi_k | \phi_k \ra_\rho \:. \]
Totally anti-symmetrizing the tensor product gives the wedge product
\[ 
\psi_1 \wedge \cdots \wedge \psi_k := \frac{1}{k!} \sum_{\sigma \in S_k}
(-1)^{\sign(\sigma)}\: \psi_{\sigma(1)} \otimes \cdots \otimes \psi_{\sigma(k)} \]
(here~$S_k$ denotes the set of all permutations and~$\sign(\sigma)$ is the sign of the permutation~$\sigma$).
The wedge product gives rise to a mapping
\[ \Lambda_k \::\: \underbrace{\H^\fermi_\rho \times \ldots \times \H^\fermi_\rho}_{\text{$k$ factors}} \rightarrow (\H^\fermi_\rho)^k
\::\: (\psi_1, \ldots, \psi_k) \mapsto \psi_1 \wedge \cdots \wedge \psi_k \:. \]
The vectors in the image of the mapping~$\Lambda_k$ are the $k$-particle {\em{Hartree-Fock states}} or {\em{factorizable states}}.
Again, these states do in general not form a vector space.
We denote the vector space generated by the $k$-particle Hartree-Fock states by
\[ \Fock^\fermi_{\rho,k} := \overline{ \bra \Lambda_k \big( (\H^\fermi_\rho)^k \big) \ket } \:\subset\:
(\H^\fermi_\rho)^k \:. \]
Taking the direct sum of these spaces gives the {\em{fermionic Fock space}} denoted by
\[ \Fock^\fermi_\rho := \bigoplus_{k=0}^\infty \Fock_{\rho, k} \:. \]

Finally, the Fock space~$\Fock_\rho$ is defined as the tensor product of the bosonic and fermionic Fock spaces,
\[ \Fock_\rho := \Fock^\bose_\rho \otimes \Fock^\fermi_\rho \:. \]
It is an important observation that the factorizable states do {\em{not}} form a vector space.
A Fock vector which is not factorizable is said to be {\em{entangled}}.
The simplest and most prominent entangled state is the spin singlet state of two observers,
usually referred to as \textsc{Alice} and \textsc{Bob},
\beq \label{alicebob}
\Phi := \frac{1}{\sqrt{2}}\: \big( \phi^\uparrow_A \otimes \phi^\downarrow_B -
\phi^\downarrow_A \otimes \phi^\uparrow_B \big)_\symm\: ,
\eeq
where~$\uparrow$ and~$\downarrow$ denote the polarizations
(for ease of presentation of bosonic particles), and~$A$ and~$B$ refer to \textsc{Alice} and \textsc{Bob}, respectively.
Entanglement is a basic phenomenon of quantum physics, with important applications
to quantum information theory and quantum computing.
We refer the reader interested in the physical background to standard textbooks
like~\cite[Section~20.4]{schwabl1}, \cite[Section~2.2.8]{nielsen}, \cite[Section~3.10]{sakurai}
or~\cite{horodecki, afriat}.

\section{Quantum States of Causal Fermion Systems and Entanglement} \label{secqintro}
In~\cite{fockfermionic}, the interacting spacetime at time~$t$ was described by a
{\em{quantum state}}, being a positive linear functional on the field algebra~$\A$
of the non-interacting spacetime (see the preliminaries in Sections~\ref{secalgebra}--\ref{secstateglobal}).
Moreover, various refinements of the constructions were given
(in particular the {\em{refined pre-state}} and the {\em{localized refined pre-state}};
see the preliminaries in Sections~\ref{secrefined} and~\ref{seclocalized}).
In this section, we reconsider these notions with the aim of determining whether
they are capable of describing entanglement.
Moreover, we shall set up the problem such as to obtain a clear mathematical setting
which can be analyzed explicitly in a concise limiting case.
In preparation, we begin with a few general considerations on the group integrals
which appear in the definition of the partition function and the state
(see for example~\eqref{Zdef} and~\eqref{Zrefined}). These considerations partly anticipate
results of the later analysis. They have the purpose of conveying the correct physical picture
in nontechnical terms.

\subsection{General Considerations on the Quantum State} \label{secgeneral}
In this section we give a few qualitative considerations on the structure of our
quantum state and its various refinements. We begin with the partition function~$Z^t$ defined in~\eqref{Zdef}.
The main task of the present paper is to unravel how the integrand behaves as a function of~$\scrU$.
In particular, we need to quantify the main contributions to the integral.
Let us now explain the qualitative picture for specific choices of~$\tilde{\rho}$.
In the case~$\tilde{\rho}=\rho$ of the Minkowski vacuum, we expect that the integrand is largest
if we choose~$\scrU=\1$. In this case, the integrand can be computed rather explicitly
in the formalism of the continuum limit (as developed in~\cite{pfp, cfs};
see also Section~\ref{sechighenergy} below). In particular, a direct computation
shows that the surface layer integral~$\gamma^t$ in~\eqref{Zdef}
has the scaling~$\gamma^t \sim \varepsilon^{-4}$
and thus diverges in the limit~$\varepsilon \searrow 0$ when the ultraviolet regularization is removed.
This divergence can be understood from the fact that in the limit~$\varepsilon \searrow 0$ the
number of physical wave functions tends to infinity, and that the contributions by all these
wave functions add up. More specifically, the contributions by all the wave functions
to the kernel of the fermionic projector~$P(x,y)$ are ``in phase'' on the light cone
(i.e.\ if~$x$ and~$y$ are lightlike separated), giving rise to singularities as~$\varepsilon \searrow 0$
(here by are ``in phase'' we mean that all the summands have the same sign,
and no relative phases appear). This explains why the integrand in~\eqref{Zdef} 
is very large if~$\scrU=\1$.
For generic~$\scrU$, however, the matrix elements of~$\scrU$ involve many phases.
As a consequence, the contributions by the physical wave functions no longer add up,
but instead we get sums of terms involving oscillatory phase factors.
Similar to the effect of destructive interference, many summands will cancel each other.
As a result, the integrand in~\eqref{Zdef} will be very small.
Here the notions ``very large'' and ``very small'' will be made more precise in terms of a
scaling behavior in~$\varepsilon$ (this is worked out in Section~\ref{sechighenergy}).

We next consider the situation when~$\tilde{\rho}$ describes Minkowski space involving
a {\em{classical bosonic field}} (for example a plane electromagnetic wave).
In this case, the unitary operator~$\scrU$ can compensate for gauge phases,
which means that the maximum of the integrand in~\eqref{Zdef} will be attained for
a certain unitary operator~$\scrU$ for which the gauge phases are compensated in
the optimal way. Then the surface layer integral can be computed again in the formalism
of the continuum limit, giving a large contribution to the partition function.
Moreover, the insertions in the quantum state~\eqref{omegatdef} (as shown in more detail in~\eqref{insertions}) become surface layer integrals which involve the electromagnetic field.
In this way, the state~$\omega^t$ makes it possible to probe the interacting spacetime
using the free fields of the vacuum spacetime.

We now move on to the situation of a general minimizing measure~$\tilde{\rho}$.
This measure can have a very complicated structure. In particular, we cannot expect
that the integrand in~\eqref{Zdef} will have a single maximum.
Instead, the integrand will have many maxima, and the integral will give rise to
a sum over contributions near the maxima
(this picture will be made precise in Section~\ref{secrefinecompute}).
In simple terms, this means that~$\tilde{\rho}$ can no longer be described by a single
spacetime involving classical fields. Instead, it could be a ``mixture'' of different spacetimes,
possibly involving many different classical fields.
Even more, the measure~$\rho$ could have a structure which can no longer be described by classical fields
or a classical spacetime.
The main point is that, even in such very non-classical situations,
the quantum state~\eqref{omegatdef} makes it possible to probe the measure~$\tilde{\rho}$
in terms of familiar objects in Minkowski space.
It turns out that this probing gives rise to a quantum state in the sense familiar from the algebraic
formulation of {\em{quantum field theory}}.
In this way, a general measure~$\tilde{\rho}$ can be described in the language of quantum
field theory. This description does not only allow for the description of quantum fields,
but it even gives words like ``quantum spacetime'' and ``quantum geometry'' a well-defined meaning
in the realm of quantum field theory.

Exactly as explained for the vacuum state, for a generic unitary operator~$\scrU$
the integrand in~\eqref{Zdef} will typically be very small due to destructive interference.
In other words, the integrand should be close to zero, except on a set of very small measure,
where the integrand should be very large.
A general conclusion from these qualitative consideration is that we must face the difficulty
that the integrand in~\eqref{Zrefloc} (and similar in the various states and pre-states)
will have a highly nonlinear dependence on~$\scrU$. In particular, we cannot expect that
a power expansion in~$\scrU$ will be sufficient for our purposes. Instead, we need to develop
methods for evaluating group integrals in the non-perturbative regime.

Before entering a more detailed discussion of the state and its various refinements,
we comment on a question which the critical reader may ask:
{\em{Why is there a freedom in introducing the quantum state? Should the quantum state as a physical object
not be given canonically in a unique way?}}
In order to answer this question, we return to our motivation and our general procedure for introducing the
quantum state: We wanted to probe the interacting measure~$\tilde{\rho}$ using the objects of the
vacuum spacetime. The fact that the measure~$\tilde{\rho}$ can have a very complicated structure
can also be expressed by saying that this measure contains a lot of information.
Only a small portion of this information should be relevant for describing macroscopic phenomena
(for example, the measure~$\tilde{\rho}$ also encodes the detailed structure of spacetime on the
Planck scale, which cannot be tested in current experiments).
Therefore, our task is to extract the portion of the information on~$\tilde{\rho}$
which is relevant for describing quantum physics.
Thinking of the problem in this way, it is no longer surprising that there is not a canonical
way to retrieve this information. Similar to the usual situation in physics that a physical quantity
can be determined alternatively in different types of experiments, there may be many different ways
to retrieve the relevant information from~$\tilde{\rho}$.
We can make use of this freedom to our advantage. Similar to setting up an optimal experiment,
we can try to come up with a definition of a quantum state which has all desirable properties and is
easiest to analyze in detail. The following considerations can be regarded as consecutive
steps towards this goal.

\subsection{Why Entanglement Requires the Refined Pre-State} \label{secnoentangle}
Clearly, entanglement is a fundamental feature of quantum physics.
Therefore, any physically sensible notion of quantum state should be capable of describing 
entanglement. We now explain why this requirement makes it necessary to consider the
{\em{refined}} pre-state.

Our arguments are very general and do not make use of the
specific form of the interacting measure. Let us assume that we want to
realize an entangled Fock state, which can be written as a linear combination of product states.
For simplicity, we consider only a bosonic Fock vector,
\beq \label{Phiansatz}
\Phi = \sum_{\alpha=1}^K \Phi_\alpha \qquad \text{with} \qquad
\Phi_\alpha = \big( \phi_{1, \alpha} \otimes \cdots \otimes \phi_{p_\alpha, \alpha} \big)_\symm \:.
\eeq
Then the expectation value of an observable~$\O$ can be written as a double sum,
\[ \la \Phi | \O | \Phi \ra_{\Fock^\bose_\rho} = \sum_{\alpha, \beta=1}^K \la \Phi_\alpha | \O | \Phi_\beta \ra_{\Fock^\bose_\rho} \:. \]
As a typical example of a $p$-particle measurement
we choose~$\O$ as a product of bosonic field operators,
\[ \O = a^\dagger(z'^{t}_1) \cdots a^\dagger(z'^t_p)\;
a(\overline{z}^t_1) \cdots a(\overline{z}^t_q) \:. \]
In this case, the expectation value can be written as
\[ \la \Phi | \O | \Phi \ra_{\Fock^\bose_\rho} = \sum_{\alpha, \beta=1}^K \big\la 
a(\overline{z}'^{t}_1) \cdots a(\overline{z}'^t_p)\, \Phi_\alpha \,\big|\,
a(\overline{z}^t_1) \cdots a(\overline{z}^t_q) \,\Phi_\beta \big\ra_{\Fock^\bose_\rho} \:. \]
Using that the~$\Phi_\alpha$ are product states~\eqref{Phiansatz},
on the left one gets products of scalar products~$( z'^{t}_\ell | \phi_{\ell', \alpha})^t_\rho$
with~$\ell, \ell' \in \{1,\ldots, p\}$.
Similarly, on the right one gets the scalar products~$( z^{t}_\ell | \phi_{\ell', \beta})^t_\rho$.
The point is that all the factors on the left involve the same index~$\alpha$
(describing the {\em{bra}}), whereas all the
factors on the right involve the same index~$\beta$ (describing the {\em{ket}}).

Our goal is to realize this expectation value by associating to the observable~$\O$
suitable insertions in~\eqref{omegareflocfinal}. These insertions are complex-valued
functions which depend on the corresponding linearized solutions~$z'^{t}_\ell$ or~$\overline{z}^t_\ell$.
Moreover, the insertions clearly depend on the unitary operator~$\scrU$.
Denoting the insertions by~$b^t(z,\scrU)$, the state takes the form
\beq \label{statebos}
\begin{split}
&\omega^t\Big( a^\dagger(z'^{t}_1) \cdots a^\dagger(z'^t_p)\;
a(\overline{z}^t_1) \cdots a(\overline{z}^t_q) \Big) \\
&:= \frac{1}{Z^t\big( \beta, \tilde{\rho} \big)} \int_\G 
b^t(z'^t_1, \scrU) \cdots b^t(z'^t_p, \scrU) \: \overline{b^t(z^t_1, \scrU) \cdots b^t(z^t_q, \scrU)} \;e^{\beta\, \gamma^t(\tilde{\rho}, \scrU \rho)}
\; d\mu_\G(\scrU) \:.
\end{split}
\eeq
In order to encode the Fock components, we write every insertion as a sum of terms indexed by~$\alpha$.
This leads us to the general ansatz
\beq \label{sumdecomp}
b^t(z, \scrU) = \sum_{\alpha=1}^K b^t_\alpha(z, \scrU) \:.
\eeq
Ultimately, the insertions with index~$\alpha$ should encode the form of the $\alpha$-component
of the Fock vector~\eqref{Phiansatz}. 
At this stage, we do not need to specify what the $\alpha$-dependence of the insertions really means
(this will be made precise in Section~\ref{secsaddlecombi} and explained in Section~\ref{secentangle}).
Instead, for explaining the basic mechanism it is preferable to work with~\eqref{sumdecomp} symbolically.
Using this ansatz in~\eqref{statebos} and multiplying out, we obtain
\begin{align*}
\omega^t\Big( &a^\dagger(z'^{t}_1) \cdots a^\dagger(z'^t_p)\;
a(\overline{z}^t_1) \cdots a(\overline{z}^t_q) \Big) = 
\frac{1}{Z^t\big( \beta, \tilde{\rho} \big)} 
\sum_{\alpha_1, \ldots \alpha_p=1}^K \sum_{\beta_1, \ldots \beta_q=1}^K \\
& \times
\int_\G b^t_{\alpha_1}(z'^t_1, \scrU) \cdots b^t_{\alpha_p}(z'^t_p, \scrU) \:
\overline{b^t_{\beta_1}(z^t_1, \scrU) \cdots b^t_{\beta_q}(z^t_q, \scrU)} \;e^{\beta\, \gamma^t(\tilde{\rho}, \scrU \rho)} \; d\mu_\G(\scrU) \:.
\end{align*}
Here the insertions and the
exponential factor depend nonlinearly on the unitary operator~$\scrU$.
As a consequence, the integral in general does not split into a product of sums.
Instead, integrating over~$\scrU$ will give rise to correlations between the different insertions.
These correlations are crucial for getting a connection to the interacting many-particle picture
of quantum field theory. More specifically, in order to describe entangled states, we
would like that all the holomorphic insertions should be in the same component,
and similarly for the anti-holomorphic insertions.
Therefore, the state should be approximately of the form
\beq \label{stateapprox}
\begin{split}
\omega^t\Big( & a^\dagger(z'^{t}_1) \cdots a^\dagger(z'^t_p)\;
a(\overline{z}^t_1) \cdots a(\overline{z}^t_q) \Big) \approx \frac{1}{Z^t\big( \beta, \tilde{\rho} \big)} \sum_{\alpha, \beta=1}^K \\
&\times 
\int_\G b^t_{\alpha}(z'^t_1, \scrU) \cdots b^t_{\alpha}(z'^t_p, \scrU) \: \overline{b^t_{\beta}(z^t_1, \scrU) \cdots b^t_{\beta}(z^t_q, \scrU)} \;e^{\beta\, \gamma^t(\tilde{\rho}, \scrU \rho)} \; d\mu_\G(\scrU) \:.
\end{split}
\eeq
Following a notion introduced in~\cite{qft}, we also say that {\em{bra}} and {\em{ket}} should be {\em{synchronized}}.

In order to clarify this construction, we work it out in more detail for the spin singlet state
\begin{Example} {\em{ For the spin singlet state~\eqref{alicebob}, we have two components~$\alpha=1,2$ and
\[ \Phi_1 = \frac{1}{\sqrt{2}} \:\big(\phi^\uparrow_A \otimes \phi^\downarrow_B \big)_\symm
\qquad \text{and} \qquad
\Phi_2 = -\frac{1}{\sqrt{2}}\: \big( \phi^\downarrow_A \otimes \phi^\uparrow_B \big)_\symm \:. \]
\textsc{Alice} and \textsc{Bob}
choose one-particle measurement operators, which we write in {\em{bra/ket}}-notation as
\[ \O_A := | z_A' )( z_A| \qquad \text{and}\qquad \O_B := | z_B')(z_B| \:. \]
Then the two-particle measurement is described by the observable~$\O = \O_A \otimes \O_B$.
We thus obtain
\begin{align*}
&\la \Phi | \O | \Phi \ra_{\Fock^\bose_\rho} = \sum_{\alpha, \beta=1}^K
\big\la \Phi_\alpha \big| z'_A \otimes z'_B \big\ra\:
\big\la z_A \otimes z_B \big| \Phi_\alpha  \big\ra \\
&= \frac{1}{8} \Big( (\phi^\uparrow_A | z'_A)^t_\rho\: (\phi^\downarrow_B | z'_B)^t_\rho
\: (z_A | \phi^\uparrow_A )^t_\rho\: (z_B | \phi^\downarrow_B)^t_\rho 
- (\phi^\downarrow_A | z'_A)^t_\rho\: (\phi^\uparrow_B | z'_B)^t_\rho
\: (z_A | \phi^\uparrow_A )^t_\rho\: (z_B | \phi^\downarrow_B)^t_\rho \\
&\quad\;\: - (\phi^\uparrow_A | z'_A)^t_\rho\: (\phi^\downarrow_B | z'_B)^t_\rho
\: (z_A | \phi^\downarrow_A )^t_\rho\: (z_B | \phi^\uparrow_B)^t_\rho
+ (\phi^\downarrow_A | z'_A)^t_\rho\: (\phi^\uparrow_B | z'_B)^t_\rho
\: (z_A | \phi^\downarrow_A )^t_\rho\: (z_B | \phi^\uparrow_B)^t_\rho\Big) \:.
\end{align*}
If a synchronization is present, these expectation values can be 
realized by the quantum state~\eqref{stateapprox} if we choose~$p=q=2$ and
\[ z'_1 = z'_A \:,\quad z'_2 = z'_B \qquad \text{and} \qquad z_1 = z_A \:,\quad z_2 = z_B \:. \]
Moreover, we need to arrange that the insertions are the corresponding one-particle
expectation values, i.e.\
\begin{align*}
b^t_{1}(z'^t_A) &= c\,(\phi^\uparrow_A | z'_A)^t_\rho \:,& \hspace*{-2cm}
b^t_{1}(z'^t_B) &= c\, (\phi^\downarrow_B | z'_B)^t_\rho \\
b^t_{2}(z'^t_A) &= -c\, (\phi^\downarrow_A | z'_A)^t_\rho \:,&\hspace*{-2cm}
b^t_{2}(z'^t_B) &= \,c(\phi^\uparrow_B | z'_B)^t_\rho
\end{align*}
with~$c=8^{-\frac{1}{4}}$, and similarly without the primes.

Without synchronization, however, the many-particle measurement reduces to a product of
one-particle measurements. A direct computation (similar to the original Einstein-Podolsky-Rosen
argument) shows that
it is indeed impossible to realize the expectation values of the spin-singlet state.
}} \QEDrem
\end{Example}

In view of this example, the
basic question is whether~\eqref{stateapprox} can hold, i.e.\ whether 
by a suitable choice of the insertions we can arrange a synchronization of {\em{bra}} and {\em{ket}}. The answer is no, as the next lemma shows.
\begin{Lemma} \label{lemmacounter}
In the case~$p=q=2$ and choosing~$z'^t_\ell=z_\ell^t=z^t$, the bosonic state~\eqref{statebos}
satisfies the inequality
\begin{align*}
&\sum_{\alpha, \beta=1}^K
\bigg| \int_\G b^t_{\alpha}(z^t) \,b^t_{\alpha}(z^t) \: \overline{b^t_{\beta}(z^t)\, b^t_{\beta}(z^t)} \;e^{\beta\, \gamma^t(\tilde{\rho}, \scrU \rho)} \; d\mu_\G(\scrU) \bigg| \\
&\leq \sum_{\alpha, \beta=1}^K \int_\G b^t_{\alpha}(z^t) \,b^t_{\beta}(z^t) \: \overline{b^t_{\alpha}(z^t)\, b^t_{\beta}(z^t)} \;e^{\beta\, \gamma^t(\tilde{\rho}, \scrU \rho)} \; d\mu_\G(\scrU)\:.
\end{align*}
\end{Lemma}
\Proof Using that the exponential factor is positive, we obtain with the help of the Schwarz inequality
\begin{align*}
&\bigg| \int_\G b^t_{\alpha}(z^t) \,b^t_{\alpha}(z^t) \: \overline{b^t_{\beta}(z^t)\, b^t_{\beta}(z^t)} \;e^{\beta\, \gamma^t(\tilde{\rho}, \scrU \rho)} \; d\mu_\G(\scrU) \bigg| \\
&\leq \int_\G \big| b^t_{\alpha}(z^t) \big|^2 \: \big| b^t_{\beta}(z^t)\big|^2 \;e^{\beta\, \gamma^t(\tilde{\rho}, \scrU \rho)} \; d\mu_\G(\scrU) \:.
\end{align*}
Using that the anti-holomorphic insertion is the complex conjugate of the holomorphic insertion,
the result follows.
\QED

This lemma shows that, no matter how the insertions are chosen,
the main contribution to the state is {\em{not}} of the form~\eqref{stateapprox}, because
the contributions with synchronizations between {\em{bra}} and {\em{ket}} dominate.
We conclude that the state~$\omega^t$ is not capable of capturing essential features of the
quantum dynamics of the causal fermion system.
The {\em{refined}} pre-state~\eqref{omegarefined}, however, does {\em{not}} suffer
from this shortcoming. This will be worked out in detail in Section~\ref{secentangle}
later in this paper. In simple terms, the double integral in~\eqref{omegarefloc}
gives rise to separate phases for the holomorphic and anti-holomorphic insertions,
so that the counter argument of Lemma~\ref{lemmacounter} no longer applies.

\subsection{Why it is Preferable to Work with Localized States} \label{seclocalize}
As explained qualitatively in Section~\ref{secgeneral}, we need to compute
integrals over the unitary group for integrands with a highly nonlinear dependence on
the unitary operator~$\scrU$. In order to get into the position where such integrals
can be analyzed explicitly, it is very helpful to work with the {\em{localized}} states,
as we now explain. Since we already know from the previous section that we must
consider refined states, we consider the localized refined pre-state 
(see~\eqref{Zrefloc} and~\eqref{omegarefloc} in Section~\ref{seclocalized}).
In this setting, the pre-state is computed in a subset~$V$ of our spacetime.
This subset can be chosen arbitrarily large with the only constraint that it should be
small compared to the size of the whole spacetime.
Clearly, if the whole system is Minkowski space, this constraint is unproblematic because
as the whole spacetime we can consider an arbitrarily large box in Minkowski space
(more details and the relevant scalings will be given in Section~\ref{secrefinecompute} below).

Associating to~$V$ a subspace~$\H^\lab \subset \H$ of the whole Hilbert space,
the integrand of the group integrals depends only on the restriction of the unitary operator
to this subspace, i.e.\ on the operator
\[ \pi_{\H^\lab} \scrU \pi_{\H^\lab} \]
(where~$\pi_{\H^\lab} : \H \rightarrow \H^\lab$ denotes the orthogonal projection).
Therefore, we can carry out the integrals over all the other matrix elements of~$\scrU$.
Doing so simplifies the structure of the remaining integrals considerably
(for details see Theorem~\ref{thmgenint} in Section~\ref{secgenint}).
Moreover, we can consider the limiting case when the dimension of~$\H$ tends to infinity,
whereas the dimension of~$\H^\lab$ remains fixed.
This simplifies the formulas further, making it possible to analyze the group integrals in detail.

We point out that working with localized states is not preferable for principal or conceptual reasons,
but it only has the computational benefit that it becomes possible to evaluate the group integrals
in a mathematically precise limiting case.

\subsection{The Precise Form of the Localized Refined Pre-State} \label{seclocdetail}
We now make the setting of the localized refined pre-state more concrete. 
In order to describe the vacuum, we 
we let~$(\H, \F, \rho)$ be a causal fermion system constructed from a vacuum Dirac system
in a three-dimensional box of length~$2L$, i.e.\ in the spacetime cylinder
\beq \label{Mcylinder}
\scrM = \R \times [-L,L]^3 \subset \R^{1,3}
\eeq
with periodic boundary conditions (similar to the standard construction in Minkowski space
as given in~\cite[Section~1.2]{cfs} or in more detail in~\cite{oppio}).
Fixing the scale~$\varepsilon>0$ of the ultraviolet regularization, the Dirac sea is built up
of a finite number~$f$ of wave functions with the scaling
\beq \label{Hdim}
\dim \H = f \sim \Big( \frac{L}{\varepsilon} \Big)^3 \:.
\eeq
The set~$V$ is chosen as
\beq \label{Vbox}
V = [-T,T] \times [-\ell,\ell]^3 \qquad \text{with} \qquad \ell \ll L \:.
\eeq
Thus~$\ell$ can be thought of as the size of the physical system of interest.
Moreover, the parameter~$T$ is the time scale on which the measurement takes place.
The three-dimensional box~$[-\ell, \ell]^3$ can be associated to a subspace of~$\H$
consisting of all wave functions which at time~$t$ are localized inside the box
(except for exponentially decaying tails which arise due to the incompatibility of
spatial localization and frequency splitting; this effect will not be of relevance for our computations).
We denote this subspace by~$\H^\lab \subset \H$. Its dimension scales like
\beq \label{Hlabdim}
\dim \H^\lab \sim \Big( \frac{\ell}{\varepsilon} \Big)^3 \:.
\eeq
We are interested in the asymptotics~$L \rightarrow \infty$
and~$\varepsilon \searrow 0$ for fixed~$\ell$ and~$T$.

We choose the group~$\G$ as the unitary group of the whole Hilbert space,
\beq \label{Gtot}
\G = \U(\H) \:.
\eeq
This is different from~\cite{fockfermionic}, where we wanted to allow for the possibility
that the Hilbert space~$\H$ is infinite dimensional, making it necessary to choose~$\G$
as the unitary group of a finite-dimensional subspace~$\H^\fermi$.
In order to have a simple and clean setting, we here decided to choose~$\H$ to be
finite-dimensional~\eqref{Hdim}. In this setting, the choice~\eqref{Gtot} is most natural.
But clearly, the dimension of~$\H$ tends to infinity if we let~$L \rightarrow \infty$.

For the later computations, we set~$N=f=\dim \H$
and often identify~$\H$ with~$\C^N$. Apart from the subspace~$\H^\lab$,
we also need to consider the subspace~$\H^\fermi \subset \H$ of
{\em{low-energy wave functions}}. Here by ``low energy'' we mean that their energy
is much smaller than the Planck energy. In technical terms, the wave functions in~$\H^\fermi$
must satisfy the conditions needed for the construction of the extended Hilbert space~$\H^\fermi_\rho$
(see the preliminaries in Section~\ref{secdynwave}). For the constructions of the present paper,
we do not need to specify~$\H^\fermi$. We recall from~\eqref{Hfermidef} that its dimension 
is denoted by~$f^\fermi$. Finally, we need to decompose the Hilbert space of the laboratory
into low- and high-energy wave functions. To this end, we set
\[ \H^\le := \H^\fermi \cap \H^\lab \qquad \text{and} \qquad \H^\he := \big( \H^\fermi \big)^\perp \cap \H^\lab \:, \]
so that
\[ \H^\lab = \H^\le \oplus \H^\he \:. \]

Before developing the methods and entering the detailed analysis, we make two remarks.
First, in order to get into the regime where the integrand depends highly nonlinearly on~$\scrU$
(as described qualitatively in Section~\ref{secgeneral}), we need to insert a factor~$N$ into
the exponents in~\eqref{Zrefloc} and~\eqref{omegarefloc}. Thus, when letting the dimension of~$\H$
tend to infinity, we also increase the nonlinearity of the exponential.
The reason for the linear scaling in~$N$ will be explained in Example~\ref{exexp};
it will also become clear in a more general setting in the model examples in Section~\ref{secmodel}.
The second remark is that we set the parameter~$\beta$ equal to zero.
This leads us to the definition of the refined localized state~$\omega^t_V \big( \cdots \big)$
as stated in the introduction (see~\eqref{omegareflocfinal} and~\eqref{Zreflocfinal}).
The precise definition of the insertions requires more preparations and is therefore
postponed to Section~\ref{secinsert}.

We conclude this section by explaining how the simplification of setting~$\beta$ to zero
is to be understood.
In preparation, we recall how the exponential factors~$\exp(\beta \gamma^t_V)$
and~$\exp(\alpha \T^t_V)$ came about. The factor~$\exp(\beta \gamma^t)$ was introduced
in~\cite{fockbosonic} motivated by the fact that it coincides with the norm of the corresponding
bosonic Fock vector (see~\cite[Sections~7 and 8]{fockbosonic}).
Moreover, in~\cite{fockfermionic} this factor has the benefit that the insertions can be understood
as variational derivatives of the partition function (see~\cite[Section~4.6]{fockfermionic}).
The factor~$\exp(\alpha \T^t_V)$, on the other hand, was introduced in the context of localized
states in order to fix the freedom to perform translations in~$M$
(see~\cite[Section~5.3]{fockfermionic} or the preliminaries in Section~\ref{seclocalized}).
The reason why this freedom cannot be fixed with the factor~$\exp(\beta \gamma^t_V)$
is that the surface layer integral~$\gamma^t_V$ is not necessarily positive.
But~$\T^t_V$ is not the only positive functional. In particular, one
could replace the spectral weight~$| x\, \scrU_> y \scrU_<^{-1} |$ in~\eqref{Tgen} by the Lagrangian,
\beq \label{Tgen2}
\bigg( \int_{\tilde{\Omega}^t \cap \tilde{V}}\!\! d\tilde{\rho}(x) \int_{\Omega^t \cap V} \!\!d\rho(y)
+ \int_{(\tilde{M} \setminus \tilde{\Omega^t}) \cap \tilde{V}} \!\!\!d\tilde{\rho}(x)
\int_{(M \setminus \Omega^t) \cap V} \!\!\!d\rho(y) \bigg)\: \L\big(x, \scrU_> \,y\, \scrU_<^{-1} \big) \:.
\eeq
Working with the Lagrangian seems more natural because of the closer connection to the causal
action principle. However, the Lagrangian is more difficult to compute, because the summands of different signs
in~\eqref{Lagrange} may and typically will cancel each other, making it rather subtle to determine
the leading contributions. 
In the spectral weight~\eqref{sw}, on the other hand, all summands are
positive, so that it suffices to compute the leading contributions to the eigenvalues.
This is the reason why we here prefer to work with the functional~$\T^t_V$.
The last consideration also shows that~$\T^t_V$ will typically be much larger than the surface layer integral~$\gamma^t_V$
constructed out of the Lagrangian. This justifies why we can leave~$\gamma^t_V$ out of the exponent
by setting~$\beta=0$.

To summarize, at present the detailed form of the functional in the exponent of the localized refined quantum state
is not completely fixed. We here choose the functional which is most suitable for the computations.
The differences to other possible choices are mainly technical and should not leave the basic mechanisms
and results of this paper unchanged.

\section{Computation of Integrals over the Unitary Group} \label{secintunit}
\subsection{Preliminary Considerations}
In general terms, our task is to analyze integrals over the unitary group~$\U(N)$
asymptotically as~$N \rightarrow \infty$.
In order to motivate our methods and to explain how they come about,
we now begin with a few elementary considerations.
We always denote the unitary group by~$\G=\U(N)$
and let~$\mu_\G$ be the normalized Haar measure on~$\G$.
For notational clarity, integrals with respect to such a normalized measure will always be denoted by~$\fint$.
The simplest integrals are obtained by choosing the integrand as a monomial in the
matrix entries and their complex conjugates, i.e.\
\beq \label{gint0}
\fint_{\G} \scrU^{i_1}_{j_1} \cdots \scrU^{i_p}_{j_p} \;
\overline{\scrU^{k_1}_{l_1} \cdots \scrU^{k_q}_{l_q} } \: d\mu_\G(\scrU)
\eeq
with $p,q \in \N_0$ and arbitrary indices in~$\{1, \ldots, N\}$.
The easiest method is to make use of the invariance of the Haar measure under transformations
\beq \label{haarinv}
\scrU \rightarrow V \scrU W \qquad \text{with} \qquad V,W \in \G\:.
\eeq
In particular, choosing~$V=e^{i \varphi}$ as a phase factor and~$W=\1$, one sees that the group integral~\eqref{gint0}
vanishes unless in the case~$p=q$ where the number of conjugated and unconjugated factors coincides.
Therefore, the simplest non-trivial integral is obtained in the case of two factors by
\beq \label{gint2}
\fint_{\G} \big| \scrU^i_j \big|^2 \: d\mu_\G(\scrU)
\eeq
for given indices~$i,j$. Applying the invariance~\eqref{haarinv} again for~$W=1$ and
a unitary operator~$V$ which
flips the components~$i$ and~$k$, i.e.\
\[ V^a_b = \delta^a_b - \delta^{ai}\, \delta_{bi} - \delta^{ak}\, \delta_{bk} 
+ \delta^{ai}\, \delta_{bk} + \delta^{ak}\, \delta_{bi} \:, \]
one sees that the integral~\eqref{gint2} is the same for every choice of~$i$. In particular,
\[ \fint_{\G} \big| \scrU^i_j \big|^2 \: d\mu_\G(\scrU) = \frac{1}{N}
\sum_{k=1}^N \fint_{\G} \big| \scrU^k_j \big|^2 \: d\mu_\G(\scrU) \:. \]
The components~$\scrU^1_j, \ldots, \scrU^N_j$ form the $j^\text{th}$ row of the matrix~$\scrU$.
For any unitary matrix, this row vector is a unit vector, i.e.\
\[ \sum_{k=1}^N \big| \scrU^k_j \big|^2 = 1 \:. \]
We thus obtain
\[ \fint_{\G} \big| \scrU^i_j \big|^2 \: d\mu_\G(\scrU) =
\frac{1}{N} \fint_{\G} d\mu_\G(\scrU) = \frac{1}{N}\:. \]
More generally, the above method yields
\[ \fint_{\G} \scrU^i_j\: \big( \scrU^{-1} \big)^l_k\: d\mu_\G(\scrU) =
\fint_{\G} \scrU^i_j\: \overline{\scrU^k_l} \: d\mu_\G(\scrU) =
\frac{1}{N} \: \delta^i_k\, \delta^l_j \:. \]

We now move on to the integral~\eqref{gint0} with four unitary factors.
Using the unitary invariance~\eqref{haarinv}, we get zero unless~$p=q=2$ and
each row and column index appears twice, once with and once without conjugation.
Thus, after commuting the factors, it suffices to consider the integral
\[ \fint_{\G} \scrU^{i_1}_{j_1} \:\scrU^{i_2}_{j_2} \;
\overline{\scrU^{k_1}_{j_1}}\: \overline{\scrU^{k_2}_{j_2} } \: d\mu_\G(\scrU) \]
with either~$i_1=k_1, i_2=k_2$ or~$i_1=k_2, i_2=k_1$.
The corresponding row vectors
\[ u^i := \scrU^i_{j_1} \qquad \text{and} \qquad v^i := \scrU^i_{j_2} \]
are again unit vectors. Moreover, in the case~$j_1 \neq j_2$ they are orthogonal.
This orthogonality condition makes the computation more difficult.
But the following qualitative argument suggests that the situation should simplify
asymptotically for large~$N$: The scalar product
\beq \label{sprodN}
\la u, v \ra_{\C^N} = \sum_{i=1}^n \overline{u^i}\: v^i
\eeq
involves a sum over the components. Each summand can be positive or negative,
so that in the sum there may be cancellations of terms with different signs.
These cancellation will be more efficient if~$N$ is large.
In other words, taking randomly chosen unit vectors~$u$ and~$v$, their scalar product~\eqref{sprodN}
typically becomes very small if~$N$ gets large.
Asymptotically, the integral~\eqref{sprodN} should factor into two integrals over~$u$ and~$v$,
both over the unit sphere in~$\C^N$.

This qualitative argument can be made precise, giving rise to the following result.
\begin{Lemma} \label{lemmagroup}
Given~$p, N \in \N$, we let~$\G=\U(N)$ be the unitary group with normalized Haar measure~$d\mu_\G$. Then, asymptotically for large~$N$,
\begin{align}
&\fint_{\G} \scrU^{i_1}_{j_1} \cdots \scrU^{i_p}_{j_p} \: \big( \scrU^{-1} \big)^{k_1}_{l_1} \cdots
\big( \scrU^{-1} \big)^{k_p}_{l_p} \:d\mu_\G(\scrU) \notag \\
&= \frac{1}{N^p} \sum_{\sigma \in S(p)}
\delta^{i_1}_{l_{\sigma(1)}} \delta^{k_{\sigma(1)}}_{j_1} \cdots \delta^{i_p}_{l_{\sigma(p)}} \delta^{k_{\sigma(p)}}_{j_p}
+ \O \bigg( \frac{1}{N^{p+1}} \bigg) \:, \label{groupint}
\end{align}
where~$S(p)$ denotes the symmetric group.
\end{Lemma} \noindent
This result can be understood in analogy to Gaussian integrals or, similarly,
to Wick's rule for free bosonic quantum fields:
Integrating over the unitary group gives rise to pairings of the unitary matrices,
and we need to sum over all possible pairings.
Each pair involves a factor~$\scrU$ and a factor~$\scrU^{-1}$. A pairing consists in replacing the two matrices
by two Kronecker deltas which connect the indices of~$\scrU$ and~$\scrU^{-1}$ crosswise.
We refer to asymptotic formulas of type~\eqref{groupint} as the {\em{Gaussian asymptotics}}.
There are various ways to prove the above lemma.
One method is to analyze properties of tensor representations of the unitary group,
as inspired by Michael Creutz~\cite{creutz}.
In order not to distract from the main ideas, the resulting proof of Lemma~\ref{lemmagroup} 
is given in Appendix~\ref{appA}.
Another method for proving formulas similar to~\eqref{groupint} is to get a direct connection to Gaussian integrals;
this will be explained in Section~\ref{secgauss} below.
Other methods can be found in~\cite{weingarten, collinsb, ipsen-kieburg}.

It is an important observation that, four our purposes, the Gaussian asymptotics is {\em{not}} sufficient.
Namely, as already described qualitatively in Section~\ref{secgeneral}, we have the 
situation in mind where the integrand is very small, except on a set of very small measure,
where the integrand is very large. Such a ``spiky landscape''
is in conflict with the Gaussian asymptotics, 
where we are only allowed to evaluate polynomials, and where the
factorization via Wick rules means that the components of the
unitary matrices are correlated only weakly in the sense that every pairing gives a small factor~$1/N$.

In order to evaluate group integrals in the strongly correlated regime,
we make essential use of the fact that the {\em{localized}} quantum state only depends
on the physical wave functions inside the laboratory as described by the subsets~$V \subset M$
and~$\tilde{V} \subset \tilde{M}$ (see~\eqref{osinlU2}).
More technically, we may choose a subspace~$I \subset \C^N$ of dimension~$q$ and write the
group integral as
\beq \label{fintro}
\fint_{\G} f(N, A)\: d\mu_\G(\scrU) \qquad \text{with} \qquad A := \pi_I \scrU \pi_I
\eeq
(and~$\pi_I : \C^N \rightarrow I$ is the orthogonal projection). Now we can take the limit
\[ N \rightarrow \infty \qquad \text{for fixed~$q = \dim I$} \:. \]
In other words, we let the whole system get larger and larger while keeping the size of our laboratory
unchanged. In view of the size of the whole universe compared to the size of a typical measurement
apparatus, this limiting case seems physically sensible for most applications in mind.
It turns out that, in this limiting case, the group integral becomes tractable, even in the case
that the function~$f$ in~\eqref{fintro} is highly nonlinear and cannot be expanded in a power series
or has an explicit $N$-dependence. 
This technical simplification was even one of our motivations for introducing the localized state.
In the case that the function~$f$ is a monomial,
we shall recover the Gaussian asymptotics similar to that in~\eqref{groupint}.

After these introductory words, we now enter the analysis of group integrals of the form~\eqref{fintro}
and work out the different limiting cases needed for our purposes.

\subsection{Integrals over Functions on a Subspace} \label{secgenint}
The following formula is the core of our method for computing the quantum states.

\begin{Thm} \label{thmgenint}
Let~$I \subset \H$ with~$\dim I=q$, $\dim \H=N$ and~$N \geq 2q$.
Let~$f \in C^0(\scrA_q, \C)$ be a function of~$A:=\pi_I \scrU|_I$ only (where $\pi_I : \H \rightarrow I$ is the orthogonal projection).
Then its group integral over~$\G=\U(N)$ with respect to the normalized Haar measure~$\mu_\G$ can be written as
\beq \label{intA}
\begin{split}
\fint_{\G} f(A)\: d\mu_\G(\scrU)
&= \frac{1}{\pi^{(q^2)}}\: \frac{(N-1)! \, (N-2)!\, \cdots \, (N-q)!}{(N-q-1)!\, (N-q-2)! \,\cdots\, (N-2q)!} \\
&\qquad \times \int_{\scrA_q} f(A)\: \big( \det(\1-A^* A) \big)^{N-2q}\:d\mu_{\C^{q \times q}}(A) \:.
\end{split}
\eeq
Here~$\scrA_q$ is the set of matrices
\beq \label{Aqdef}
\scrA_q := \big\{ A \in \C^{q \times q} \:\big|\: A^* A < \1_{\C^q} \big\} \subset \C^{q \times q} \:,
\eeq
and~$\mu_{\C^{q \times q}}$ is the Lebesgue measure on~$\C^{q \times q} \simeq \R^{2q^2}$, i.e.\
\[ d\mu_{\C^{q \times q}}(A) = \prod_{j,k=1}^q dx^j_k\: dy^j_k \qquad \text{with} \qquad
x^j_k := \re A^j_k,\quad y^j_k := \im A^j_k
 \:. \]
\end{Thm}
Similar results can be found in the literature (see~\cite[Theorem~1.3]{neretin}
or~\cite{friedman-mello,fyodorov, akemann-kieburg}). For completeness, and in order to
get the prefactors right, we shall give a detailed proof of this theorem.
Before beginning, we state two Corollaries. The first is the special case when~$f$ is chosen as the
constant function one.
\begin{Corollary} \label{cor2} Under the assumptions of Theorem~\ref{thmgenint},
\[ \int_{\scrA_q} \big( \det(\1-A^* A) \big)^{N-2q}\:d\mu_{\C^{q \times q}}(A)
= \pi^{(q^2)}\: \frac{(N-q-1)!\, (N-q-2)! \,\cdots\, (N-2q)!}{(N-1)! \, (N-2)!\, \cdots \, (N-q)!} \:. \]
\end{Corollary} \noindent
Next, expanding the factorials in~\eqref{intA}, one immediately gets the following result.
\begin{Corollary} \label{cor3}
The integral~\eqref{intA} has the leading asymptotics for large~$N$
\begin{align*}
\fint_{\G} f(A)\: d\mu_\G(\scrU)
&= \bigg( \frac{N}{\pi} \bigg)^{(q^2)} \!\!\!\int_{\scrA_q} f(A)\: \big( \det(\1-A^* A) \big)^{N-2q}\:
d\mu_{\C^{q \times q}}(A) \: \bigg(1 + \O \Big( \frac{1}{N} \Big) \bigg) \:.
\end{align*}
\end{Corollary}

The remainder of this section is devoted to the proof of Theorem~\ref{thmgenint}.
We let~$e_1, \ldots, e_q$ be an orthonormal basis of~$I$ and extend it to an orthonormal basis of~$\H$.
Then for the evaluation of the integral in~\eqref{intA} it suffices to restrict attention to the first~$q$ columns
of the unitary matrix~$\scrU$. The corresponding column vectors, denoted by
\[ \scrU_i \in \C^N \qquad \text{with~$i \in \{1,\ldots, q \}$} \:, \]
are orthonormal. For computational purposes, it is most convenient to realize the orthonormality
by integrating over all matrix entries and inserting $\delta$-distributions.
This leads us to introduce on~$\C^{N\times q}$ the measure
\beq \label{mudelta}
d\mu_{\G|_I} := \prod_{j=1}^q 
\delta\big( \|\scrU_j\|^2-1 \big) \:\prod_{i < j} \delta \big( \re \la \scrU_i, \scrU_j \ra \big)\:
\delta \big( \im \la \scrU_i, \scrU_j \ra \big)\: d\scrU_i\:,
\eeq
where~$d\scrU_i \in \C^N \simeq \R^{2N}$ is the Lebesgue measure on the $i^\text{th}$ column,
\[ \prod_{k=1}^N  d\re \scrU^k_i\; d\im \scrU^k_i \:. \]
Since the measure~\eqref{mudelta} is invariant under unitary transformations,
and the Haar measure is uniquely determined by unitary invariance and the normalization,
these two measures coincide up to a constant, i.e.\
\beq \label{Haarrel}
\fint_{\G} f(A)\: d\mu_\G(\scrU) = \frac{1}{c} \int_{\C^{N \times q}} f(A)\: d\mu_{\G|_I} \:.
\eeq
We first compute the normalization constant.
\begin{Lemma} \label{lemmanorm}
The constant~$c$ in~\eqref{Haarrel} is given by
\[ c = \int_{\C^{N \times q}} d\mu_{\G|_I} = \frac{\pi^{\frac{q\, (2N+1-q)}{2}}}{(N-1)!\, \cdots \, (N-q)!}\:. \]
\end{Lemma}
\Proof
We carry out the integrals iteratively
choosing polar coordinates. The integral over the first column is computed by
\begin{align}
\int_{\C^N} \delta\big( \|\scrU_1\|^2-1 \big)\: d\scrU_1
&= \int_0^\infty R^{2N-1}\: dR \int_{S^{2N-1}} d\omega\: \delta\big(R^2-1 \big) \notag \\
&= \frac{\mu(S^{2N-1})}{2} = \frac{\pi^N}{(N-1)!} \:. \label{laststep}
\end{align}
Here we carried out the $R$-integral using the well-known formula
\beq \label{deltaformula}
\int f(x)\: \delta\big( g(x) \big)\: dx = \sum_x \frac{f(x)}{|g'(x)|} \:,
\eeq
where the sum goes over all the zeros of the function~$g$. This identity will also be used frequently
in the following computations. Moreover, in the last step in~\eqref{laststep} we
used that the volume of an odd-dimensional unit sphere is given by
(see for example~\cite{huber-sphere})
\[ 
\mu \big( S^{2n-1} \big) = \frac{2\, \pi^n}{(n-1)!} \]
for~$n \in \N$.

Having carried out the integral over~$\scrU_1$, in the next step we can assume
by spherical symmetry of the measure that~$\scrU_1 = (1,0,\ldots, 0)$. Hence
\begin{align*}
&\int_{\C^N} \delta\big( \|\scrU_2\|^2-1 \big)\: \delta \big( \re \la \scrU_1, \scrU_2 \ra \big)\:
\delta \big( \im \la \scrU_1, \scrU_2 \ra \big)\: d\scrU_2 \\
&=\int_{\C^N} \delta\big( \|\scrU_2\|^2-1 \big)\: \delta \big( \re \scrU^1_2\big)\:
\delta \big( \im \scrU^1_2\big)\: d\scrU_2
= \int_{\C^{N-1}} \delta\big( \|u\|^2-1 \big)\: du \\
&= \int_0^\infty r^{2N-3}\: dr \int_{S^{2N-3}} d\omega\: \delta\big(r^2-1 \big)
= \frac{\mu(S^{2N-3})}{2} = \frac{\pi^{N-1}}{(N-2)!} \:.
\end{align*}
Proceeding inductively, we conclude that
\[ \int_{\C^{N \times q}} d\mu_{\scrU|_I}
= \frac{\pi^N \, \pi^{N-1} \,\cdots\, \pi^{N+1-q}}
{(N-1)! \, (N-2)!\, \cdots \, (N-q)!}
= \frac{\pi^{\frac{q\, (2N+1-q)}{2}}}{(N-1)!\, \cdots \, (N-q)!} \:, \]
giving the result.
\QED

We next compute the integral on the right side of~\eqref{Haarrel}.
\begin{Lemma} \label{lemmaIweight}
\begin{align*} 
\int_{\C^{N \times q}} f(A)\: d\mu_{\G|_I}
&= \frac{\pi^{\frac{q (2N+1-3q)}{2}}}{(N-q-1)!\, (N-q-2)! \,\cdots\, (N-2q)!} \\
&\quad\: \times \int_{\scrA_q} f(A)\: \big( \det (\1_{\C^q} - A^* A) \big)^{N-2q} \:d\mu_{\C^{q \times q}}(A) \:.
\end{align*}
\end{Lemma}
\Proof Given~$\scrU \in \U(N)$ we choose an ``adapted'' orthonormal
basis~$f_1, \ldots, f_{N-q}$ of~$J:= I^\perp$ such that the lower left block matrix entry of~$\scrU$
is an upper triangular matrix, i.e.
\beq \label{Bmatrix}
B:= \pi_J \,\scrU|_I =
\begin{pmatrix}
R_1 & m^1_2 & m^1_3 & \cdots & m^1_q \\[0.3em]
0 & R_2 & m^2_3 & \cdots & m^2_q \\[0.3em]
0 & 0 & R_3 & \cdots & m^3_q \\
\vdots & \vdots & \vdots & \ddots & \vdots \\
0 & 0 & 0 & \cdots & R_q \\
0 & 0 & 0 & \cdots & 0 \\
\vdots & \vdots & \vdots & \cdots & \vdots \\
0 & 0 & 0 & \cdots & 0
\end{pmatrix}
\eeq
with~$R_k \geq 0$ (in other words, $B$ is the matrix~$R$ in the $QR$-decomposition of the lower
left block matrix entry of~$\scrU$). It suffices to consider the case that all the~$R_k$ are strictly positive,
because the unitary matrices~$\scrU$ for which one of the~$R_k$ is zero form a set of measure zero.
All the matrix entries of this matrix can be computed uniquely from the matrix~$A$
using the fact that the columns of a unitary matrix are orthonormal. For example,
\begin{align*}
R_1 &= \sqrt{1 - \|A e_1\|^2 } \\
m^1_2 = -\frac{1}{R_1} \: \la A e_1, A e_2 \ra_{\C^q} \:,\qquad
R_2 &= \sqrt{1 - \|A e_2\|^2 - |m^1_2|^2} \:,
\end{align*}
and so on.

Let us verify that this procedure of constructing~$B$ from~$A$ works if and only
if the matrix~$\1_{\C^q}-A^* A$ is positive definite.
Clearly, starting from a unitary matrix~$\scrU$, the corresponding matrix~$A=\pi_I \scrU|I$ has the property
\beq \label{ABrel}
\1_{\C^q} = \pi_I \scrU^* \scrU|I = A^* A + B^* B \:,
\eeq
showing that~$A^*A \leq \1_{\C^q}$. Moreover, the condition that all~$R_k$ are strictly positive
is equivalent to~$B^* B$ being positive definite.
Conversely, assume that~$\1_{\C^q} -A^* A >0$. Then its square root~$\sqrt{\1_{\C^q} -A^* A}$
is again positive definite. We make the ansatz
\[ B = V \, \sqrt{\1_{\C^q} -A^* A} \]
with a unitary matrix~$V$ which is chosen such that~$B$ is an upper triangular matrix.
Choosing
\beq \label{ABmatrix}
\scrU|_I = \begin{pmatrix} A \\ B \end{pmatrix} \:,
\eeq
the columns of this matrix are orthonormal.
Extending this orthonormal system to an orthonormal basis of~$\H$,
one can extend~\eqref{ABmatrix} to a unitary operator~$\scrU$
simply by choosing the additional columns as the additional basis vectors.

It remains to compute the integral. We again work with the block representation~\eqref{ABmatrix} and integrate out
the columns of~$B$ after each other. In the first column, we obtain
\begin{align*}
&\int_{\C^{N-q}} \delta\big( \|\scrU_1\|^2-1 \big)\: dB_1 = 
\int_{\C^{N-q}} \delta\big( \|A_1\|^2 + \|B_1\|^2 -1 \big)\: dB_1 \\
&= \int_0^\infty r^{2N-2q-1}\: dr \int_{S^{2N-2q-1}} d\omega\: \delta\big(\|A_1\|^2 + r^2-1 \big) \\
&= \frac{\mu(S^{2N-2q-1})}{2}\: R_1^{2N-2q-2} = \frac{\pi^{N-q}}{(N-q-1)!}\: R_1^{2N-2q-2} \:,
\end{align*}
where~$A_1:= A e_1$ and~$B_1 := B e_1$ are the first column vectors of the corresponding matrices,
and~$R_1$ coincides with the corresponding matrix entry in~\eqref{Bmatrix}.
In the following computation we can assume by symmetry that~$B_1 = (R_1,0,\ldots, 0)$
(in other words, we work in a basis where the first column of~$B$ has the same form
as in~\eqref{Bmatrix}). We then obtain in the second column
\begin{align*}
&\int_{\C^{N-q}} \delta\big( \|\scrU_2\|^2-1 \big)\: \delta \big( \re \la \scrU_1, \scrU_2 \ra \big)\:
\delta \big( \im \la \scrU_1, \scrU_2 \ra \big)\: dB_2 \\
&=\int_{\C^{N-q}} \delta\big( \|A_2\|^2 + \|B_2\|^2-1 \big) \\
&\qquad \times \delta \big( \re \la A_1, A_2 \ra 
+ R_1\, \re B^1_2 \big)\: \delta \big( \im \la A_1, A_2 \ra + R_1 \,\im B^1_2 \big)\: dB_2 \\
&= \frac{1}{R_1^2} \int_{\C^{N-q-1}} \delta\big( \|A_2\|^2 + |B^1_2|^2 + \|u\|^2-1 \big)\: du \\
&= \frac{1}{R_1^2} \int_0^\infty r^{2N-2q-3}\: dr \int_{S^{2N-2q-3}} d\omega\: \delta\big(r^2-R_2^2 \big) \\
&= \frac{\mu(S^{2N-2q-3})}{2 R_1^2}\: R_2^{2N-2q-4} = \frac{\pi^{N-q-1}}{(N-q-2)!\, R_1^2}\:
R_2^{2N-2q-4}\:,
\end{align*}
where we decomposed $B_2$ as~$B_2 = (B^1_2, u)$ and carried out the integrals
over the real and imaginary parts of~$B^1_2$ with the help of~\eqref{deltaformula}
(the parameter~$R_2$ is again the corresponding matrix entry in~\eqref{Bmatrix}).
Proceeding inductively, we conclude that
\begin{align*}
\int_{\scrU|_I} f(A)\: d\mu_{\scrU|_I} &= \int_{\scrA_q} d\mu_{\C^{q \times q}}(A)\: f(A)\:
\frac{1}{R_1^{2N-4q} \cdots R_q^{2N-4q}} \\
&\qquad \times
\frac{\pi^{N-q} \, \pi^{N-q-1}\, \cdots\, \pi^{N+1-2q}}{(N-q-1)!\, (N-q-2)! \,\cdots\, (N-2q)!} \:.
\end{align*}
Now, we apply~\eqref{ABrel} to obtain
\[ R_1^2 \cdots R_q^2 = |\det B|^2 = \det (B^* B) = \det(\1 - A^* A\big) \:. \]
We finally use that
\[ \pi^{N-q} \, \pi^{N-q-1}\, \cdots\, \pi^{N+1-2q} = \pi^{\frac{1}{2}\, q (2N+1-3q)}\:. \]
This gives the result.
\QED
Combining the formulas of Lemmas~\ref{lemmanorm} and~\ref{lemmaIweight},
one immediately gets the statement of Theorem~\ref{thmgenint}.

\subsection{A Few Simple Examples} \label{secnorm}
In this section we illustrate and verify the result of Corollary~\ref{cor2} in the
cases~$q=1$ and~$q=2$.

\begin{Example} {\bf{(q=1)}} {\em{ In the case~$q=1$, Corollary~\ref{cor2} states that
\[ \int_{\scrA_1} \big( \det(\1-A^* A) \big)^{N-2}\:d\mu_{\C^{q \times q}}(A) = \frac{\pi}{N-1}\:. \]
On the other hand, the integral on the left can be computed directly,
\begin{align*}
&\int_{\scrA_1} \big( \det(\1-A^* A) \big)^{N-2}\:d\mu_{\C}(A) \\
&= \int_{-\infty}^\infty dx \int_{-\infty}^\infty dy \: \Theta(1-x^2-y^2)\: \big( 1-x^2-y^2 \big)^{N-2}
= 2\pi \int_0^1 r\: dr \big( 1-r^2 \big)^{N-2} \\
&= \pi \int_0^1 d(r^2) \big( 1-r^2 \big)^{N-2} = -\pi\: \frac{\big( 1-r^2 \big)^{N-1}}{N-1} \bigg|_{r^2=0}^{r^2=1}
= \frac{\pi}{N-1} \:,
\end{align*}
as desired.
}} \QEDrem
\end{Example}

\begin{Example} {\bf{(q=2)}} {\em{ In the case~$q=2$, we write~$A$ as
\[ A = \alpha_0 \,\1 + \vec{\alpha} \vec{\sigma} + i \beta_0\, \1 + i \vec{\beta} \vec{\sigma} \]
with~$\alpha=(\alpha_0, \vec{\alpha}), \beta=(\beta_0, \vec{\beta}) \in \R^4$. Then
\begin{align*}
A^* &= \alpha_0 \,\1 + \vec{\alpha} \vec{\sigma} - i \beta_0\, \1 - i \vec{\beta} \vec{\sigma} \\
A^* A &= \big(\alpha_0^2 + \beta_0^2 + |\vec{\alpha}|^2 + |\vec{\beta}|^2 \big) \1
+ 2 \alpha_0 \, \vec{\alpha} \vec{\sigma} + 2 \beta_0\, \vec{\beta} \vec{\sigma} 
- 2 \,(\vec{\alpha} \wedge \vec{\beta}) \,\vec{\sigma} \:.
\end{align*}
We want to compute for~$p=0,1,2,\ldots$ the integral
\[ \int_{\scrA_2} \big( \det(\1-A^*A) \big)^p\: d\mu_{\C^{2\times 2}}(A) \:. \]
To this end, we first transform to the integration variables~$\alpha, \beta \in \R^4$,
\[ \int_{\scrA_2} \big( \det(\1-A^*A) \big)^p\: d\mu_{\C^{2\times 2}}(A)
= 16 \int_{\R^4} d^4\alpha \int_{\R^4} d^4 \beta \:\big( \det(\1-A^*A) \big)^p \:. \]
Introducing polar coordinates in the plane,
\[ \alpha_0 = r\, \cos \phi\:,\qquad \beta_0 = r\, \sin \phi \:, \]
we obtain
\[ \int_{\scrA_2} \big( \det(\1-A^*A) \big)^p\: d\mu_{\C^{2\times 2}}(A)
= 32\, \pi \int_0^\infty r\: dr
\int_{\R^3} d^3\alpha \int_{\R^3} d^3 \beta \; \chi_{\scrA_2}(A)\:\big( \det(\1-A^*A) \big)^p\:. \]
In view of rotational symmetry in the angle~$\phi$, we may assume
that~$\alpha_0=r$ and~$\beta_0=0$. Moreover, we introduce spherical coordinates with
\[ R := |\vec{\alpha}|\:,\qquad \hat{R} := |\vec{\beta}|\:, \]
and denote the angle between~$\vec{\alpha}$ and~$\vec{\beta}$ by~$\vartheta$. We thus obtain
\begin{align*}
&\int_{\scrA_2} \big( \det(\1-A^*A) \big)^p\: d\mu_{\C^{2\times 2}}(A) \\
&= 32\, \pi \int_0^\infty r\: dr\; 4 \pi \int_0^\infty R^2\: dR \; 2 \pi \int_0^\infty \hat{R}^2\: d\hat{R}
\int_{-1}^1 d\cos \vartheta\;\chi_{\scrA_2}(A)\:\big( \det(\1-A^*A) \big)^p \\
&= 256\, \pi^3 \int_0^\infty dr \int_0^\infty dR \int_0^\infty d\hat{R}
\int_{-1}^1 d\cos \vartheta\;\chi_{\scrA_2}(A) \: r\: R^2\: \hat{R}^2\: \big( \det(\1-A^*A) \big)^p \:,
\end{align*}
where, due to the symmetries, the matrix~$A$ can be arranged to be of the form
\[ A = \begin{pmatrix} r+R+ i r \cos \vartheta & i r \sin \vartheta \\ i r \sin \vartheta & r+R - i r \cos \vartheta \end{pmatrix} \:. \]
A direct computation yields for the eigenvalues~$\lambda_\pm$ of the matrix~$A^* A$
\[ \lambda_\pm = R^2 + r^2 + \hat{R}^2 \pm 2R\, \sqrt{r^2 + \hat{R}^2\, \sin^2 \vartheta } \]
and thus
\[ \det(\1 - A^* A) = ( 1 - \lambda_+)(1-\lambda-)
= \big( 1 - R^2 + r^2 + \hat{R}^2 \big)^2 - 4 R^2 \big( r^2 + \hat{R}^2\, \sin^2 \vartheta \big) \:. \]

Next, we need to parametrize the set~${\scrA_2}$. To this end, we need to ensure that the largest eigenvalue
of~$A^* A$ is at most one, i.e.\
\[ f(r,R,\hat{R}, \vartheta) := R^2 + r^2 + \hat{R}^2 + 2R\, \sqrt{r^2 + \hat{R}^2\, \sin^2 \vartheta } \leq 1 \:. \]
Obviously, the function~$f$ is monotone increasing in~$R$. Its minimal value is
attained at~$R=0$,
\[ f(r,0,\hat{R}, \vartheta) = r^2+\hat{R}^2\:. \]
Therefore, we may parametrize the integration domain~${\scrA_2}$ as follows,
\begin{align*}
&\int_0^\infty dr \int_0^\infty dR \int_0^\infty d\hat{R}
\int_{-1}^1 d\cos \vartheta\;\chi_{\scrA_2}(A) \:\cdots \\
&= \int_0^1 dr \int_0^{\sqrt{1-r^2}} d\hat{R} \int_{-1}^1 d\cos \vartheta \int_0^{R_{\max}} dR 
\:\cdots \:,
\end{align*}
where~$R_{\max}$ is determined by the equation~$f(r,R_{\max},\hat{R}, \vartheta)=1$ to be
\[ R_{\max} = \sqrt{1-\hat{R}^2 \cos^2 \vartheta} - \sqrt{r^2+\hat{R}^2 \sin^2 \vartheta} \:. \]
A straightforward computation (which we carried with the help of \textsc{Maple}) gives
\[ \int_{\scrA_2} \big( \det(\1-A^*A) \big)^p\: d\mu_{\C^{2\times 2}}(A) = \pi^4\: \frac{(p+1)!\: p!} {(p+3)!\, (p+2)!} \]
for~$p=0, \ldots, 3$. This is in agreement with the general formula from Corollary~\ref{cor2} if we choose~$N=4, \ldots, 7$.
}} \QEDrem
\end{Example}

We conclude this section with another example, where the function~$f$ in the integrand
is chosen as an exponential. This example serves several purposes:
First, it illustrates the Gaussian asymptotics of Lemma~\ref{lemmagroup},
which will be developed further in the following section (Section~\ref{secgauss}).
Second, it shows that by adding a factor~$N$ into the exponent, one can get
into a nonlinear regime where the Gaussian asymptotics no longer applies,
but where the group integral can be computed instead using saddle point methods.
It is a simple example of $N$-dependent integrands as will be analyzed in Section~\ref{secgaussN}.
Moreover, it is a preparation for the saddle point methods to be used in greater generality in Section~\ref{secmodel}.

\begin{Example} {\bf{(an exponential integrand)}} \label{exexp} 
{\em{ Let~$I \subset \H$ be a one-dimensional subspace of~$\H$.
We denote the orthogonal projection to~$I$ by~$\pi_I$. We choose~$\G = \U(\H) \simeq \U(N)$ as the
group of all unitary transformations on~$\H$ and consider the integral
\beq \label{exint1}
{\mathscr{J}} := \fint_\G \exp \Big( \beta\, \tr_\H \big(\pi_I \scrU \pi_I \scrU^{-1} \big) \Big) \: d\mu_\G(\scrU) \:,
\eeq
where~$\beta$ is a real parameter.
This is indeed a special case of the Harish-Chandra integral~\cite{mcswiggen}.
In the present one-dimensional situation, the integral can be computed explicitly as follows.
Applying Theorem~\ref{thmgenint}, we obtain
\[ {\mathscr{J}} = \frac{1}{\pi}\: \frac{(N-1)!}{(N-2)!} \int_{\scrA_1}
e^{\beta\,|A|^2}\:  \big( \det(\1-A^* A) \big)^{N-2}\:d\mu_{\C}(A) \]
Using a polar decomposition~$A = r e^{i \varphi}$ and integrating out the angular dependence, we obtain~$d\mu_\C(A) = r\, d\varphi\, dr$ and thus
\beq \label{Isaddle}
{\mathscr{J}} = 2\, (N-1) \int_0^1 e^{\beta r^2}\: (1-r^2)^{N-2}\: r\: dr \:.
\eeq
Introducing the integration variable~$u:= r^2$, we obtain
\[ {\mathscr{J}} = (N-1) \int_0^1 e^{\beta u}\: (1-u)^{N-2}\: du \:. \]
Asymptotically for large~$N$, the weight function~$(1-u)^{N - 2}$ is very small except near~$u=0$.
Therefore, we may expand the exponential in a power series and carry out the integral term by term,
\begin{align*}
{\mathscr{J}} &= (N-1) \sum_{p=0}^\infty \int_0^1 \frac{(\beta u)^p}{p!}\: (1-u)^{N - 2}\: du
=(N-1) \sum_{p=0}^\infty \beta^p\: \frac{(N-2)!}{(N+p-1)!} \\
&= \sum_{p=0}^\infty \beta^p\: \frac{(N-1)!}{(N+p-1)!} 
= \sum_{p=0}^\infty \frac{\beta^p}{N^p}  \Big( 1 + \O \big( N^{-1}\big) \Big)\:.
\end{align*}
This formula is obtained alternatively if we expand the exponential in~\eqref{exint1}
in a power series and compute term by term with the help of Lemma~\ref{lemmagroup}.
In particular, we conclude that~${\mathscr{J}}$ tends to one as~$N \rightarrow \infty$.

In order to obtain a nontrivial limit as~$N \rightarrow \infty$,
one must increase~$\beta$ with~$N$.
The right scaling is to choose~$\beta$ as a linear function in~$N$. In order to see how this comes about,
we now insert a factor~$N$
into the exponent in~\eqref{exint1},
\begin{align}
&\fint_\G \exp \Big( \beta\,N \tr_\H \big(\pi_I \scrU \pi_I \scrU^{-1} \big) \Big) \: d\mu_\G(\scrU)\notag
= (N-1) \int_0^1 e^{ \beta N u}\: ( 1-u )^{N - 2}\: du \notag \\
&= (N-1) \int_0^1 e^{N \big( \beta u + \log(1-u) \big)} \: (1-u)^{-2}\: du \:. \label{exint3}
\end{align}
We point out that, due to this additional factor~$N$, it is not clear how the the large-$N$-asymptotics
can be obtained from the Harish-Chandra integral.
But we can use that, for large~$N$, the integrand is peaked near the maximum of the function
\[ 
\beta u + \log(1-u) \:. \]
If~$\beta \leq 1$, this function has its maximum at~$u_0=0$, whereas in the case~$\beta>1$
it has a maximum at some~$u_0>0$. In both cases, the function has a unique maximum,
which means that the main contribution to the integral~\eqref{exint3} comes from a neighborhood
of one point~$u_0$.

We finally compute the integral~\eqref{exint3} asymptotically as~$N \rightarrow \infty$,
for brevity only in the case~$\beta<1$.
In this case, the main contribution to the integral comes from the region near~$u=0$. Expanding the logarithm as
\[ \log (1-u) = -u + \O\big( u^2 \big) \:, \]
we obtain
\begin{align}
(N&-1) \int_0^1 e^{-N (1-\beta) u } \: (1-u)^{-2}\: du \;\bigg( 1 + \O \Big( \frac{1}{N} \Big) \bigg) \notag \\
&= (N-1)  \int_0^\infty e^{-N (1-\beta) u } \: du \;\bigg( 1 + \O \Big( \frac{1}{N} \Big) \bigg)  \label{beta0} \\
&= (N-1) \: \frac{1}{N (1-\beta)} \;\bigg( 1 + \O \Big( \frac{1}{N} \Big) \bigg) 
= \frac{1}{1-\beta} + \O \bigg( \frac{1}{N} \bigg) \:, \label{geombeta}
\end{align}
valid if~$\beta<1$.
More systematically, integrals of this type can be computed asymptotically for large~$N$
by applying the saddle point method at the maximum of the integrand
(to this end, it is preferable to work as in~\eqref{Isaddle} with the coordinate~$r$,
in which case the exponential in~\eqref{beta0} becomes a Gaussian).
The saddle point method will be developed systematically in Section~\ref{secmodel}.
\QEDrem }}
\end{Example}

\subsection{The Gaussian Asymptotics} \label{secgauss}
In the next proposition it is shown that for large~$N$, the integral
over the unitary group goes over to a Gaussian integral.
Choosing orthonormal bases, we use the identification~$I \simeq \C^q$.
\begin{Prp} \label{prpgauss} Let~$I \subset \H$ with~$\dim I=q$ and~$\dim \H=N$.
Assume that~$f \in C^\infty(\C^{q \times q}, \C)$ is homogeneous of degree~$d \in \N_0$. Then,
asymptotically for large~$N$,
\beq \label{fgauss}
\fint_{\G} f(A)\: d\mu_\G(\scrU)
=\frac{1}{\pi^{(q^2)}}\: \frac{1}{N^\frac{d}{2}} \int_{\C^{q \times q}} f \big( \tilde{A}\big)\: e^{-\Tr \big( \tilde{A}^* \tilde{A} \big)}\: d\mu_{\C^{q \times q}}\big( \tilde{A} \big)+ \O \bigg(\frac{1}{N^{\frac{d}{2}+1}} \bigg) \:.
\eeq
\end{Prp}
Related results on the Gaussian asymptotics can be found in the literature; see for
example~\cite[Section~II]{ipsen-kieburg}.
We remark that, if~$f$ is a homogeneous polynomial, the Gaussian can be computed
in a straightforward way. Since the covariance is the identity matrix, we get back the
formula in Lemma~\ref{lemmagroup} (the condition that~$f$ is a function of~$A$ only
is not a restriction because in order to deduce the formula~\eqref{groupint} 
from~\eqref{fgauss} one simply chooses~$I$
as the subspace spanned by the basis vectors corresponding to all the indices in~\eqref{groupint}).

We also remark for clarity that the formula~\eqref{fgauss} applies more generally if~$f$ is real analytic
or smooth. This can be proved by approximation. However,
in these cases it is more subtle to specify the error term.
Working with homogeneous functions will be sufficient for the purpose of this paper.
\Proof[Proof of Proposition~\ref{prpgauss}.] According to Corollary~\ref{cor3},
\begin{align*}
\fint_{\G} f(A)\: d\mu_\G(\scrU)
&= \bigg( \frac{N}{\pi} \bigg)^{(q^2)} \int_{\scrA_q} f(A)\: \big( \det(\1-A^* A) \big)^{N-2q}\:d\mu_{\C^{q \times q}}(A)
\Big( 1 + \O\big(N^{-1} \big) \Big) \:.
\end{align*}
We transform to the new integration variable~$\tilde{A}= \sqrt{N} A$,
\begin{align*}
& \fint_{\G} f(A)\: d\mu_\G(\scrU) \\
&= \frac{1}{\pi^{(q^2)}} \int_{\sqrt{N}\,\scrA_q} f\Big( \frac{\tilde{A}}{\sqrt{N}} \Big)
\bigg( \det \Big( \1 - \frac{\tilde{A}^* \tilde{A}}{N} \Big) \bigg)^{N-2q}\:d\mu_{\C^{q \times q}}(\tilde{A})
\Big( 1 + \O\big(N^{-1} \big) \Big) \:.
\end{align*}
We now expand the determinant using the well-known formula
\beq \label{detexp}
\frac{d}{d\tau} \det B(\tau) = \det B(\tau)\: \tr \big( B(\tau)^{-1} \: \dot{B}(\tau) \big)
\eeq
(this formula can be derived for example by differentiating the Laplace expansion).
\[ \det \Big( \1 - \frac{\tilde{A}^* \tilde{A}}{N} \Big) = 1 - \frac{1}{N}\: \Tr\big( \tilde{A}^* \tilde{A} \big) 
+ \O\big(N^{-2} \big) \:, \]
we obtain
\begin{align*}
& \fint_{\G} f(A)\: d\mu_\G(\scrU) \\
&= \frac{1}{\pi^{(q^2)}} \int_{\sqrt{N}\,\scrA_q} f\Big( \frac{\tilde{A}}{\sqrt{N}} \Big)
\bigg( 1 - \frac{1}{N}\: \Tr\big( \tilde{A}^* \tilde{A} \big) \bigg)^{N-2q}\:d\mu_{\C^{q \times q}}(\tilde{A})
\Big( 1 + \O\big(N^{-1} \big) \Big) \:,
\end{align*}
where, from now on, the error term denotes the leading order as specified in the
statement of the proposition above.
Applying Gauss' formula
\beq \label{zinseszins}
\bigg( 1 - \frac{x}{N} \bigg)^N = e^{-x} \:\bigg( 1 + \O \Big(\frac{1}{N} \Big) \bigg)
\eeq
gives
\begin{align*}
& \fint_{\G} f(A)\: d\mu_\G(\scrU) \\
&= \frac{1}{\pi^{(q^2)}} \int_{\sqrt{N}\,\scrA_q} f\Big( \frac{\tilde{A}}{\sqrt{N}} \Big)\:
e^{-\Tr \big( \tilde{A}^* \tilde{A} \big)}\:d\mu_{\C^{q \times q}}(\tilde{A})
\Big( 1 + \O\big(N^{-1} \big) \Big) \:.
\end{align*}
Using that~$f$ is homogeneous of degree~$d$ gives the result.
\QED

\subsection{Treating Integrands which Depend on the Dimension} \label{secgaussN}
If the integrand~$f$ depends on~$N$, the behavior of the integral for large~$N$ may change completely.
In this case, the asymptotics will depend sensitively on the detailed form of the $N$-dependence.
But we can treat the case when the integrand is a product of two functions, one of which
is independent of~$N$. More precisely, we choose two orthogonal subspaces~$I$ and~$J$ of~$\H$,
\beq \label{ortho}
I \perp J \qquad \text{with} \qquad \dim I=q, \dim J =p \quad \text{and} \quad
2(p+q) \leq N\:.
\eeq
Choosing orthonormal bases, we use the identifications
\[ I \simeq \C^q \qquad \text{and} \qquad J \simeq \C^p \:. \]
Moreover, we use the notation
\beq \label{ADdef}
A:= \pi_I \scrU|_I \qquad \text{and} \qquad D := \pi_J \scrU|_J \:.
\eeq
Here and in what follows, we always compute Gaussian integrals with the well-known formula
\beq \label{gaussian}
\int_{-\infty}^\infty e^{-\alpha x^2} \: dx = \sqrt{\frac{\pi}{\alpha}} \qquad \text{(for~$\alpha>0$)}
\eeq

\begin{Prp} \label{prpfact}
Given~$p,q \in \N$, we assume that~$g \in C^\infty(\C^{p \times p}, \C)$ is smooth
and homogeneous of degree~$d \in \N_0$ and that~$f \in C^\infty(\R \times \scrA_q, \C)$
is smooth. Then, asymptotically for large~$N$,
\begin{align*}
&\fint_{\G} f(N, A)\, g(D)\: d\mu_\G(\scrU)
=\bigg( \frac{N}{\pi} \bigg)^{(q^2)} \int_{\scrA_q} f(N,A) \:\big(\det ( \1-A^* A)\big)^{N-2q}\:d\mu_{\C^{q \times q}}(A) \\
&\times \bigg\{
\frac{1}{\pi^{(p^2)}}\: \frac{1}{N^\frac{d}{2}} \int_{\C^{p \times p}} g\big(\tilde{D} \big) \: e^{-\Tr \big( \tilde{D}^* \tilde{D} \big)}\: d\mu_{\C^{p \times p}}(\tilde{D})\: 
+ \O \bigg(\frac{1}{N^{\frac{d}{2}+1}} \bigg) \bigg\}  \bigg( 1 + \O \Big(\frac{1}{N} \Big) \bigg) \:.
\end{align*}
\end{Prp}
In words, this formula shows that the integral factorizes into an integral over~$A$
(which is similar to the general formula of Corollary~\ref{cor3}) and an integral over~$D$
(where the Gaussian asymptotics of Proposition~\ref{prpgauss} applies).
In order to verify the prefactors, it is instructive to consider the case that~$g$ is the constant function one
(being homogeneous of degree~$d=0$). Then, carrying
out the $2q^2$-dimensional Gaussian integral with the help of~\eqref{gaussian}
gives us back the formula in Corollary~\ref{cor3}.

\Proof[Proof of Proposition~\ref{prpfact}]
We choose the subspace~$\hat{I} = I \oplus J$. For the matrix~$\hat{A} := \pi_{\hat{I}} \scrU |_{\hat{I}}$
we use the block matrix notation
\beq \label{Ablock}
\hat{A} = \begin{pmatrix} A & B \\ C & D \end{pmatrix} \:.
\eeq
We apply Corollary~\ref{cor3} for the subspace~$\hat{I}$
and the integrand~$\hat{f}(\hat{A}) = f(N, A) \,g(D)$,
\begin{align}
&\fint_{\G} f(N, A)\, g(D)\: d\mu_\G(\scrU)
= \bigg( \frac{N}{\pi} \bigg)^{(p+q)^2} \notag \\
&\times \int f(N, A)\, g(D)\:
\big( \det(\1-\hat{A}^* \hat{A}) \big)^{N-2p-2q}\:d\mu_{\C^{p+q \times p+q}}(\hat{A})
\:\bigg(1 + \O \Big( \frac{1}{N} \Big) \bigg) \:. \label{ADform}
\end{align}
 For ease in notation, we write the integration measure as
\[ d\mu_{\C^{{p+q} \times {p+q}}}(\hat{A}) =  dA\, dB\, dC\, dD \:. \]

Our next task is to compute the factor~$\det (\1 - \hat{A}^* \hat{A})$. Again in block matrix notation,
\begin{align*}
\hat{A}^* &= \begin{pmatrix} A^* & C^* \\ B^* & D^* \end{pmatrix} \\
\hat{A}^* \hat{A} &= \begin{pmatrix} A^* A + C^* C & A^* B + C^* D \\ B^* A + D^* C & B^* B + D^* D \end{pmatrix} \\
\1 - \hat{A}^* \hat{A} 
&= \begin{pmatrix} \1-A^* A - C^* C & -A^* B - C^* D \\ -B^* A - D^* C & \1 - B^* B - D^* D \end{pmatrix} \\
&= \begin{pmatrix} \1-A^* A & 0 \\ 0 & \1 \end{pmatrix}
- \begin{pmatrix} C^* C & A^* B + C^* D \\ B^* A + D^* C & B^* B + D^* D \end{pmatrix} \\
&= \begin{pmatrix} \1-A^* A & 0 \\ 0 & \1 \end{pmatrix} \:\big( \1 - X \big)
\end{align*}
with the matrix~$X$ given by
\[ X := \begin{pmatrix} (\1-A^* A)^{-1} \,C^* C & (\1-A^* A)^{-1} \, \big( A^* B + C^* D \big) \\ B^* A + D^* C & B^* B + D^* D \end{pmatrix} \:. \]
Hence we can factorize the determinant as
\beq \label{detfact}
\det \big( \1 - \hat{A}^* \hat{A} \big) = \det \big( \1 - A^* A \big)\: \det(\1 - X) \:.
\eeq

Similar as in the proof of Proposition~\ref{prpgauss}, the integrals over~$B$, $C$ and~$D$
can be computed as Gaussian integrals,
\begin{align*}
&\int g(D) \: \det \big( \1 - X \big)^{N-2p-2q}\: dB\,dC\,dD
= \left\{ B,C,D = \frac{1}{\sqrt{N}}\: (\tilde{B},\tilde{C}, \tilde{D}) \right\} \\
&=\frac{1}{N^{2pq+p^2}} \int g\Big(\frac{\tilde{D}}{\sqrt{N}} \Big) \:
\det \bigg\{ \1 - \frac{1}{N} \begin{pmatrix} (\1-A^* A)^{-1} \,\tilde{C}^* \tilde{C} & (\1-A^* A)^{-1} \, \tilde{C}^* \tilde{D} \big) \\ \tilde{D}^* \tilde{C} & \tilde{B}^* \tilde{B} + \tilde{D}^* \tilde{D} \end{pmatrix} \\
&\qquad\qquad\qquad\qquad\qquad\qquad\; -\frac{1}{\sqrt{N}} \begin{pmatrix} 0 & (\1-A^* A)^{-1} \, A^* \tilde{B} \\ \tilde{B}^* A  & 0 \end{pmatrix}
 \bigg\}^{N-2p-2q}\: d\tilde{B}\,d\tilde{C}\,d\tilde{D} \:.
\end{align*}
Again using~\eqref{detexp}, we now expand the determinant to second order,
\begin{align*}
\frac{d}{d\tau} \det(\1- \tau Y) &= -\det(\1-\tau Y)\: \tr\big((\1-\tau Y)^{-1} Y \big) \\
\frac{d}{d\tau} \det(\1- \tau Y) \Big|_{\tau=0} &= -\tr (Y) \\
\frac{d^2}{d\tau^2} \det(\1- \tau Y) \big|_{\tau=0} &= \tr(Y)^2 - \tr\big( Y^2 \big) \\
\Longrightarrow \quad\det(1-\tau Y) &= 1 - \tau\,  \tr (Y) + \frac{\tau^2}{2}\: \Big( \tr(Y)^2 - \tr\big( Y^2 \big) \Big)
+ \O\big( \tau^3 \big) \:.
\end{align*}
We thus obtain
\begin{align*}
&\int g(D) \: \det \big( \1 - X \big)^{N-2p-2q}\: dB\,dC\,dD \\
&=\frac{1}{N^{2pq+p^2}} \bigg( 1 + \O \Big(\frac{1}{N} \Big) \bigg) \int g\Big(\frac{\tilde{D}}{\sqrt{N}} \Big) \:
\bigg\{ 1 - \frac{1}{N} \:\Tr \begin{pmatrix} (\1-A^* A)^{-1} \,\tilde{C}^* \tilde{C} & 0  \\ 0 & \tilde{B}^* \tilde{B} + \tilde{D}^* \tilde{D} \end{pmatrix} \\
&\quad\:-\frac{1}{2N} \:\Tr\begin{pmatrix} (\1-A^* A)^{-1} \, A^* \tilde{B} \tilde{B}^* A & 0 \\ 
0  &  \tilde{B}^* A (\1-A^* A)^{-1} \, A^* \tilde{B} \end{pmatrix}
\bigg\}^{N-2p-2q}\: d\tilde{B}\,d\tilde{C}\,d\tilde{D} \\
&=\frac{1}{N^{2pq+p^2}} \bigg( 1 + \O \Big(\frac{1}{N} \Big) \bigg) \int d\tilde{B}\,d\tilde{C}\,d\tilde{D}\:
g\Big(\frac{\tilde{D}}{\sqrt{N}} \Big) \\
&\quad\:\times
\exp \bigg( - \Tr \Big( (\1-A^* A)^{-1} \,\tilde{C}^* \tilde{C} + \tilde{B}^* \tilde{B} + \tilde{D}^* \tilde{D}
+(\1-A^* A)^{-1} \, A^* \tilde{B} \tilde{B}^* A \Big) \bigg)\:,
\end{align*}
where in the last step we again used~\eqref{zinseszins}.
The trace of the last summand can be simplified as follows,
\begin{align*}
&\Tr \Big( (\1-A^* A)^{-1} \, A^* \tilde{B} \tilde{B}^* A \Big)
= \Tr \Big( A (\1-A^* A)^{-1} \, A^* \tilde{B} \tilde{B}^* \Big) \\
&= \Tr \Big( A A^* (\1-A A^*)^{-1} \, \tilde{B} \tilde{B}^* \Big)
= -\Tr \big( \tilde{B} \tilde{B}^* \big)
+ \Tr \Big( (\1-A A^*)^{-1} \, \tilde{B} \tilde{B}^* \Big) \:.
\end{align*}
We conclude that
\begin{align*}
&\int g(D) \: \det \big( \1 - X \big)^{N-2p-2q}\: dB\,dC\,dD \\
&=\frac{1}{N^{2pq+p^2}} \bigg( 1 + \O \Big(\frac{1}{N} \Big) \bigg) \int d\tilde{B}\,d\tilde{C}\,d\tilde{D}\:
g\Big(\frac{\tilde{D}}{\sqrt{N}} \Big) \\
&\quad\:\times
\exp \bigg( - \Tr \Big( (\1-A^* A)^{-1} \,\tilde{C}^* \tilde{C} + (\1-A A^*)^{-1} \, \tilde{B} \tilde{B}^*
+ \tilde{D}^* \tilde{D} \Big) \bigg)\:.
\end{align*}

Now we can carry out the integrals over~$\tilde{B}$ and~$\tilde{C}$, which are both Gaussian.
We begin with the integral of~$\tilde{C}$. Using~\eqref{gaussian} we obtain
\begin{align*}
\int&  \exp \bigg( - \Tr \Big( (\1-A^* A)^{-1} \,\tilde{C}^* \tilde{C} \Big)\: d\tilde{C} \\
&= \frac{\pi^{pq}}{\det\big( (\1-A^* A)^{-1} \big)^{p}}
= \pi^{pq}\: \big( \det ( \1-A^* A)\big)^{p} \:.
\end{align*}
The power~$-p$ of the determinant can be derived as follows: The Gaussian integral
is of dimension~$2pq$. If we diagonalize the $q\times q$-matrix $(\1-A^*A)^{-1}$, every eigenspace
gives rise to a Gaussian integral of dimension~$2pq/q=2p$. Therefore, Gaussian integration
yields the corresponding eigenvalue to the power~$-p$. In total, we get the factor~$\det (\1-A^*A)^{-p}$.

The integral over~$\tilde{B}$ is computed similarly. We conclude that
\begin{align*}
&\int g(D) \: \det \big( \1 - X \big)^{N-2p-2q}\: dB\,dC\,dD \\
&=\frac{\pi^{2pq}}{N^{2pq+p^2}} \:\big( \det ( \1-A^* A)\big)^{2p}
\int g\Big(\frac{\tilde{D}}{\sqrt{N}} \Big) \:\exp^{-\Tr \big( \tilde{D}^* \tilde{D} \big)}\: d\tilde{D}\:
\bigg( 1 + \O \Big(\frac{1}{N} \Big) \bigg) \:.
\end{align*}
After applying~\eqref{detfact} in~\eqref{ADform}, we can use the last formula to obtain the result.
\QED

\subsection{Asymptotic Decoupling on Orthogonal Subspaces}
In this section we again consider the situation of two orthogonal subspaces~\eqref{ortho}
and two function~$f(A)$ and~$g(B)$ depending on the sub-matrices in~\eqref{ADdef}.
But now both functions may depend on~$N$. Clearly, we no longer get the Gaussian asymptotics.
But the resulting formula makes it possible to analyze whether and in which sense the integral factorizes
into separate integrals over~$A$ and~$B$. We first state our result and briefly discuss it afterward.

\begin{Prp} \label{prpfact2}
Given~$p,q \in \N$, we 
assume that~$g \in C^\infty(\R \times \scrA_p, \C)$ and~$f \in C^\infty(\R \times \scrA_q, \C)$
are smooth. Then, asymptotically for large~$N$,
\begin{align}
\fint_{\G} & f(N, A)\, g(N, D)\: d\mu_\G(\scrU)
=\bigg( \frac{N}{\pi} \bigg)^{(p^2+q^2)} \notag \\
& \times \int_{\scrA_q} d\mu_{\C^{q \times q}}(A)
\int_{\scrA_p} d\mu_{\C^{p \times p}}(D)\:f(N,A) \: g(N, D) \notag \\
&\qquad \times \:\frac{\big( \det ( \1-A^* A)\big)^{N-2q}\: \big( \det ( \1-D^* D)\big)^{N-2p}}{\det \big( \1 - A^* A \otimes D^* D \big)} \;\bigg( 1 + \O \Big(\frac{1}{N} \Big) \bigg) \:. \label{fgform}
\end{align}
\end{Prp} \noindent
Here the tensor product is a $(pq \times pq)$-matrix with entries
\[ (A^* A \otimes D^* D)^{ik}_{jl} = \big( A^* A  \big)^i_j\, \big( D^* D  \big)^k_l\:. \]
These matrix entries are very small, provided that either~$A$ or~$D$ is small.
In these cases, we know that
\beq \label{AADDdet}
\det \big( \1 - A^* A \otimes D^* D \big) \approx 1 \:,
\eeq
and the integral factorizes. In particular, this will be the case if~$g$ is homogeneous of degree~$d$,
and applying Proposition~\ref{prpgauss}, we get back the formula of Proposition~\ref{prpfact}.
The main improvement of Proposition~\ref{prpfact2} is that, by estimating the determinant in~\eqref{AADDdet}
from above, one can quantify the error of the factorization.

\Proof[Proof of Proposition~\ref{prpfact2}] We begin similar as in the proof of Proposition~\ref{prpfact}.
Using again the notation~\eqref{Ablock}, we obtain
\begin{align*}
\hat{A}^* \hat{A} &= \begin{pmatrix} A^* A + C^* C & A^* B + C^* D \\ B^* A + D^* C & B^* B + D^* D \end{pmatrix} \\
\1 - \hat{A}^* \hat{A} 
&= \begin{pmatrix} \1-A^* A - C^* C & -A^* B - C^* D \\ -B^* A - D^* C & \1 - B^* B - D^* D \end{pmatrix} \\
&= \begin{pmatrix} \1-A^* A & 0 \\ 0 & \1-D^* D \end{pmatrix}
- \begin{pmatrix} C^* C & A^* B + C^* D \\ B^* A + D^* C & B^* B \end{pmatrix} \\
&= \begin{pmatrix} \1-A^* A & 0 \\ 0 & \1-D^* D \end{pmatrix} \:\big( \1 - X \big)
\end{align*}
where now the matrix~$X$ is given by
\[ X := \begin{pmatrix} (\1-A^* A)^{-1} \,C^* C & (\1-A^* A)^{-1} \, \big( A^* B + C^* D \big) \\ 
(\1-D^* D)^{-1}\, \big( B^* A + D^* C \big) & (\1-D^* D)^{-1}\, B^* B \end{pmatrix} \:. \]
Hence we can factorize the determinant as
\[ \det \big( \1 - \hat{A}^* \hat{A} \big) = \det \big( \1 - A^* A \big)\: \det \big( \1 - D^* D \big)\: \det(\1 - X) \:. \]

Our task is to carry out the integrals over~$B$ and~$C$. Compared to the proof of Proposition~\ref{prpfact},
the formulas are more involved. For this reason, it is useful to simplify the formulas as follows.
We take polar decompositions of~$A$ and~$D$,
\[ A = U_L \D_A U_R \qquad \text{and} \qquad D = V_L \D_B V_R \]
with unitary matrices~$U_L, U_R \in \C^{q \times q}$ and~$V_L, V_R \in \C^{p \times p}$.
Here~$\D_A$ and~$\D_B$ are diagonal matrices
\[ \D_A = \diag (a_1, \ldots, a_q) \qquad \text{and} \qquad \D_D = \diag(d_1, \ldots, d_p) \:, \]
whose entries take values in the interval~$(0,1)$. It follows that
\[ \hat{A} = \begin{pmatrix} U_L & 0 \\ 0 & V_L \end{pmatrix} 
\begin{pmatrix} \D_A & U_L^{-1} B V_R^{-1} \\ V_L^{-1} C U_R^{-1} & \D_D \end{pmatrix}
\begin{pmatrix} U_R & 0 \\ 0 & V_R \end{pmatrix} \:. \]
The unitary transformations at the very left and right drop out when forming~$\det(\1 - \hat{A}^* \hat{A})$.
The unitary transformations in the off-diagonal entries
\beq \label{BCtrans}
B \rightarrow U_L^{-1} B V_R^{-1} \qquad \text{and} \qquad C \rightarrow V_L^{-1} C U_R^{-1} \:,
\eeq
on the other hand, do not change the integration measures~$dB$ and~$dC$.
Indeed, these integration measures transform with the Jacobian determinant of the
transformation matrices, if we write out the real and imaginary parts separately.
As is shown in Lemma~\ref{lemmadet} below, the resulting Jacobian determinants are equal to one.
Therefore, the unitary transformations in~\eqref{BCtrans} indeed preserve the integration measures.

After these transformations, we can assume without loss of generality that~$A$ and~$D$
coincide with the diagonal matrices~$\D_A$ and~$\D_D$, respectively.
After this simplification, the integrals over~$B$ and~$C$ can be computed as follows.
We first rewrite the integrals as
\begin{align*}
&\int \det \big( \1 - X \big)^{N-2p-2q}\: dB\,dC = \left\{ B,C = \frac{1}{\sqrt{N}}\: (\tilde{B},\tilde{C}) \right\} \\
&=\frac{1}{N^{2pq}} \int 
\det \bigg\{ \1 - \frac{1}{N} 
\begin{pmatrix} (\1-A^2)^{-1} \,\tilde{C}^* \tilde{C} & 0 \\ 
0 & (\1-D^2)^{-1}\, \tilde{B}^* \tilde{B} \end{pmatrix} \\
&\quad\:-\frac{1}{\sqrt{N}} 
\begin{pmatrix} 0 & (\1-A^2)^{-1} \, \big( A \tilde{B} + \tilde{C}^* D \big) \\ 
(\1-D^2)^{-1}\, \big( \tilde{B}^* A + D \tilde{C} \big) & 0 \end{pmatrix} \bigg\}^{N-2p-2q}\: d\tilde{B}\,d\tilde{C} \:.
\end{align*}
Now we can again expand the determinant with the help of~\eqref{detexp} to second order to obtain
\begin{align*}
&\int \det \big( \1 - X \big)^{N-2p-2q}\: dB\,dC \\
&=\frac{1}{N^{2pq}} \bigg( 1 + \O \Big(\frac{1}{N} \Big) \bigg) \int
\bigg\{ 1 - \frac{1}{N} \:\Tr \begin{pmatrix} (\1-A^2)^{-1} \,\tilde{C}^* \tilde{C} & 0 \\ 
0 & (\1-D^2)^{-1}\, \tilde{B}^* \tilde{B} \end{pmatrix} \\
&\quad\:-\frac{1}{2N} \:\Tr
\begin{pmatrix} 0 & \!\!\!\!\!\!\!\!\!\!(\1-A^2)^{-1} \, \big( A \tilde{B} + \tilde{C}^* D \big) \\ 
(\1-D^2)^{-1}\, \big( \tilde{B}^* A + D \tilde{C} \big) & 0 \end{pmatrix}^2
\bigg\}^{N-2p-2q}\!\!\!\! d\tilde{B}\,d\tilde{C} \\
&=\frac{1}{N^{2pq}} \bigg( 1 + \O \Big(\frac{1}{N} \Big) \bigg) \int d\tilde{B}\,d\tilde{C} \\
&\quad\:\times
\exp \bigg( - \Tr \Big( (\1-A^2)^{-1} \,\tilde{C}^* \tilde{C} +(\1-D^2)^{-1}\, \tilde{B}^* \tilde{B} \\
&\qquad\qquad\qquad\qquad + (\1-A^2)^{-1} \, \big( A \tilde{B} + \tilde{C}^* D \big)\, 
(\1-D^2)^{-1}\, \big( \tilde{B}^* A + D \tilde{C} \big) \Big)
 \bigg)\:.
\end{align*}
The trace can be simplified using the formulas
\begin{align*}
&\Tr \Big( (\1-A^2)^{-1} \, \big( A \tilde{B} \big) \, (\1-D^2)^{-1} \, \big( \tilde{B}^* A  \big) \Big) \\
&=\Tr \Big( A^2 (\1-A^2)^{-1} \, \tilde{B}  \, (\1-D^2)^{-1} \, \tilde{B}^* \Big) \\
&= - \Tr \Big( \tilde{B}  \, (\1-D^2)^{-1} \, \tilde{B}^* \Big)
+ \Tr \Big( (\1-A^2)^{-1} \, \tilde{B}  \, (\1-D^2)^{-1} \, \tilde{B}^* \Big) \:.
\end{align*}
and similarly
\begin{align*}
&\Tr \Big( (\1-A^2)^{-1} \, \big( \tilde{C}^* D \big) \, (\1-D^2)^{-1} \, \big( D \tilde{C}  \big) \Big) \\
&= - \Tr \Big( (\1-A^2)^{-1} \, \tilde{C}^* \tilde{C} \Big)
+ \Tr \Big( (\1-A^2)^{-1} \, \tilde{C}^* (\1-D^2)^{-1} \tilde{C} \Big) \:.
\end{align*}
This gives
\begin{align*}
&\int \det \big( \1 - X \big)^{N-2p-2q}\: dB\,dC \\
&=\frac{1}{N^{2pq}} \bigg( 1 + \O \Big(\frac{1}{N} \Big) \bigg) \int d\tilde{B}\,d\tilde{C} \\
&\quad\:\times
\exp \bigg( - \Tr \Big( (\1-A^2)^{-1} \, \tilde{B}  \, (\1-D^2)^{-1} \, \tilde{B}^*
+  (\1-A^2)^{-1} \, \tilde{C}^*\, (\1-D^2)^{-1} \tilde{C} \\[-0.3em]
&\qquad\qquad\qquad\qquad +
 (\1-A^2)^{-1} \, A \tilde{B} \, 
(\1-D^2)^{-1}\, D \tilde{C} \\
&\qquad\qquad\qquad\qquad +
 (\1-A^2)^{-1} \, \tilde{C}^* D\, 
(\1-D^2)^{-1}\, \tilde{B}^* A \Big)
 \bigg)\:.
\end{align*}

Using that~$A$ and~$D$ are diagonal, we obtain
\begin{align*}
&\int \det \big( \1 - X \big)^{N-2p-2q}\: dB\,dC \\
&=\frac{1}{N^{2pq}} \bigg( 1 + \O \Big(\frac{1}{N} \Big) \bigg) \int d\tilde{B}\,d\tilde{C} \\
&\quad\:\times \exp \bigg( - \sum_{i=1}^q \sum_{j=1}^p
\Big( \frac{1}{1-a_i^2} \: \frac{1}{1-d_j^2}\: \big( | \tilde{B}_{ij} |^2 + |\tilde{C}^*_{ij}|^2 \big) \\
&\qquad\qquad\qquad\qquad\qquad + \frac{a_i}{1-a_i^2} \: \frac{d_j}{1-d_j^2}
\:\big( \tilde{B}_{ij}\, \overline{\tilde{C}^*_{ij}} + 
\overline{\tilde{B}_{ij}}\, \tilde{C}^*_{ij} \big) \Big) \bigg) \:.
\end{align*}
This integral can be computed separately for every~$i$ and~$j$, giving
\[ X:= \int_{\C} db \int_{\C} dc \exp \bigg( - \bigg\la
\begin{pmatrix} b \\ c \end{pmatrix},
Y \begin{pmatrix} b \\ c \end{pmatrix}
 \bigg\ra_{\C^2} \bigg) \]
 with
\[ Y := \frac{1}{1-a_i^2} \: \frac{1}{1-d_j^2}\: 
\begin{pmatrix} 1 & a_i d_j \\ a_i d_j & 1 \end{pmatrix} \:. \]
Next,
\begin{align*}
\det Y &= \bigg( \frac{1}{1-a_i^2} \: \frac{1}{1-d_j^2} \bigg)^2 (1- a_i^2 d_j^2)
= \frac{1- a_i^2 d_j^2}{(1-a_i^2)^2\, (1-d_j^2)^2} \:, \\
\intertext{and carrying out the Gaussian integrals with the help of~\eqref{gaussian} gives}
X &= \frac{\pi^2}{\det Y} = \pi^2\: \frac{(1-a_i^2)^2\, (1-d_j^2)^2}{1- a_i^2 d_j^2} \\
\int \det & \big( \1 - X \big)^{N-2p-2q}\: dB\,dC \\
&= \frac{1}{N^{2pq}} \bigg( 1 + \O \Big(\frac{1}{N} \Big) \bigg) 
\prod_{i=1}^q \prod_{j=1}^p \pi^2\: \frac{(1-a_i^2)^2\, (1-d_j^2)^2}{1- a_i^2 d_j^2} \\
&= \frac{\pi^{2pq}}{N^{2pq}}\: \det \big(\1 - A^2)^{2p}\:  \det \big(\1 - D^2)^{2q}\: \bigg( 1 + \O \Big(\frac{1}{N} \Big) \bigg) 
\prod_{i=1}^q \prod_{j=1}^p \frac{1}{1- a_i^2 d_j^2} \:.
\end{align*}
The last term tells us about the ``coupling'' of~$A$ and~$B$. It can be written as
\[ \prod_{i=1}^q \prod_{j=1}^p \frac{1}{1- a_i^2 d_j^2}
= \frac{1}{\det \big( \1 - A^2 \otimes D^2 \big)} \:. \]
This gives the result.
\QED

It remains to show that the unitary transformations in~\eqref{BCtrans} are measure-preserving.
This follows immediately from the following lemma.
\begin{Lemma} \label{lemmadet} Let~$X$ be an invertible complex quadratic matrix. Then
\[ \det \begin{pmatrix} \re X & -\im X \\ \im X & \re X \end{pmatrix} = |\det X|^2 \:. \]
\end{Lemma}
\Proof By direct computation, one finds that
\[ \begin{pmatrix} \re X & -\im X \\ \im X & \re X \end{pmatrix} 
\begin{pmatrix} \re Y & -\im Y \\ \im Y & \re Y \end{pmatrix} 
= \begin{pmatrix} \re (X Y) & -\im (X Y) \\ \im (X Y) & \re (X Y) \end{pmatrix} \:. \]
Proceeding inductively, we deduce that for any polynomial~$g$,
\[ g \bigg( \begin{pmatrix} \re X & -\im X \\ \im X & \re X \end{pmatrix} \bigg)
= \begin{pmatrix} \re \big (g(X) \big) & -\im \big( g(X) \big) \\[0.2em] \im \big( g(X) \big) & \re \big( g(X) \big) \end{pmatrix} \:. \]
The continuous functional calculus allows us to extend this formula to any continuous function~$g$.
In particular,
\beq \label{faformula}
\log \begin{pmatrix} \re X & -\im X \\ \im X & \re X \end{pmatrix}  = 
\begin{pmatrix} \re \log X & -\im \log X \\ \im \log X & \re \log X \end{pmatrix} \:.
\eeq
We now take the trace,
\begin{align*}
\log \det \begin{pmatrix} \re X & -\im X \\ \im X & \re X \end{pmatrix}
&= \Tr \bigg( \log \begin{pmatrix} \re X & -\im X \\ \im X & \re X \end{pmatrix} \bigg) \\
&\!\!\!\overset{\eqref{faformula}}{=} \Tr \begin{pmatrix} \re \log X & -\im \log X \\ \im \log X & \re \log X \end{pmatrix}
= 2\: \Tr \re \log X \:.
\end{align*}
Taking the exponential gives
\[ \det \begin{pmatrix} \re X & -\im X \\ \im X & \re X \end{pmatrix}
= e^{2\:\Tr \re \log X} = e^{2\:\re \Tr \log X}  = \big| e^{\Tr \log X} \big|^2 = |\det X|^2 \:. \]
This concludes the proof.
\QED

\section{Model Examples} \label{secmodel}
In this section we compute group integrals for two model examples in which the
integrand is the exponential of a specific potential. As we shall see in Section~\ref{secrefinecompute},
the localized refined pre-state can be computed by suitably combining
the results of these two model examples.

\subsection{An Exponential of a Linear Functional} \label{seclinear}
We let~$\H=\C^N$ with the canonical scalar product and~$I \subset \H$
a $q$-dimensional subspace. Choosing an orthonormal basis, we identify~$I$ with~$\C^q$.
As in~\eqref{ADdef} we use the notation
\[ A:= \pi_I \scrU|_I \:. \]
We let~$Y \in \Lin(I)$ be a Hermitian~$q \times q$-matrix. In this section, we shall compute the integral
\[ \fint_\G e^{N \re \Tr(A Y)}\: d\mu_\G(\scrU) \]
asymptotically for large~$N$. We first state our main result and discuss and prove it afterward.
\begin{Prp} \label{prpsaddlelin}
Given a symmetric operator~$Y \in \Lin(I)$, we define the operator~$A_0$ by
\beq \label{A0def}
A_0 := \frac{Y}{\sqrt{ Y^2 +\1 } + \1} \:.
\eeq
Then, asymptotically for large~$N$,
\beq \label{saddlelin}
\fint_{\G} e^{N \re \Tr(A Y)}\: d\mu_\G(\scrU)
= e^{N \Tr(A_0 Y)}\:  \frac{\big( \det(\1-A_0^2) \big)^N}{\sqrt{
\det \big (\1 - A_0^2 \otimes A_0^2 \big)}}\: \Big( 1 + \O\big(N^{-1} \big) \Big) \:.
\eeq
If~$Y \neq 0$, the inequality
\beq \label{linstrict}
\Tr(A_0 Y) + \log \det (\1-A_0^2) > 0
\eeq
holds, showing that,  for large~$N$, the right side of~\eqref{saddlelin} grows exponentially in~$N$.
\end{Prp}

Before coming to the proof, we illustrate the statement of this proposition by proving a simple special case.
\begin{Example} {\bf{(q=1)}} {\em{ In the case~$q=1$, by applying Corollary~\eqref{cor3}, we obtain the leading contribution
\begin{align*}
&\fint_{\G} e^{N \re \Tr(A Y)}\: d\mu_\G(\scrU) \Big( 1 + \O\big(N^{-1} \big) \Big) \\
&= {\mathscr{J}} := \frac{N}{\pi} \int_{B_1} e^{N y \re a}\: \big( 1-|a|^2 \big)^{N-2}\:d\mu_{\C}(a) \:,
\end{align*}
where we write~$Y = y \1$ (with~$y \in \R$), and~$B_1 \subset \C$ denotes the unit disc. 
Choosing polar coordinates, we obtain
\[ {\mathscr{J}} = \frac{N}{\pi} \int_0^1 r\: dr \int_0^{2 \pi} d\varphi \:  e^{N y \,r \cos \varphi}\: \big( 1-r^2 \big)^{N-2}
= \frac{N}{\pi}\int_0^1 dr \int_0^{2 \pi} d\varphi \: e^{N\, g(r, \varphi)} \: \frac{r}{(1-r^2)^2} \]
with the function~$g$ defined by
\[ g(r, \varphi) := y \,r \cos \varphi + \log(1-r^2)\:. \]
Our strategy for evaluating this integral asymptotically for large~$N$ is to compute the maxima
of the function~$g$ and to use the saddle point approximation.
Without loss of generality, it suffices to consider the case~$y>0$.
Then the maximum of~$g$ is attained at~$\varphi=0$ and the radius~$r$ with
\[ 0 = \frac{\partial}{\partial r} g(r,0) = y - \frac{2 r}{1-r^2} \:. \]
This equation has the two solutions
\[ r= \lambda_\pm := -\frac{1}{y} \pm \sqrt{1 + \frac{1}{y^2} } \:. \]
Clearly, only the solution~$\lambda_+$ lies in the interval~$[0,1]$. We can simplify it to
\[ \lambda_+ =  \frac{\sqrt{y^2+1} - 1}{y} = \frac{\sqrt{y^2+1} - 1}{y} \:
\frac{\sqrt{y^2+1} + 1}{\sqrt{y^2+1} + 1} = \frac{y}{\sqrt{y^2+1} + 1} \:, \]
giving agreement with~\eqref{A0def}.
Since~$g(0,0)=0$ and~$\lim_{r \nearrow 1} g(r,0) = -\infty$, the point $(\lambda_+, \varphi=0)$
is the unique maximum.

Having a unique maximum, we can compute the integral asymptotically for large~$N$
with the saddle point approximation: The Hessian of~$g$ at the maximum is the diagonal matrix
\[ D^2g(\lambda_+, 0) = - 2 \diag \bigg( 
\frac{1+\lambda_+^2}{(1-\lambda_+^2)^2}, \frac{\lambda_+^2}{1-\lambda_+^2} \bigg) \:. \]
Its determinant is computed by
\[ \det D^2 g(\lambda_+, 0) = 4\, \frac{\lambda_+^2 \,(1+\lambda_+^2)}{(1-\lambda_+^2)^3}
= 4\, \frac{\lambda_+^2 \,(1-\lambda_+^4)}{(1-\lambda_+^2)^4} \:. \]
Hence, using~\eqref{gaussian}, the saddle point approximation gives
\begin{align*}
{\mathscr{J}} &= \frac{e^{g(\lambda_+, 0)}}{\sqrt{\det D^2g(\lambda_+, 0)/2}}\: \frac{\lambda_+}{(1-\lambda_+^2)^2}\:\Big( 1 + \O\big(N^{-1} \big) \Big) \\
&= e^{N y \lambda_+}\: (1-\lambda_+^2)^N\: 
\sqrt{\frac{(1-\lambda_+^2)^4}{\lambda_+^2 \,(1-\lambda_+^4)}}
\: \frac{\lambda_+}{(1-\lambda_+^2)^2}\:\Big( 1 + \O\big(N^{-1} \big) \Big) \\
&= e^{N y \lambda_+}\:
\frac{(1-\lambda_+^2)^N}{\sqrt{1-\lambda_+^4}}
\:\Big( 1 + \O\big(N^{-1} \big) \Big) \:.
\end{align*}
This agrees with~\eqref{saddlelin} in the special case~$q=1$.
}} \QEDrem
\end{Example}

We now enter the proof of Proposition~\ref{prpsaddlelin}, which will be completed at the end of this section.
Applying Corollary~\eqref{cor3}, we obtain the leading contribution
\begin{align*}
&\fint_{\G} e^{N \re \Tr(A Y)}\: d\mu_\G(\scrU)\:\Big( 1 + \O\big(N^{-1} \big) \Big) \\
&= \bigg( \frac{N}{\pi} \bigg)^{(q^2)} 
\int_{\scrA_q} e^{N \re \Tr(A Y)}\: \big( \det(\1-A^* A) \big)^{N-2q}\:d\mu_{\C^{q \times q}}(A) \:.
\end{align*}
We rewrite the integrand as
\[ e^{N \re \Tr(A Y)}\: \big( \det(\1-A^* A) \big)^{N-2q} = e^{N g(A)}\: \big( \det(\1-A^* A) \big)^{-2q} \]
with
\beq \label{gdeflin}
g : \scrA_q \rightarrow \R\:,\qquad g(A) = \re \Tr(A Y) + \log \det(\1-A^* A) \:.
\eeq

\begin{Lemma} \label{lemmamaxlin} The function~$g(A)$ has a unique maximum at~$A=A_0$
with~$A_0$ according to~\eqref{A0def}.
\end{Lemma}
\Proof At the boundary of~$\scrA_q$, the operator~$A^* A$ has eigenvalues which are equal to one, implying that
the factor~$\log \det(\1-A^* A)$ tends to minus infinity on this boundary.
Therefore, it suffices to consider the interior of~$\scrA_q$.

The determinant can be expanded linearly again using the formula~\eqref{detexp}.
We thus obtain for first variations of~$g$
\begin{align*}
\delta g(A) &= \re \Tr\big( (\delta A)\, Y \big) - \Tr \Big( (\1-A^* A)^{-1}\: \big( (\delta A^*)\, A + A^* (\delta A) \big) \Big) \\
&= \Tr \bigg( \delta A^* \Big( \frac{Y}{2} - A \,(\1-A^* A)^{-1} \Big) \bigg) 
+ \Tr \bigg( \delta A \Big( \frac{Y}{2} - (\1-A^* A)^{-1} \, A^* \Big) \bigg) \:.
\end{align*}
At the maximum, this variation vanishes for any choice of~$\delta A$.
The freedom in inserting a phase factor~$\delta A \rightarrow e^{i \varphi} \delta A$ in the above equation
shows that the summands involving~$\delta A$ and~$(\delta A)^*$ must vanish separately. Thus
\[ \Tr \bigg( \delta A^* \Big\{ \frac{Y}{2} - A \,(\1-A^* A)^{-1} \Big\} \bigg) = 0 \:. \]
Choosing~$\delta A$ equal to the matrix in the curly brackets, we conclude that
\beq \label{Acondlin}
A \,(\1-A^* A)^{-1} = \frac{Y}{2} = (\1-A^* A)^{-1} \, A^* \:.
\eeq

We now form a polar decomposition of~$A$, i.e.\
\[ A = U S \]
with~$U \in \U(\C^q)$ unitary and~$S \in \Lin(\C^q)$ positive semi-definite with spectrum in the
half-open interval~$[0,1)$.
Then~\eqref{Acondlin} can be written as
\beq \label{UTrel}
U\, T = Y = T\, U^{-1} \:,
\eeq
where the operator~$T$ is defined via the spectral calculus as the positive semi-definite operator
\beq \label{TSdef}
T := 2 S \, \big(\1-S^2)^{-1} = \frac{2 S}{\1-S^2} \geq 0 \:.
\eeq
Applying~\eqref{UTrel} iteratively, it follows that~$U^p T = T U^{-p}$. The continuous functional calculus yields
\beq \label{fspec}
f(U)\:T = T\: f \big( U^{-1} \big)
\eeq
for any continuous function~$f$. Forming a spectral decomposition of~$U$,
\[ U = \sum_{k} \lambda_k \: E_k \qquad \text{with} \qquad \lambda_k \in \C,\: |\lambda_k|=1 \:, \]
and evaluating~\eqref{fspec} for functions~$f$ which vanish on the spectrum except at single eigenvalues, we conclude
that for every eigenvalue~$\lambda \in \sigma(U)$, also its complex conjugate~$\overline{\lambda}$ is 
an eigenvalue of~$U$ and
\[ E_{\lambda} T = T E_{\overline{\lambda}} \:. \]

Let~$\lambda$ with~$\lambda \neq \overline{\lambda}$ be a non-real eigenvalue of~$U$. Then
\[ \Tr \big( E_\lambda\, T \big) = \Tr \big( E_\lambda\, (E_\lambda T) \big)
= \Tr \big( E_\lambda\, (T E_{\overline{\lambda}} ) \big) = \Tr \big( E_{\overline{\lambda}} \,E_\lambda\, T \big) 
= 0 \:. \]
Since~$T$ is positive semi-definite, it follows that
\[ E_\lambda\, T = 0 = T\, E_\lambda \:. \]
In other words, the operator $T$ vanishes on all the eigenspaces of~$U$ corresponding to non-real eigenvalues.
Since the spectral calculus~\eqref{TSdef} maps the kernel of~$T$ to the kernel of~$S$,
the same is true for the operator~$S$.
Therefore, by changing~$U$ to be the identity on the kernel of~$T$, we can arrange the
polar decomposition~\eqref{UTrel} with a unitary operator~$U$ with purely real eigenvalues.

After this construction, the operator~$U$ in~\eqref{UTrel} has the properties
\[ U = U^* \qquad \text{and} \qquad [U, T ] = 0 = [U,S] \:. \]
Moreover, the operator~$A$ is symmetric. Therefore, the equations~\eqref{Acondlin} simplify to
\[ \frac{2 A}{\1-A^2} = Y \:. \]
The corresponding scalar quadratic equation (which can be viewed as an equation for the
eigenvalues of the corresponding matrices) takes the form
\[ a^2 + \frac{2 a}{y} - 1 = 0 \:; \]
it has a unique root in the interval~$(0,1)$ provided that~$y \neq 0$ given by
\[ a = -\frac{1}{y} + \epsilon(y)\: \sqrt{ 1 + \frac{1}{y^2}} = \frac{y}{\sqrt{y^2+1} + 1} \:, \]
where~$\epsilon$ is the sign function $\epsilon(x)=1$ for $x \geq 0$ and $\epsilon(x)=-1$ otherwise.
The last formula also applies if~$y=0$.
Therefore, the spectral calculus determines~$A$ uniquely by~\eqref{A0def}.
\QED

We next compute the Hessian at the maximum. It is most convenient to work in an
eigenvector basis of the matrix~$A_0$.
\begin{Lemma} \label{lemmaexpandlin} Assume that the matrix~$A_0$ in~\eqref{A0def} is diagonal,
\[ A_0 = \diag \big( \lambda_1, \ldots, \lambda_q \big) \:. \]
Then the function~$g(A)$ defined in~\eqref{gdeflin} has the quadratic expansion
\begin{align*}
g(A) &= g(A_0) - \frac{1}{2} \sum_{i,j=1}^q \frac{1}{1-\lambda_i^2}\:\frac{1}{1-\lambda_j^2}\: 
\bigg\la \begin{pmatrix} \delta A^i_j \\[0.4em] \overline{\delta A^j_i} \end{pmatrix}, 
\begin{pmatrix} 1 & \lambda_i \lambda_j \\ \lambda_i \lambda_j & 1 \end{pmatrix}
\begin{pmatrix} \delta A^i_j \\[0.4em] \overline{\delta A^j_i} \end{pmatrix} \bigg\ra_{\C^2} \\
&\quad\: + \O \big( (\Delta A)^3 \big) \:,
\end{align*}
where~$\delta A$ denotes a linear variation, whereas~$\Delta A := A - A_0$ is a
small but finite perturbation.
\end{Lemma}
\Proof The determinant can be expanded iteratively using the formula~\eqref{detexp} together with
\[ \frac{d}{d\tau}\, A(\tau)^{-1} = - A(\tau)^{-1} \: \dot{A}(\tau)\: A(\tau)^{-1} \:. \]
We thus obtain
\begin{align*}
&A^* A = A_0^2 + (\Delta A^*) \,A_0 + A_0 \,(\Delta A) + (\Delta A^*) (\Delta A) \\
&\log \det (\1_{\C^q} - A^* A) = \log \det \big( \1 - A_0^2 \big) \\
&\quad\: - \tr\Big( (\1 - A_0^2)^{-1}\: \big( (\Delta A^*) A_0  + A_0 (\Delta A) \big) \Big)
- \tr\Big( (\1 - A_0^2)^{-1}\: (\Delta A^*)(\Delta A) \Big) \\
&\quad\: - \frac{1}{2}\:\tr\Big( (\1 - A_0^2)^{-1}\: 
\big( (\Delta A^*) A_0  + A_0 (\Delta A) \big)\: (\1 - A_0^2)^{-1}\: \big( (\Delta A^*) A_0  + A_0 (\Delta A) \big) \Big) \\
&\quad\: + \O \big( (\Delta A)^3 \big) \:.
\end{align*}
Using~\eqref{A0def}, the term linear in~$\Delta A$ cancels with the variation~$\re \Tr(Y\,\Delta A)$
of the first summand in~\eqref{gdeflin}. Using that~$A_0$ is diagonal, we obtain
\begin{align*}
& g(A) - g(A_0) + \O \big( (\delta A)^3 \big) \\
&=- \sum_{i,j=1}^q \frac{1}{1-\lambda_j^2}\: |\delta A^i_j|^2
-\frac{1}{2} \sum_{i,j=1}^q \frac{1}{1-\lambda_i^2}\:\frac{1}{1-\lambda_j^2}\:
\big( \overline{\delta A^j_i} \,\lambda_j + \lambda_i \, \delta A^i_j \big)
\big( \overline{\delta A^i_j} \,\lambda_i + \lambda_j \, \delta A^j_i \big) \\
&=- \frac{1}{2} \sum_{i,j=1}^q \frac{1}{1-\lambda_i^2}\:\frac{1}{1-\lambda_j^2}\:
\bigg( 2\, |\delta A^i_j|^2 + \lambda_i \lambda_j \, \Big( (\delta A^i_j) (\delta A^j_i) + 
\overline{(\delta A^i_j) (\delta A^j_i)} \Big) \bigg) \:.
\end{align*}
Writing the bilinear terms as a matrix expectation value gives the result.
\QED

\Proof[Proof of Proposition~\ref{prpsaddlelin}.]
We now compute the Gaussian integrals. In the case~$i=j$, setting~$z=\delta A^i_i$,
the matrix expectation value becomes
\begin{align*}
&\bigg\la \begin{pmatrix} z \\[0.4em] \overline{z} \end{pmatrix}, 
\begin{pmatrix} 1 & \lambda_i^2 \\ \lambda_i^2 & 1 \end{pmatrix}
\begin{pmatrix} z \\[0.4em] \overline{z} \end{pmatrix} \bigg\ra_{\C^2}
= 2\, |z|^2 + \lambda_i^2\, \big( z^2 + \overline{z}^2 \big) \\
&= 2 \big( \re^2 z + \im^2 z \big) + \lambda_i^2\, \big( 2 \re^2 z - 2 \im^2 z \big)
= 2\, (1+\lambda_i^2)\: \re^2 z + 2\, (1-\lambda_i^2)\: \im^2 z \:.
\end{align*}
Thus
\begin{align*}
&\int_\C \exp \bigg( -
\frac{N}{2} \frac{1}{(1-\lambda_i^2)^2}\:
\bigg\la \begin{pmatrix} z \\[0.4em] \overline{z} \end{pmatrix}, 
\begin{pmatrix} 1 & \lambda_i^2 \\ \lambda_i^2 & 1 \end{pmatrix}
\begin{pmatrix} z \\[0.4em] \overline{z} \end{pmatrix} \bigg\ra_{\C^2} \bigg)\: dz \\
&= \int_{-\infty}^\infty d\re z \int_{-\infty}^\infty d\im z \:
\exp \bigg( - \frac{N}{(1-\lambda_i^2)^2} \: \Big( (1+\lambda_i^2)\: \re^2 z - (1-\lambda_i^2)\: \im^2 z \Big) \bigg) \\
&= \frac{\pi}{N}\: \frac{(1-\lambda_i^2)^2}{\sqrt{(1+\lambda_i^2) (1-\lambda_i^2)}} 
= \frac{\pi}{N}\: \frac{(1-\lambda_i^2)^2}{\sqrt{1-\lambda_i^4}} \:,
\end{align*}
where we again carried out the Gaussian integrals with the help of~\eqref{gaussian}.

Given~$i < j$, we get a four-dimensional Gaussian integral over~$z:=\delta A^i_j$
and~$\hat{z} := \overline{\delta A^j_i}$,
\begin{align*}
&\int_\C dz \int_\C d\hat{z} \exp \bigg( -
N \:\frac{1}{1-\lambda_i^2}\:\frac{1}{1-\lambda_j^2}\: 
\bigg\la \begin{pmatrix} \delta A^i_j \\[0.4em] \overline{\delta A^j_i} \end{pmatrix}, 
\begin{pmatrix} 1 & \lambda_i \lambda_j \\ \lambda_i \lambda_j & 1 \end{pmatrix}
\begin{pmatrix} \delta A^i_j \\[0.4em] \overline{\delta A^j_i} \end{pmatrix} \bigg\ra_{\C^2}
 \bigg) \\
&= \frac{\pi^2}{N^2}\: \frac{(1-\lambda_i^2)^2 (1-\lambda_j^2)^2}{1-\lambda_i^2 \lambda_j^2} \:.
\end{align*}

Putting these results together, we obtain
\begin{align*}
&\fint_{\G} e^{N \re \Tr(A Y)}\: d\mu_\G(\scrU) \\
&= e^{N \Tr(A_0 Y)}\: \big( \det(\1-A_0^2) \big)^{N-2q} \: 
\prod_{i,j=1}^q \frac{(1-\lambda_i^2) (1-\lambda_j^2)}{\sqrt{1-\lambda_i^2 \lambda_j^2}} \:
 \Big( 1 + \O\big(N^{-1} \big) \Big)
\end{align*}
(note that~$\Tr(A_0 Y)$ is real, making it possible to leave out the real part in the exponent).
Writing the products with determinants gives~\eqref{saddlelin}.

In order to prove~\eqref{linstrict}, we recall that in Lemma~\ref{lemmamaxlin} we showed
that the function~$g$ has a unique maximum at~$A_0$.
Since~$g(0)=0$, this uniqueness statement gives rise to the implication
\[ A_0 \neq 0 \quad \Longrightarrow \quad g(A_0) > 0 \:. \]
We finally note that, in view of~\eqref{A0def}, $A_0$ vanishes only if~$Y=0$.
This concludes the proof of~\eqref{linstrict}.
\QED

\subsection{An Exponential of a Quartic Functional} \label{secquartic}
We let~$\H=\C^N$ with the canonical scalar product.
We consider the finite one-dimensional set~$X$ formed of~$q$ points
\beq \label{Xdef}
X := \{1, \ldots, q\} \:.
\eeq
We let~$I\simeq \C^q$ be the space of complex-valued functions on~$X$.
We choose the canonical orthonormal basis~$(e_y)_{y \in X}$ by
\[ e_y : X \rightarrow \C \:,\qquad (e_y)(x) = \delta_{x,y} \:. \]
We regard~$I$ as a subspace of the Hilbert space~$\H$.
The wave evaluation operator is introduced for any~$x \in X$ by
\beq \label{wopex}
\Psi(x) : I \rightarrow \C \:,\quad e_y \mapsto e_y(x) \:.
\eeq
The computation
\[ \la \Psi(x)^* z \,|\, e_y \ra = \overline{z} \,\big( \Psi(x) \,e_y \big) = \overline{z}\, e_y(x) 
= \Big\la \sum_{a=1}^q \overline{e_a(x)}\: e_a \:z \:\Big|\: e_y \Big\ra \]
(with~$z \in \C$)
shows that the adjoint of the wave evaluation operator is given by
\beq \label{wopad}
\Psi(x)^* = \sum_{a=1}^q \overline{e_a(x)}\: e_a \:.
\eeq
We want to model the surface layer integral~\eqref{Tgen}, keeping~$\xi=y-x$ fixed,
in the simplified situation that all the wave functions are ``in phase'' along the light cone, meaning that~$\Psi(x)$
and~$\Psi(x+\xi)$ coincide. This leads us to considering the expression
\beq \label{gammaex}
\T(A_<, A_>) := \sum_{x \in X}
\Big| \tr_I \big( \scrU_> \,\Psi(x)^* \,\Psi(x) \,\scrU_<^{-1} \,\Psi(x)^* \,\Psi(x) \big) \Big|^2 \:.
\eeq
Substituting the form of the wave evaluation operator~\eqref{wopex} and its adjoint~\eqref{wopad},
we obtain
\begin{align}
\T(A_<, A_>) &= \sum_{x \in X} \Big| \tr_I \big( \scrU_> \,\Psi(x)^*\, \Psi(x)\, \scrU_<^*\, \Psi(x)^*\, \Psi(x) \big) \Big|^2
\label{ggauge} \\
&= \sum_{x \in X} 
\big(A_>\big)^x_x \,\big(A_<^*\big)^x_x \,\big(A_>^* \big)^x_x \,\big(A_<\big)^x_x \:. \label{fform}
\end{align}

We want to compute the integral
\beq \label{groupintegral}
\fint_\G d\mu_\G(\scrU_>) \fint_\G d\mu_\G(\scrU_<) \:e^{\beta N \T(A_<, A_>)} \:.
\eeq
Before entering the detailed computation, we illustrate the structure of this expression
by considering the case of one point.
\begin{Example} {\bf{(q=1)}} \label{exquarticq1}
{\em{ In the case~$q=1$, the functional~$\T$ in~\eqref{fform}
simplifies to
\[ \T = |a_<|^2\, |a_>|^2 \:. \]
Hence, using Theorem~\ref{thmgenint} in the case~$q=1$, we obtain
\begin{align*}
&\fint_\G d\mu_\G(\scrU_>) \fint_\G d\mu_\G(\scrU_<) \:e^{\beta N \T(A_<, A_>)} \\
&= \frac{(N-1)^2}{\pi^2} \int_{B_1} da_< \int_{B_1} da_>\:
e^{\beta N \:|a_<|^2\, |a_>|^2} \: \big( 1 - |a_<|^2 \big)^{N-2}\: \big( 1 - |a_>|^2 \big)^{N-2} \:,
\end{align*}
where~$B_1 \subset \C$ denotes the unit disc. Choosing polar coordinates, we obtain
\begin{align*}
&\fint_\G d\mu_\G(\scrU_>) \fint_\G d\mu_\G(\scrU_<) \:e^{\beta N \T(A_<, A_>)} \\
&= 4\, (N-1)^2 \int_0^1 r_< \: dr_< \int_0^1 r_>\: dr_>\:
e^{\beta N \:r_<^2\, r_>^2} \: \big( 1 - r_<^2 \big)^{N-2}\: \big( 1 - r_>^2 \big)^{N-2} \\
&= 4\, (N-1)^2 \int_0^1 r_< \: dr_< \int_0^1 r_>\: dr_>\: e^{N g(r_<, r_>)}\:
\: \frac{1}{(1 - r_<^2)^2}\: \frac{1}{( 1 - r_>^2)^2} \:,
\end{align*}
where
\[ g(r_<, r_>) := \beta \:r_<^2\, r_>^2 + \log \big( 1 - r_<^2 \big) + \log \big( 1 - r_<^2 \big) \:. \]

Our strategy for evaluating this integral asymptotically for large~$N$ is to compute the maxima
of the function~$g$ and to use the saddle point approximation. In order to compute the maxima,
we set the gradient of~$g$ to zero. This gives the equations
\[ \beta \, r_<^2 = \frac{1}{1-r_>^2} \qquad \text{and} \qquad \beta \, r_>^2 = \frac{1}{1-r_<^2} \:. \]
Solving the first equation for~$r_>^2$ and substituting it into the second equation,
we obtain the quadratic equation
\[ \big( r_>^2 \big)^2 - r_>^2 + \frac{1}{\beta} = 0 \:. \]
If~$\beta < 4$, this quadratic polynomial has no roots. This means that the function~$g$ has no
interior maxima. Its global maximum is at the origin,
\[ g(0,0) = 0 \:. \]
In the case~$\beta \geq 4$, however, the quadratic polynomial has two roots. We thus obtain the local
extrema at
\[ r_>^2 = r_<^2 = \lambda_- := \frac{1}{2} - \frac{1}{2} \:\sqrt{1 - \frac{4}{\beta}} \qquad \text{and} \qquad
r_>^2 = r_<^2 = \lambda_+ := \frac{1}{2} + \frac{1}{2} \:\sqrt{1 - \frac{4}{\beta}} \:. \]
The first extremum is a minimum, whereas the second is a local maximum.
The value of~$g$ at the local maximum is given by
\[ g(c,c) = \frac{c^2}{1-c^2} + 2 \log(1-c^2) \qquad \text{with} \qquad c := \sqrt{\lambda_+} \:. \]

For sufficiently large~$\beta$ (more precisely, if~$\beta > 4.911\ldots$), the local maximum is strictly
positive and thus the global maximum. This makes it possible to compute the integral
with the saddle point approximation. To this end, one computes the Hessian of~$g$ to
\[ D^2g(c,c) = -\frac{4}{(1-c^2)^2} \begin{pmatrix} c^2 & -(1-c^2) \\ -(1-c^2) & c^2 
\end{pmatrix} \:. \]
Hence the saddle point approximation gives
\begin{align*}
&\fint_\G d\mu_\G(\scrU_>) \fint_\G d\mu_\G(\scrU_<) \:e^{\beta N \T(A_<, A_>)} \:\Big( 1 + \O\big(N^{-1} \big) \Big) \\
&= 4\, (N-1)^2 \,\frac{c^2}{(1-c^2)^4}\, \int_0^1  dr_< \int_0^1 dr_>\: e^{N g(c,c)}
\:\frac{\pi}{N}\: \frac{1}{\sqrt{\det (-D^2g(c,c)/2)}} \\
&= 4 \pi N \,\frac{c^2}{(1-c^2)^4}\: e^{N g(c,c)}\: \frac{(1-c^2)^2}{2}
\: \frac{1}{\sqrt{2c^2-1}} \\
&= 2 \pi N \,\frac{1}{(1-c^2)^2}\: \frac{c^2}{\sqrt{2c^2-1}} \: \exp \bigg\{ N \Big( \frac{c^2}{1-c^2} + 2 \log(1-c^2) \Big)
\bigg\} \:.
\end{align*}
In this way, we have computed the double integral asymptotically for large~$N$.
}} \QEDrem
\end{Example}

We now state our main result for general~$q$. Exactly as in the last example, we only
get local maxima if~$\beta>4$. In this case, we can again use the saddle point approximation
to obtain the following result.
\begin{Thm} \label{thmsaddle} Assume that~$\beta>4$.
Then the integrand of the group integral~\eqref{groupintegral} with~$\T$ according to~\eqref{gammaex}
has saddle points at
\beq \label{Aphase}
A_< =c\: \diag \big( e^{i \varphi_<^1}, \ldots, e^{i \varphi_<^q} \big)\:,\qquad
A_> =c\: \diag \big( e^{i \varphi_>^1}, \ldots, e^{i \varphi_>^q} \big) \:,
\eeq
where~$\varphi_<^1, \ldots, \varphi_<^q$
and~$\varphi_>^1, \ldots, \varphi_>^q$ are arbitrary real phase angles, and~$c$ is given in terms of~$\beta$ by
\beq \label{cval}
c^2 = \frac{1}{2} + \frac{1}{2} \:\sqrt{1 - \frac{4}{\beta}} \:.
\eeq
The corresponding contribution to the group integral~\eqref{groupintegral}
has the large-$N$-asymptotics
\begin{align}
\fint_\G &d\mu_\G(\scrU_>) \fint_\G d\mu_\G(\scrU_<) \:e^{\beta N \T(A_<, A_>)} \notag \\
&\asymp (2 \pi N)^q\: \big( 1-c^4 \big)^{-q^2} \: \bigg( 
\frac{1+c^2}{1-c^2} \:\frac{c^2}{\sqrt{2c^2-1}} \bigg)^{q} \notag \\
&\qquad\: \times \exp \Big\{ N q \Big( \frac{c^2}{1-c^2} + 2 \log \big(1-c^2 \big) \Big) \Big\}\:
 \bigg(1 + \O \Big( \frac{1}{N} \Big) \bigg) \label{formulathmsaddle}
\end{align}
(here the symbol~$\asymp$ means that we restrict attention to the contribution near the
saddle points).
\end{Thm}

The remainder of this section is devoted to the proof of this theorem.
Applying Theorem~\ref{thmgenint}, the double integral~\eqref{groupintegral} can be rewritten as
\begin{align}
&\fint_\G d\mu_\G(\scrU_>) \fint_\G d\mu_\G(\scrU_<) \:e^{\beta N \T(A_<, A_>)} \notag \\
&= C \int_{\scrA_q}  d\mu_{\C^{q \times q}}(A_<) \int_{\scrA_q}  d\mu_{\C^{q \times q}}(A_>) \notag \\
&\qquad \qquad \times e^{\beta N \T(A_<, A_>)}
\big( \det (\1 - A_<^* A_<) \big)^{N-2q}\: \big( \det (\1 - A_>^* A_>) \big)^{N-2q} \notag \\
&= C \int_{\scrA_q}  d\mu_{\C^{q \times q}}(A_<) \int_{\scrA_q}  d\mu_{\C^{q \times q}}(A_>) \notag \\
&\qquad \qquad \times e^{N g(A_<, A_>)}
\big( \det (\1 - A_<^* A_<) \big)^{-2q}\: \big( \det (\1 - A_>^* A_>) \big)^{-2q} \label{dint}
\end{align}
with the function~$g(A_<, A_>)$ and the normalization constant~$C$ given by
\begin{gather}
g(A_<, A_>) := \beta \T(A_<, A_>) + \log \det (\1 - A_<^* A_<) + \log \det (\1 - A_>^* A_>)
\label{gAAdef} \\
C := \bigg( \frac{1}{\pi^{(q^2)}}\: \frac{(N-1)! \, \cdots \, (N-q)!}{(N-q-1)!\, \cdots\, (N-2q)!} \Bigg)^2 = \bigg( \frac{N}{\pi} \bigg)^{2q^2} \:\Big( 1 + \O\big(N^{-1} \big) \Big) \label{Cval} \:.
\end{gather}

The first step is to determine the maxima of the function~$g$.
\begin{Lemma} \label{lemmamax}
Assume that~$\beta>4$. Then the func\-tion~$g$ has a global maximum at~$(A_<, A_>)$
if and only if both matrices~$A_<$ and~$A_>$ are, up to phase factors on the diagonal,
the same multiple of the identity matrix, i.e.\
\beq \label{Aansatz}
A_< =c\: \diag \big( e^{i \varphi_<^1}, \ldots, e^{i \varphi_<^q} \big)\:,\qquad
A_> =c\: \diag \big( e^{i \varphi_>^1}, \ldots, e^{i \varphi_>^q} \big)
\eeq
with phase angles~$\varphi_{<,>}^x \in \R$. Moreover, the parameter~$c>0$ is
related to~$\beta$ by
\beq \label{cval2}
c^2 = \frac{1}{2} + \frac{1}{2} \:\sqrt{1 - \frac{4}{\beta}} \qquad \text{or} \qquad
\beta = \frac{1}{c^2\,(1-c^2)} \:.
\eeq
The maximum of~$g$ is given by
\beq \label{gmax}
g\big( c \1_{\C^q}, c \1_{\C^q} \big) = \frac{q\, c^2}{1-c^2} + 2q \log \big( 1-c^2 \big) \:.
\eeq
\end{Lemma}
\Proof Obviously, the functional~$\det (\1 - A^* A)$ does not change if~$A$ is multiplied
from the left or right by a unitary matrix, i.e.\ it is invariant under the transformation
\beq \label{Atrans}
A \rightarrow U A V \qquad \text{with} \qquad U,V \in \U(I) \:.
\eeq
With this in mind, we first compute the maximum of the function~$g$ under such transformations.
In view of~\eqref{gAAdef}, we need to maximize the functional~$\T(U A_< V, A_>)$.
We begin by varying~$A_<$, keeping~$A_>$ fixed. We set
\[ s_x := \big| (A_>)^x_x \big|^2 \geq 0 \:. \]
By renaming the points in~$X$ we can arrange that
\[ 0 \leq s_1 \leq \cdots \leq s_q \:. \]
Using this notation in~\eqref{fform}, we obtain
\begin{align}
&\T(U A_< V, A_>) = \sum_{x \in X} s_x\: \Big| \big(U A_< V \big)^x_x \big|^2 \leq
\sum_{x,y \in X} s_x\: \Big| \big(U A_< V \big)^y_x \big|^2 \label{fineq} \\
&= \sum_{x \in X} s_x\: \Big( \big(U A_< V \big) \big(V^* A_<^* U^* \big) \Big)^x_x
= \sum_{x \in X} s_x\: \big( U A_<  A_<^* U^* \big)^x_x =: G(U)
\end{align}
(in the inequality in~\eqref{fineq} we increased the sum by taking more non-negative summands).

Let us prove the estimate
\beq \label{Geigen}
G(U) \leq \sum_{\ell \in X} s_\ell\: \lambda_<^\ell \:,
\eeq
where~$\lambda^1_<, \ldots, \lambda^q_<$ are the eigenvalues of the symmetric matrix~$A_<  A_<^*$
in increasing order, i.e.
\beq \label{lamorder}
0 \leq \lambda_<^1 \leq \cdots \leq \lambda_<^q \:.
\eeq
To this end, let~$U$ be a maximum of~$G$ (this maximum is attained due to continuity of~$G$ and
compactness of the group~$\U(I)$). We consider a first order variation~$\delta U = i B U$
with a symmetric operator~$B$. Rewriting~$G(U)$ as a trace according to
\[ G(U) = \tr_I \big( S  \:U A_<  A_<^* U^* \big) \]
with~$S:=\diag(s_1, \ldots, s_q)$, maximality implies that
\[ 0 = \delta G(U) = i \tr_I \big( S\: B\:U A_<  A_<^* U^* - S\: U A_<  A_<^* U^* \:B^* \big) 
= \tr_I \big( B\:i \big[ U A_<  A_<^* U^*,\:S \big] \Big)\:. \]
Since~$B$ is arbitrary, it follows that the commutator vanishes,
\[ \big[ U A_<  A_<^* U^*,\:S \big] = 0 \:. \]
This shows that the matrix~$U A_<  A_<^* U^*$ is invariant on the eigenspaces of~$S$.
This in turns implies that~$G$ has the form~\eqref{Geigen} with~$\lambda_\ell$ the eigenvalues
of~$A_<  A_<^*$, in an arbitrary order. For the maximum, we obviously need to reorder the eigenvalues
according to~\eqref{lamorder}. This completes the proof of~\eqref{Geigen} and~\eqref{lamorder}.

We next study how to arrange equality in all these estimates. To this end, we form a polar decomposition
of~$A_<$ by choosing unitary operators~$U$ and~$V$ such that
\[ U A_< V = \diag \Big( \sqrt{\lambda_<^1}, \ldots, \sqrt{\lambda_<^q} \:\Big) \:. \]
In this way, the upper bound~\eqref{Geigen} is realized.
This diagonal matrix also realizes equality in~\eqref{fineq}.
Therefore, it is not only a maximizer of the functional~$G$, but it is even
a desired maximizer of the functional~$\T(U A_< V, A_>)$.

It remains to determine the general maximizer. To this end, we note that the
inequality in~\eqref{fineq} becomes an equality if and only if~$\big(U A_< V \big)^y_x$ vanishes for all~$x \neq y$.
In other words, the matrix~$U A_< V$ must be diagonal. By direct computation, one sees that
each diagonal entry must be a phase factor times the square root of the~$\lambda^\ell_<$. We conclude that
the maximum of~$\T(U A_< V, A_>)$ under variations of~$U$ and~$V$ is attained if and only if
\beq \label{Apolar}
U A_< V = \diag \Big( e^{i \varphi_<^1} \,\sqrt{\lambda_<^1}, \ldots, 
e^{i \varphi_<^q} \,\sqrt{\lambda_<^q} \:\Big)
\eeq
with arbitrary phase angles~$\varphi_<^1, \ldots, \varphi_<^q$.

We next interchange the roles of~$A_<$ and~$A_>$ and vary~$A_>$ while keeping~$A_<$ fixed.
We thus obtain that the maximum of~$g$ is attained for matrices~$U A_> V$ again of the form~\eqref{Apolar},
but with~$<$ replaced by~$>$. We conclude that, using the unitary invariance
of the functional~$\det (\1 - A^* A)$ under the transformation~\eqref{Atrans},
the maxima of~$g$ are attained if the matrices~$A_>$ and~$A_<$ are both diagonal.
Restricting attention to such diagonal matrices, the functional~$g(A_<, A_>)$ simplifies to
\beq \label{gsum}
g(A_<, A_>) = \sum_{x=1}^q \Big( \beta \lambda_<^x \lambda_>^x + \log \big(1 - \lambda_<^x \big) 
+ \log \big(1 - \lambda_>^x \big) \Big) \:.
\eeq
Now we can maximize for every~$x$ separately.
Similar as in Example~\ref{exquarticq1}, the criticality conditions are
\[ \beta \lambda_>^x = \frac{1}{1-\lambda_<^x} \:,\qquad \beta \lambda_<^x = \frac{1}{1-\lambda_>^x} \:. \]
Solving the first equation for~$\lambda^x_>$ and substituting it into the second equation,
we obtain a quadratic equation for~$\lambda_<^x$, having the two roots
\[ \lambda_\pm = \frac{1}{2} \pm \frac{1}{2} \:\sqrt{1 - \frac{4}{\beta}} \:. \]
By direct computation one sees that the plus sign gives the maximum.
Furthermore, one readily finds that~$\lambda_>^x=\lambda_<^x=\lambda_+=c^2$ with~$c^2$ according to~\eqref{cval2}.
Moreover, each summand in~\eqref{gsum} takes the value
\begin{align*}
&\beta \lambda_<^x \lambda_>^x + \log \big(1 - \lambda_<^x \big) + \log \big(1 - \lambda_>^x \big) \\
&= \beta\, c^4 + 2 \log \big(1-c^2\big) 
= \frac{c^2}{1-c^2} + 2 \log \big(1-c^2\big) \:,
\end{align*}
where in the last step we used that~$\beta$ can be expressed in terms of~$c$ by
the right equation in~\eqref{cval2}. Using this formula in~\eqref{gsum} gives~\eqref{gmax}.
This concludes the proof.
\QED

We next expand the function~$g$ near its maximum. In view of the gauge freedom
in~\eqref{Aansatz}, it suffices to consider the maximum at
\beq \label{Amax}
A_< = A_> = c\, \1
\eeq
with~$c$ as given by~\eqref{cval2}. It suffices to expand the matrices~$A_<$ and~$A_<$ to first order, i.e.\
\[ A_{<,>} = c\, \1 + \Delta A_{<,>} \qquad \text{with} \qquad \Delta A_{<,>} = \tau \:\delta A_{<,>} + \O \big( \tau^2 \big) \:, \]
where~$\tau$ denotes the perturbation parameter.
\begin{Lemma} \label{lemmaexpand}
Near the maximum~\eqref{Amax}, the function~$g(A_<, A_>)$ has the quadratic expansion
\begin{align}
&g(A_<, A_>) = \frac{q c^2}{1-c^2} + 2q \log (1-c^2) \\
&\quad\: + \frac{4 \tau^2}{1-c^2} \sum_{x \in X} \bigg(
\re \Big( \big(\delta A_>\big)^x_x \Big) \:\re \Big( \big(\delta A_<\big)^x_x \bigg) \label{positive} \\
&\quad\:-\frac{2 \tau^2 c^2}{(1-c^2)^2} \sum_{x \in X} \Big( \big| \re (\delta A_<)^x_x \big|^2 
+ \big| \re (\delta A_>)^x_x \big|^2 \Big) \label{negative} \\
&\quad\:- \frac{\tau^2}{1-c^2} \sum_{x \neq y} \Big( \big| (\delta A_<)^x_y \big|^2
+ \big| (\delta A_>)^x_y \big|^2 \Big) \label{qu1} \\
&\quad\:
- \frac{\tau^2 c^2}{2\,(1-c^2)^2} \sum_{x \neq y} \bigg( \Big| (\delta A_<)^x_y + \overline{(\delta A_<)^y_x} \Big|^2 
+  \Big| (\delta A_>)^x_y + \overline{(\delta A_>)^y_x} \Big|^2 \bigg) + \O \big( \tau^3 \big) \:. \label{qu2}
\end{align}
\end{Lemma}
\Proof We first expand the function~$\T(A_<, A_>)$ in~\eqref{fform},
\begin{align}
\T(A_<, A_>) &= \sum_{x \in X} 
\Big| \big(A_<\big)^x_x \Big|^2 \: \Big| \big(A_>^* \big)^x_x \Big|^2 \notag \\
&= q c^4 + 2 c^3 \sum_{x \in X} \re \Big( \big(\Delta A_<\big)^x_x + \big(\Delta A_>\big)^x_x \Big) \label{dT1} \\
&\quad\: + \tau^2 c^2 \sum_{x \in X} \Big( \big| (\delta A_<)^x_x \big|^2 +
\big| (\delta A_>)^x_x \big|^2 \Big) \\
&\quad\: + 4 \tau^2 c^2 \sum_{x \in X} \bigg(
\re \Big( \big(\delta A_>\big)^x_x \Big) \:\re \Big( \big(\delta A_<\big)^x_x \Big) \bigg)
+ \O \big( \tau^3 \big) \:.
\end{align}

The last two summands in~\eqref{gAAdef}, on the other hand, can be expanded
as in the proof of Lemma~\ref{lemmaexpandlin}. We thus obtain
\begin{align*}
&A^* A = c^2\,\1 + c\, (\Delta A^* + \Delta A) + (\Delta A^*) (\Delta A) \\
&\log \det (\1 - A^* A) = \log \det \big( (1-c^2)\,\1_{\C^q} \big) \\
&\quad\: - \tr\Big( (\1 - A^* A)^{-1}\: c \,\big( \Delta A^* + \Delta A \big) \Big)
- \tr\Big( (\1 - A^* A)^{-1}\: (\Delta A^*)(\Delta A) \Big) \\
&\quad\: - \frac{1}{2}\:\tr\Big( (\1 - A^* A)^{-1}\: c \,\big( \Delta A^* + \Delta A \big) (\1 - A^* A)^{-1}\: c \,\big( \Delta A^* + \Delta A \big) \Big) + \O \big( (\Delta A)^3 \big) \\
&=q \log (1-c^2) - \frac{c}{1-c^2}\: \tr\big( \Delta A^* + \Delta A \big) - \frac{\tau^2}{1-c^2}\:
\tr \big( (\delta A^*)(\delta A) \big) \\
&\quad\: - \frac{\tau^2 c^2}{2\,(1-c^2)^2}
\tr\Big( \big( \delta A^* + \delta A \big) \big( \delta A^* + \delta A \big) \Big)  + \O \big( \tau^3 \big) \\
&=q \log (1-c^2) - \frac{2c}{1-c^2} \sum_{x \in X} \re (\Delta A )^x_x
- \frac{\tau^2}{1-c^2} \sum_{x \in X} \big| (\delta A)^x_x \big|^2
- \frac{\tau^2}{1-c^2} \sum_{x \neq y} \big| (\delta A)^x_y \big|^2 \\
&\quad\:- \frac{2 \tau^2 c^2}{(1-c^2)^2} \sum_{x \in X} \big| \re (\delta A)^x_x \big|^2
- \frac{\tau^2 c^2}{2\,(1-c^2)^2} \sum_{x \neq y} \Big| (\delta A)^x_y + \overline{(\delta A)^y_x} \Big|^2 
+ \O \big( \tau^3 \big) \:,
\end{align*}
where in the last step we wrote the contributions by the diagonal and off-diagonal matrix entries separately.

Adding all the contributions and using the formula for~$\beta$ in~\eqref{cval2}
gives the result.
\QED

One complication in applying the saddle point approximation is that there is an underlying
{\em{local gauge symmetry}}.
This gauge symmetry is apparent in~\eqref{ggauge}, showing that~$\T$ is invariant under local phase transformations of~$\scrU_<$ and~$\scrU_>$,
\[ ( A_< )^x_y \rightarrow e^{i \tau \varphi_<(x)}\: ( A_< )^x_y \qquad \text{and} \qquad
( A_> )^x_y \rightarrow e^{i \tau \varphi_>(x)}\: ( A_> )^x_y \:, \]
where~$\varphi_<$ and~$\varphi_>$ are two arbitrary real-valued functions on~$X$.
Likewise, the global maximum constructed in Lemma~\ref{lemmamax} is unique only up to
the gauge phases in~\eqref{Aansatz}. Moreover, the gauge invariance becomes apparent
in the quadratic expansion in Lem\-ma~\ref{lemmaexpand} in that the imaginary parts of the diagonal matrix entries
\[ \im \delta A^x_x \qquad \text{with} \qquad x \in X \]
do not enter in the expansion.
The method to deal with the gauge invariance is to integrate over the resulting compact gauge group
before applying the saddle-point approximation. Then we can
work with arbitrary representatives of the gauge orbits. 
This justifies the following {\em{gauge-fixing procedure}}. 
Perturbing to first order around the maximum at~$A_{<,>}=c\1_{\C^q}$, we obtain
\[ A_{<,>} = c\,\1_{\C^q} + i \tau\:c\: \varphi_{<,>}(x) + \O(\tau^2) \:. \]
Thus we can use the linear gauge freedom in every order of perturbation theory
in order to arrange that the finite perturbations~$\Delta A_{<,>}$ satisfy for all~$k\in \{1,\ldots, q\}$
the constraints
\[ \im (\Delta A_{<,>})^x_x = 0 \qquad \text{for all~$a \in \{1, \ldots, q\}$}\:. \]

After this gauge fixing, we are ready to compute the Gaussian integral in the saddle point approximation.
\Proof[Proof of Theorem~\ref{thmsaddle}]
Our strategy is to carry out the Gaussian integral step by step using the relation~\eqref{gaussian}.
Before beginning, we integrate over the gauge phases of~$A_<$. This amounts to $q$ integrals over
a circle of radius~$c$, giving rise to the factor
\[ \big( 2 \pi c \big)^q \:. \]
Clearly, integrating over the gauge phases of~$A_>$ gives the same factor.
The terms in~\eqref{positive} and~\eqref{negative} give, for each~$x$, a two-dimensional
Gaussian integral with the integrand
\[ \exp \bigg(-N \:\bigg\la \begin{pmatrix} \Re (\delta A_>)^x_x \\ \Re (\delta A_<)^x_x \end{pmatrix},
E \begin{pmatrix} \Re (\delta A_>)^x_x \\ \Re (\delta A_<)^x_x \end{pmatrix} \bigg\ra_{\C^2}  \bigg)\]
and with the covariance matrix
\[ E := \frac{2}{(1-c^2)^2} \begin{pmatrix} c^2 & -(1-c^2) \\ -(1-c^2) & c^2 \end{pmatrix} 
\:. \]
Gaussian integration gives a factor
\[ \frac{\pi}{\sqrt{\det(N E)}} = \frac{\pi}{N}\:  \frac{(1-c^2)^2}{\sqrt{8c^2-4}}\:. \]
In this way, for the resulting $2q$-dimensional Gaussian integrals over~$\re (\delta A_<)^x_x$
and $\re (\delta A_>)^x_x$
we obtain
\[ \bigg( \frac{\pi}{N}\:  \frac{(1-c^2)^2}{\sqrt{8c^2-4}} \bigg)^{q}\:. \]

We next consider the off-diagonal matrix entries~$(\delta A_<)^x_y$ for~$x \neq y$.
Multiplying out~\eqref{qu2}, one sees that for each~$x \neq y$ one gets a Gaussian integral
with the covariance matrix
\[ \exp \bigg(-N \:\bigg\la \begin{pmatrix} \Re (\delta A_>)^x_y \\[0.2em] \Re (\delta A_<)^y_x \end{pmatrix},
F \begin{pmatrix} \Re (\delta A_>)^x_y \\[0.2em] \Re (\delta A_<)^y_x \end{pmatrix} \bigg\ra_{\C^2}  \bigg)\]
with the covariance matrix
\[ F := \frac{1}{(1-c^2)^2} \begin{pmatrix} 1 & c^2 \\ c^2 & 1 \end{pmatrix} 
\:. \]
Thus Gaussian integration gives a factor
\[ \frac{\pi}{\sqrt{\det(N F)}} = \frac{\pi}{N}\:  \frac{(1-c^2)^2}{\sqrt{1-c^4}}\:. \]
The integral over~$\Im (\delta A_<)^y_x$ and~$\Im (\delta A_<)^y_x$ gives the same factor.
Moreover, the integrals over the off-diagonal terms of~$A_>$ can be treated in the same way.
Keeping in mind that we have~$q(q-1)/2$ ways to choose indices~$x,y$ with~$x<y$, we end up
with the factor
\[ \bigg( \frac{\pi}{N}\:  \frac{(1-c^2)^2}{\sqrt{1-c^4}} \bigg)^{2q(q-1)}\:. \]

Finally, the additional factors in~\eqref{dint} need to be evaluated at the maximum. 
We thus obtain the factor
\[ \big( \det (\1 - A^* A) \big)^{-2q} \big|_{A = c\1} = \big( \big(1-c^2 \big)^q \big)^{-2q} = 
(1-c^2)^{-2q^2} \:, \]
again for~$A_<$ and~$A_>$.

Combining all the terms gives
\begin{align*}
&\fint_\G d\mu_\G(\scrU_>) \fint_\G d\mu_\G(\scrU_<) \:e^{\beta N \T(A_<, A_>)} \\
&= C\: \big( 2 \pi c \big)^{2q} \: \bigg( \frac{\pi}{N}\:  \frac{(1-c^2)^2}{\sqrt{8c^2-4}} \bigg)^{q}\:
\bigg( \frac{\pi}{N}\:  \frac{(1-c^2)^2}{\sqrt{1-c^4}} \bigg)^{2q(q-1)}\:(1-c^2)^{-4q^2} \\
&\quad\: \times \exp \Big\{ N q \Big( \frac{c^2}{1-c^2} + 2 \log \big(1-c^2 \big) \Big) \Big\}\:
 \bigg(1 + \O \Big( \frac{1}{N} \Big) \bigg) \\
 &= C\: 2^q \:c^{2q} \:\frac{\pi^{2q^2+q}}{N^{2q^2-q}} \bigg( \frac{1}{\sqrt{2c^2-1}} \bigg)^{q}\:
\bigg( \frac{1}{\sqrt{1-c^4}} \bigg)^{2q(q-1)}\:(1-c^2)^{-2q} \\
&\quad\: \times \exp \Big\{ N q \Big( \frac{c^2}{1-c^2} + 2 \log \big(1-c^2 \big) \Big) \Big\}\:
 \bigg(1 + \O \Big( \frac{1}{N} \Big) \bigg) \\
 &= C\: 2^q \:c^{2q} \:\frac{\pi^{2q^2+q}}{N^{2q^2-q}} \bigg( \frac{1}{\sqrt{2c^2-1}} \bigg)^{q}\:
\big(1-c^4 \big)^{-q^2} \:\frac{(1-c^4)^q}{(1-c^2)^{2q}} \\
&\quad\: \times \exp \Big\{ N q \Big( \frac{c^2}{1-c^2} + 2 \log \big(1-c^2 \big) \Big) \Big\}\:
 \bigg(1 + \O \Big( \frac{1}{N} \Big) \bigg)
\end{align*}
with~$C$ according to~\eqref{Cval}. We conclude that
\begin{align*}
&\fint_\G d\mu_\G(\scrU_>) \fint_\G d\mu_\G(\scrU_<) \:e^{\beta N \T(A_<, A_>)} \\
&= (2 \pi N)^q \:c^{2q} \:\bigg( \frac{1}{\sqrt{2c^2-1}} \bigg)^{q}\:
\big(1-c^4 \big)^{-q^2} \:\frac{(1-c^4)^q}{(1-c^2)^{2q}} \\
&\quad\: \times \exp \Big\{ N q \Big( \frac{c^2}{1-c^2} + 2 \log \big(1-c^2 \big) \Big) \Big\}\:
 \bigg(1 + \O \Big( \frac{1}{N} \Big) \bigg) \:.
\end{align*}
This gives the result.
\QED

\section{Computation of the Localized Refined Pre-State} \label{secrefinecompute}
\subsection{The  Saddle Point Asymptotics for the High-Energy Wave Functions} \label{sechighenergy}
We now want to extend the construction of the previous section to the situation in
four-dimensional Minkowski space. Our goal is to compute the localized refined pre-state
as introduced in Section~\ref{seclocdetail}.
We choose sets~$\tilde{\Omega} \subset \tilde{M}$ and~$\Omega \subset M$
as well as two sets~$\tilde{V} \subset \tilde{M}$ and~$V \subset M$ of the same finite volume,
\[ \rho(V) = \tilde{\rho}(\tilde{V}) < \infty \:, \]
and consider the partition function~\eqref{Zreflocfinal}, where
the functional~$\T^t_V( \tilde{\rho}, T_{\scrU_<, \scrU_>} \rho)$ is given by~\eqref{Tgen}.
This functional has the advantage that it is relatively easy to compute. Therefore,
it is a good starting point for the detailed computation of quantum states.

In order to find the saddle point and to determine their scaling behavior,
it suffices to consider the case that both~$\rho$ and~$\tilde{\rho}$
describe the Minkowski vacuum with a regularization on the scale~$\varepsilon$.
We again work in finite spatial volume as introduced in Section~\ref{seclocalize}.
Thus, instead of Minkowski space, as in~\eqref{Mcylinder}
we consider again the spacetime cylinder~$\scrM = \R \times [-L,L]^3$.
We consider solutions of the Dirac equation~$(i \Pdd-m)\psi=0$ in this cylinder with periodic boundary conditions
with the usual scalar product
\[ ( \psi | \phi)_m := \int_{[-L,L]^3} \Sl \psi(t,\vec{x})\,|\, \gamma^0\, \phi(t,\vec{x}) \Sr \: d^3x \:. \]
We denote the resulting Hilbert space of Dirac solutions by~$(\H_m, (.|.)_m)$.
Next we choose~$\H=\tilde{\H}$ as the subspace of all negative energy solutions with a
 cutoff on the scale~$1/\varepsilon$, i.e.\
\[ -\frac{1}{\varepsilon} < \omega < 0 \:. \]
Denoting the local correlation map by (see also~\eqref{Fid})
\[ F^\varepsilon \::\: \scrM \rightarrow \F \:,\qquad x \mapsto -\Psi^\varepsilon(x)^* \Psi^\varepsilon(x) \:, \]
we choose
\[ \tilde{\rho} = \rho = (F^\varepsilon)_* \mu \:, \]
where~$d\mu = d^4x$ is the volume measure on~$\scrM$. The local correlation map
makes it possible to identify Minkowski space~$\scrM$ with the spacetimes~$M$ and~$\tilde{M}$.
For convenience, in what follows we shall work in Minkowski space.
Next, we choose the sets~$\Omega$ and~$\tilde{\Omega}$ as the past of the Cauchy surface
of time zero, i.e.\
\[ \tilde{\Omega} = \Omega = \{(t, \vec{x}) \in \scrM \,|\, t <0 \} \:. \]
Next, we choose the sets~$\tilde{V} =V=[-T,T] \times [-\ell,\ell]^3$ which localize
the state again as in~\eqref{Vbox}.
The wave functions which are localized mainly inside the three-dimensional box~$[-\ell, \ell]^3$
can be associated to a subspace~$\H^\lab \subset \H$
whose dimension has the scaling behavior~\eqref{Hlabdim}.
In this setting, the functional~$\T^t_V$ can be written as
\begin{align}
\T^t_V( \tilde{\rho}, T_{\scrU_<, \scrU_>} \rho) \notag 
&= \bigg( \int_{\Omega \cap V}\!\! d^4x \int_{\Omega \cap V} \!\!d^4y
+ \int_{(M \setminus \Omega) \cap V} \!\!\!\!\!d^4x
\int_{(M \setminus \Omega) \cap V} \!\!\!\!\!d^4y \bigg) \\
&\qquad \times 
\big| F^\varepsilon(x)\, \scrU_> \,F^\varepsilon(y)\, \scrU_<^{-1} \big|^2 \:.
\label{TMink}
\end{align}

In order to understand the structure of the functional~$\T^t_V$ in~\eqref{TMink}, we note that the
unitary operators~$\scrU_>$
and~$\scrU_<$ involve many phases, typically leading to destructive interference.
If~$\scrU_<$ and~$\scrU_>$ are chosen as a multiple of the identity, then no 
destructive interference occurs, making the functional~$\T^t_V$ large.
On the other hand, the set of unitary operators which are close to a multiple of the identity
has a very small measure, making the contribution to the partition function smaller.
This suggests that, in order to identify the leading contributions to the partition function,
one should consider configurations where the unitary operators~$\scrU_<$ and~$\scrU_>$ are close to
multiples of the identity on a suitable {\em{subspace}} of~$\H^\lab$, which we denote by~$\H^\sp$.
Using a notation similar to~\eqref{ABmatrix}, we set
\[ A_{<,>} := \pi_{\H^\sp} \,\scrU_{<,>} \big|_{\H^\sp} \:. \]
We expect a saddle point if these matrices are multiples of the identity matrix, e.g.\
\beq \label{Aid}
A_< = A_> = c \,\1_{\H^\sp}
\eeq
for suitable~$c \in \R$. We make the simplifying assumption that, on the orthogonal complement
of~$\H^\sp$, destructive interference does occur, making it possible to 
restrict attention in the computation to the subspace~$\H^\sp$. Proceeding in this way, we can 
compute the group integrals again with saddle point methods.
Since there are many possible choices of~$\H^\sp$, there will also be many saddle points.
Our task is to determine their scaling behavior and combinatorics.

The main step in the quantitative analysis is to compute how the contribution
of the saddle point to the partition function depends on the dimension of the subspace~$\H^\sp$.
We begin this analysis by restricting attention to the wave functions with energy on the Planck scale,
\[ |\omega| \simeq \frac{1}{\varepsilon} \:. \]
We also refer to these wave functions as the {\em{high-energy wave functions}}. We denote all the high-energy
wave functions of our system by
\[ \H^\he \subset \H^\lab \:. \]
This analysis will make it possible to determine the scaling behavior in~$\varepsilon$
(the low-energy wave functions will not change this scaling behavior, but they will modify the structure
of the saddle points; this will be worked out in Section~\ref{seclowenergy} below.)

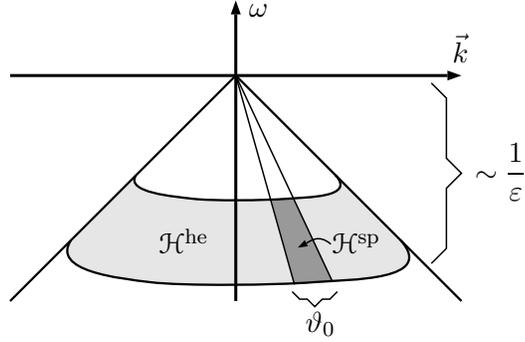
\begin{figure}
\psscalebox{1.0 1.0} 
{
\begin{pspicture}(0,27.770296)(6.060337,31.909168)
\definecolor{colour2}{rgb}{0.9019608,0.9019608,0.9019608}
\definecolor{colour1}{rgb}{0.6,0.6,0.6}
\pspolygon[linecolor=colour2, linewidth=0.02, fillstyle=solid,fillcolor=colour2](4.4239388,29.459436)(5.263939,28.619436)(5.303939,28.499435)(5.323939,28.419436)(5.2572722,28.319435)(5.077272,28.239435)(4.837272,28.199436)(4.343939,28.15277)(3.9239388,28.126102)(3.4972723,28.099436)(3.0439389,28.099436)(2.477272,28.099436)(1.9172721,28.119436)(1.4639388,28.19277)(1.0972722,28.27277)(0.85727215,28.386103)(0.79727215,28.472769)(0.7906055,28.579435)(0.8172721,28.65277)(1.6772721,29.499435)(1.6706054,29.439436)(1.7372721,29.366102)(1.8439388,29.319435)(2.0439389,29.27277)(2.3506055,29.226103)(2.783939,29.199436)(3.3306055,29.199436)(3.7039387,29.219437)(4.0039387,29.239435)(4.2506056,29.279436)(4.377272,29.326103)(4.4239388,29.392769)
\psline[linecolor=colour1, linewidth=0.02, fillstyle=solid,fillcolor=colour1](3.4956055,29.209436)(3.7756054,29.219437)(4.2906055,28.144436)(3.7956054,28.119436)
\psline[linecolor=black, linewidth=0.04, arrowsize=0.05291667cm 2.0,arrowlength=1.4,arrowinset=0.0]{->}(3.0206056,27.869436)(3.0206056,31.869436)
\psline[linecolor=black, linewidth=0.04, arrowsize=0.05291667cm 2.0,arrowlength=1.4,arrowinset=0.0]{->}(0.02060547,30.869436)(6.0206056,30.869436)
\psline[linecolor=black, linewidth=0.03](3.0206056,30.869436)(6.0206056,27.869436)
\psline[linecolor=black, linewidth=0.03](3.0206056,30.869436)(0.02060547,27.869436)
\psbezier[linecolor=black, linewidth=0.03](0.82949436,28.678326)(0.56060547,28.267214)(1.4593775,28.1716)(1.8906054,28.11943603515625)(2.3218334,28.067272)(3.5506055,28.079435)(4.0506053,28.119436)(4.5506053,28.159435)(5.6918335,28.137272)(5.2106056,28.679436)
\psline[linecolor=black, linewidth=0.02](5.6206055,30.769436)(5.8206053,30.569435)(5.8206053,29.769436)(6.0206056,29.569435)(5.8206053,29.369436)(5.8206053,28.569435)(5.6206055,28.369436)
\psline[linecolor=black, linewidth=0.02](3.0206056,30.869436)(3.8006055,28.089436)
\psline[linecolor=black, linewidth=0.02](3.0206056,30.869436)(4.3006053,28.139437)
\psbezier[linecolor=black, linewidth=0.03](1.6944944,29.5438)(1.5409896,29.30888)(2.2090852,29.259243)(2.455267,29.22943603515625)(2.7014484,29.199629)(3.4029362,29.20658)(3.6883788,29.229437)(3.9738214,29.252293)(4.6203313,29.245295)(4.3456054,29.544436)
\psline[linecolor=black, linewidth=0.02](3.7706056,27.994436)(3.8706055,27.894436)(3.9706054,27.894436)(4.0706053,27.794436)(4.1706057,27.894436)(4.2706056,27.894436)(4.3706055,27.994436)
\psbezier[linecolor=black, linewidth=0.02, arrowsize=0.05291667cm 2.0,arrowlength=1.4,arrowinset=0.0]{->}(4.283939,28.626102)(4.123939,28.766102)(4.0106053,28.666103)(3.8439388,28.579436035156274)
\rput[bl](3.2,31.65){$\omega$}
\rput[bl](5.9,31.05){$\vec{k}$}
\rput[bl](3.95,27.4){$\vartheta_0$}
\rput[bl](6.2,29.2){$\displaystyle \sim \frac{1}{\varepsilon}$}
\rput[bl](4.3,28.5){$\H^\sp$}
\rput[bl](2,28.5){$\H^\he$}
\end{pspicture}
}
\caption{The subsystems~$\H^\sp \subset \H^\he \subset \H^\lab$ in momentum space.}
\label{figIcone}
\end{figure}

Next, in order to maximize the contribution to~$\T^t_V$,
we choose~$\H^\sp$ as the subset of wave functions whose spatial momenta lie in a cone with
opening angle~$\vartheta_0 \in [0, \pi)$, i.e.\
\beq \label{Ccone}
p=(\omega, \vec{k}) \qquad \text{with} \qquad \vec{k} \in C_{\vartheta_0}(\vec{k}_0)
:= \big\{ \vec{k} \in \R^3 \:\big|\: \sphericalangle (\vec{k}, \vec{k}_0) \leq \vartheta_0 \big\} \:.
\eeq
This subset is illustrated in Figure~\ref{figIcone}.
The dimension of~$\H^\sp$ scales like
\beq \label{Idim}
\dim \H^\sp \sim \sin^2 \vartheta_0\: \dim \H^\he \sim \vartheta_0^2 \:\Big( \frac{\ell}{\varepsilon} \Big)^3\:.
\eeq

Having chosen~$\H^\sp$, we next compute the form of the resulting saddle point.
To this end, we can use methods and results of the light-cone expansion
and the continuum limit analysis
(see the preliminaries in Section~\ref{seccfsmink} and~\cite{cfs})
and combine them with methods introduced in~\cite{action}
for the computation of surface layer integrals.
We begin with the regularized kernel of the fermionic projector in the vacuum, for simplicity
with $i \varepsilon$-regularization,
\beq \label{Peps}
P^\varepsilon(x,y) = \int_{\R^4} \frac{d^4p}{(2 \pi)^4} \: (\slashed{p}+m)\:
\delta(p^2 - m^2)\: \Theta(-p^0)\: e^{\varepsilon p^0} e^{-ip (x-y)} \:.
\eeq
In the limit~$\varepsilon \searrow 0$, this distribution develops singularities on the 
light cone. Likewise, for small~$\varepsilon>0$, this distribution has large contributions near the 
light cone. These contributions can be analyzed in detail with the so-called
regularized light-cone expansion and the formalism of the continuum limit
(for details see the preliminaries in Section~\ref{seccfsmink} and~\cite[Sections~2.2. and~2.4]{cfs}
or~\cite{reghadamard}). Here we are interested only in the scaling behavior in~$\varepsilon$,
making it possible to work with the following simple argument.
Let~$\xi:=y-x$ be a vector on the light cone (i.e.\ $\xi \neq 0$ and $\xi^2=0$).
The Fourier integral~\eqref{Peps} involves the oscillating phase factor~$\exp(i p \xi)$.
The oscillations give rise to destructive interference. This is why the leading contribution to the
Fourier integral is obtained simply by restricting attention to those momenta for which~$p \xi$ is small.
For determining the scaling behavior in~$\varepsilon$, it suffices to integrate over the set
\[ \Phi(\xi) := \big\{ p \in \R^4 \:\big|\:  |p \xi| \leq 1 \big\} \]
and leave out the phase factor. We thus obtain
\beq \label{Pphase}
P^\varepsilon(x,y) \simeq \int_{\Phi(\xi)} \frac{d^4p}{(2 \pi)^4} \: (\slashed{p}+m)\:
\delta(p^2 - m^2)\: \Theta(-p^0)\: e^{\varepsilon p^0} \big( 1+ \O(\varepsilon/t) \big) \:.
\eeq
We compute this integral in order to illustrate that this computation is indeed compatible with
the scaling obtained in the formalism of the continuum limit.
Without loss of generality, we only consider the case that~$\xi$ is on the upper light cone, i.e.\
\[ t := \xi^0 = |\vec{\xi}| \:. \]
Moreover, we set~$p=(\omega, \vec{k})$ and introduce polar coordinates~$(k, \vartheta, \varphi)$ with
\[ k := |\vec{k}| \qquad \text{and} \qquad \vec{\xi} \,\vec{k} = -t k\: \cos \vartheta \]
(here the minus sign has the advantage that for small~$\vartheta$, the vector~$p$ 
is in~$\Phi(\xi)$).
Then the set~$\Phi(\xi)$ is characterized by the inequality
\[ -\omega t + k \,|\vec{\xi}|\, \cos \vartheta \leq 1 \:. \]
Again restricting attention to large energies, we can set~$\omega \approx -k$. Moreover, we can
expand for small angles to obtain
\[ k t - k t\, \Big( 1 - \frac{\vartheta^2}{2} \Big) \leq 1 + kt\: \O \big(\vartheta^4 \big) \]
and thus
\beq \label{varthetaprelim}
\vartheta \leq \hat{\vartheta}\: \big( 1 + \O(\hat{\vartheta}^2) \big) \qquad \text{with}
\qquad \hat{\vartheta}^2 := \frac{1}{k t} \:.
\eeq
Using this formula in~\eqref{Pphase} gives
\begin{align*}
P^\varepsilon(x,y) &\simeq \int_0^\infty k^2\: dk \int_0^{\hat{\vartheta}} 
\sin \vartheta\: d\vartheta \:\frac{1}{|\omega|} (\slashed{p}+m)\:
e^{\varepsilon \omega} \Big|_{\omega=-k} \:\big( 1 + \O(\hat{\vartheta}^2) \big) \\
&\simeq \int_0^\infty k^2\: dk \: \frac{1}{kt} \frac{1}{|\omega|} (\slashed{p}+m)\:
e^{\varepsilon \omega} \Big|_{\omega=-k} \:\big( 1 + \O(\hat{\vartheta}^2) \big) \\
&\simeq \frac{1}{t} \int_0^\infty k\: e^{-\varepsilon k} dk\:\big( 1 + \O(\hat{\vartheta}^2) \big)
\simeq \frac{1}{\varepsilon^2 t}\:\big( 1 + \O(\hat{\vartheta}^2) \big) \:.
\end{align*}
This is compatible with the formalism of the continuum limit, where
\beq \label{Pcont}
P^\varepsilon(x,y) \simeq \xi\slsh \, T^{(-1)} + (\deg<2)
\simeq \frac{\xi\slsh}{\varepsilon^2 t^2} + (\deg<2) \simeq \frac{1}{\varepsilon^2 t} + (\deg<2) \:.
\eeq

Before going on, we slightly simplify the setup. Using that the
leading terms for small~$\varepsilon$ come from the wave functions of energy~$k \sim \omega \sim \varepsilon^{-1}$,
we may replace the factor~$k$ in the definition of~$\hat{\vartheta}$ in~\eqref{varthetaprelim}
by~$1/\varepsilon$,
\beq \label{varthetamax}
\hat{\vartheta}^2 := \frac{\varepsilon}{t} \:.
\eeq
This changes the computation of the Fourier integral to
\begin{align}
P^\varepsilon(x,y) &\simeq \int_0^\infty k^2\: dk \int_0^{\hat{\vartheta}} 
\sin \vartheta\: d\vartheta \:\frac{1}{|\omega|} (\slashed{p}+m)\:
e^{\varepsilon \omega} \Big|_{\omega=-k} \:\big( 1+ \O(\varepsilon/t) \big) \notag \\
&\simeq \frac{\varepsilon}{t} \int_0^\infty k^2\: e^{-\varepsilon k} dk\:\big( 1+ \O(\varepsilon/t) \big) 
\simeq \frac{1}{\varepsilon^2 t}\:\big( 1+ \O(\varepsilon/t) \big) \:, \label{Fourier1}
\end{align}
without any effect on the scalings. The choice~\eqref{varthetamax} will simplify the
following con\-side\-ra\-tions.

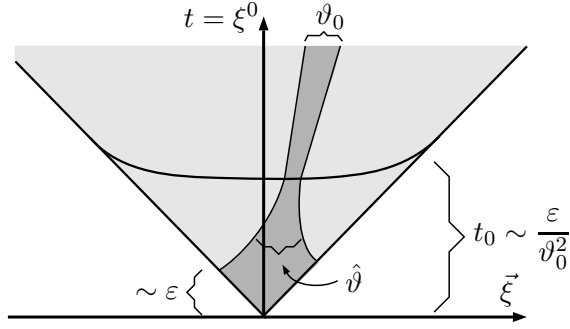
\begin{figure}[tb]
\psscalebox{1.0 1.0} 
{
\begin{pspicture}(0,27.774542)(6.9497313,31.904922)
\definecolor{colour0}{rgb}{0.9019608,0.9019608,0.9019608}
\definecolor{colour1}{rgb}{0.7019608,0.7019608,0.7019608}
\pspolygon[linecolor=colour0, linewidth=0.02, fillstyle=solid,fillcolor=colour0](0.11,31.26519)(0.11,31.46519)(6.91,31.46519)(3.41,27.86519)
\pspolygon[linecolor=colour1, linewidth=0.01, fillstyle=solid,fillcolor=colour1](3.97,31.46519)(4.43,31.46519)(4.3233333,31.065191)(4.19,30.698524)(3.97,29.96519)(3.8966668,29.691856)(3.8766668,29.34519)(3.9166667,28.978523)(4.01,28.731857)(4.136667,28.591858)(3.4166667,27.885191)(2.8233333,28.485191)(3.0366666,28.645191)(3.2566667,28.871857)(3.47,29.171858)(3.6166666,29.471857)(3.6966667,29.698524)(3.8566666,30.671858)(3.9233334,31.131857)
\psline[linecolor=black, linewidth=0.04, arrowsize=0.05291667cm 2.0,arrowlength=1.4,arrowinset=0.0]{->}(3.41,27.86519)(3.41,31.86519)
\psline[linecolor=black, linewidth=0.04, arrowsize=0.05291667cm 2.0,arrowlength=1.4,arrowinset=0.0]{->}(0.01,27.86519)(6.91,27.86519)
\psline[linecolor=black, linewidth=0.03](0.11,31.26519)(3.4433334,27.848524)
\psline[linecolor=black, linewidth=0.03](6.91,31.46519)(3.41,27.881857)
\psbezier[linecolor=black, linewidth=0.03](1.1033334,30.251858)(1.5083805,29.759476)(2.2827008,29.736568)(2.6655414,29.721857096354174)(3.048382,29.707146)(3.8777707,29.68757)(4.2832613,29.721857)(4.688752,29.756144)(5.2883897,29.797146)(5.75,30.271856)
\psline[linecolor=black, linewidth=0.02](5.6766667,29.931856)(5.8766665,29.731857)(5.8766665,29.131857)(6.076667,28.931856)(5.8766665,28.731857)(5.8766665,28.131857)(5.6766667,27.931856)
\psline[linecolor=black, linewidth=0.02](3.31,28.90019)(3.41,28.80019)(3.51,28.80019)(3.61,28.70019)(3.71,28.80019)(3.81,28.80019)(3.91,28.90019)
\psline[linecolor=black, linewidth=0.02](3.6833334,29.678524)(3.9633334,31.471857)
\psline[linecolor=black, linewidth=0.02](3.9033334,29.70519)(4.43,31.471857)
\psbezier[linecolor=black, linewidth=0.02](3.9033334,29.70519)(3.83,29.098524)(3.95,28.70519)(4.13,28.6051904296875)
\psbezier[linecolor=black, linewidth=0.02](3.6833334,29.698524)(3.5233333,29.098524)(3.07,28.645191)(2.8233333,28.491857096354167)
\psline[linecolor=black, linewidth=0.02](4.4566665,31.53519)(4.3966665,31.610685)(4.2566667,31.609447)(4.21,31.66019)(4.1566668,31.609447)(3.9966667,31.610685)(3.9566667,31.53519)
\psbezier[linecolor=black, linewidth=0.02, arrowsize=0.05291667cm 2.0,arrowlength=1.4,arrowinset=0.0]{->}(4.3933334,28.348524)(4.193333,28.241858)(3.82,28.268524)(3.66,28.561857096354164)
\psline[linecolor=black, linewidth=0.02](2.59,28.493525)(2.49,28.393524)(2.49,28.293524)(2.39,28.193523)(2.49,28.093523)(2.49,27.993525)(2.59,27.893524)
\rput[bl](2.35,31.65){$t=\xi^0$}
\rput[bl](6.55,28){$\vec{\xi}$}
\rput[bl](6.2,28.5){$\displaystyle t_0 \sim \frac{\varepsilon}{\vartheta_0^2}$}
\rput[bl](4.1,31.7){$\vartheta_0$}
\rput[bl](4.5,28.2){$\hat{\vartheta}$}
\rput[bl](1.7,28.1){$\sim \varepsilon$}
\end{pspicture}
}
\caption{The support of the closed chain as a function of~$\xi$.}
\label{figsuppA}
\end{figure}

The next step is to adapt the above method to the case where we only take into account the wave functions
in the subspace~$\H^\sp \subset \H^\he$. This simply corresponds to replacing the integration domain~$\Phi(\xi)$
in~\eqref{Pphase} by
\[ \Phi_\sp(\xi) := \big\{ p \in \R^4 \:\big|\:  |p \xi| \leq 1 \text{ and } \vec{k} \in C_{\vartheta_0}(\vec{k}_0) \big\} \:, \]
where~$C_{\vartheta_0}(\vec{k}_0)$ is again the cone in~\eqref{Ccone}.
For the computation of the resulting Fourier integral, it is convenient to distinguish the cases
where~$\hat{\vartheta}$ is larger or smaller than~$\vartheta_0$.
This corresponds to the two cases
\[ t< t_0 \qquad \text{respectively} \qquad t \geq t_0 \]
with
\[ t_0 := \frac{\varepsilon}{\vartheta_0^2} \:. \]
In the case~$t \gg t_0$, the computation~\eqref{Fourier1} as well as~\eqref{Pcont} remain valid, but only
if~$\vec{\xi}$ lies inside the cone around~$\vec{k}_0$ with opening angle~$\vartheta_0$, i.e.\
\[ P^\varepsilon(x,y) \simeq \frac{\xi\slsh}{\varepsilon^2 t^2}\: \chi_{C_{\vartheta_0}(\vec{k}_0)}(\vec{\xi}) 
\qquad \text{if~$t \gg t_0$} \:. \]
In the opposite case~$t \ll t_0$, we get a contribution whenever~$\vec{\xi}$ lies
inside the cone around~$\vec{k}_0$ with opening angle~$\hat{\vartheta}$. In this case, all the
vectors in~$\H^\sp$ are ``in phase.'' However, these are fewer wave functions than in~\eqref{Fourier1}.
Since the number of wave functions scales quadratically in the opening angle of the cone, we obtain
\beq \label{Pepssmall}
P^\varepsilon(x,y) \simeq \frac{\xi\slsh}{\varepsilon^2 t^2}\: \frac{\vartheta_0^2}{\hat{\vartheta}^2}\:
\chi_{C_{\hat{\vartheta}}(\vec{k}_0)}(\vec{\xi}) \simeq \frac{\xi\slsh}{\varepsilon^2 t \,t_0}\:
\chi_{C_{\hat{\vartheta}}(\vec{k}_0)}(\vec{\xi}) \qquad \text{if~$t \ll t_0$} \:.
\eeq
We next compute the closed chain~$A_{xy} := P^\varepsilon(x,y)\, P^\varepsilon(y,x)$.
In the case~$t \gg t_0$, we can use the formalism of the
continuum limit to obtain (more precisely, we apply~\eqref{cont0} with~$L=3$)
\[ A_{xy} \simeq \frac{1}{\varepsilon^3 t^3}\: \chi_{C_{\vartheta_0}(\vec{k}_0)}(\vec{\xi}) 
\qquad \text{if~$t \gg t_0$} \:. \]
In the case~$t \ll t_0$, however, the kernel of the fermionic projector no longer depends on~$t$
(see~\eqref{Pepssmall}; intuitively speaking, this is because all the wave functions of~$\H^\sp$ are ``in phase'').
Therefore, the close chain is obtained by taking the closed chain at~$t=t_0$,
\[ A_{xy} \simeq \frac{1}{\varepsilon^3 t_0^3}\: \chi_{C_{\vartheta_0}(\vec{k}_0)}(\vec{\xi}) 
\qquad \text{if~$t \ll t_0$} \:. \]
In order to determine the scaling behavior, it suffices to interpolate between the asymptotics
for large and small~$t$. We thus obtain
\[ 
A_{xy} \simeq \frac{1}{\varepsilon^3 t^3\,t_0^3}\: \big( 1+ \O(\varepsilon/t) \big) 
\times \left\{ \begin{array}{cl} t_0^3\:\chi_{C_{\vartheta_0}(\vec{k}_0)}(\vec{\xi})
& \text{if~$t \geq t_0$} \\
t^3\:\chi_{C_{\hat{\vartheta}}(\vec{k}_0)}(\vec{\xi}) 
& \text{if~$t < t_0$}\:. \end{array} \right. \]
The support of this function is shown in Figure~\ref{figsuppA}.
 
Now we can proceed in the formalism of the continuum limit.
More precisely, using~\eqref{cont1}, we obtain
\begin{align}
|\lambda^{xy}_i|^2
&\simeq \frac{1}{\varepsilon^6 \,t^6\, t_0^6}\:\big( 1+ \O(\varepsilon/t) \big) 
\times
\left\{ \begin{array}{cl} t_0^6\:\chi_{C_{\vartheta_0}(\vec{k}_0)}(\vec{\xi})
& \text{if~$t \geq t_0$} \notag \\
t^6\:\chi_{C_{\hat{\vartheta}}(\vec{k}_0)}(\vec{\xi}) 
& \text{if~$t < t_0$} \end{array} \right. \notag \\
&\simeq \frac{1}{\varepsilon^5 \,t^6\, t_0^6} \: \delta \big( t-|\vec{\xi}| \big)\:
\big( 1+ \O(\varepsilon/t) \big) \times
\left\{ \begin{array}{cl} t_0^6\:\chi_{C_{\vartheta_0}(\vec{k}_0)}(\vec{\xi})
& \text{if~$t \geq t_0$} \notag \\
t^6\:\chi_{C_{\hat{\vartheta}}(\vec{k}_0)}(\vec{\xi}) 
& \text{if~$t < t_0\:.$} \end{array} \right. \notag \\
\intertext{We thus obtain the following contribution to the functional~$\T^t_V$ in~\eqref{TMink} (in the
special case~$\tilde{\rho}=\rho$ and~$\scrU_< = \scrU_> = \1$ for the states in~$\H^\sp$),}
\T^t_V(\rho, \rho) &\simeq T \ell^3 \int_{\R^4} |\lambda^{xy}_i|^2\: d^4y 
\label{Trange} \\
&\simeq T \ell^3 \int_0^\infty r^2\:dr\:
\frac{1}{\varepsilon^5 r^6} \:\frac{1}{t_0^6}\:\big( 1+ \O(\varepsilon/r) \big) 
\times \left\{ \begin{array}{cl} t_0^6\:\vartheta_0^2
& \text{if~$r \geq t_0$} \notag \\
r^6\:\hat{\vartheta}^2
& \text{if~$r < t_0$} \end{array} \right. \notag \\
&\simeq T \ell^3 \int_0^\infty r^2\:dr\:
\frac{1}{\varepsilon^4 r^6} \:\frac{1}{t_0^6}\:\big( 1+ \O(\varepsilon/r) \big) 
\times \left\{ \begin{array}{cl} t_0^5 & \text{if~$r \geq t_0$} \notag \\[0.2em]
r^5 & \text{if~$r < t_0$} \end{array} \right. \notag \\
&\simeq T \ell^3 
\frac{1}{\varepsilon^4 t_0^3} \:\frac{1}{t_0^6}\: t_0^5\:\big( 1+ \O(\vartheta_0^2) \big) 
= T \ell^3 \frac{1}{\varepsilon^4 t_0^4}\:\big( 1+ \O(\vartheta_0^2) \big) 
= \frac{T \ell^3}{\varepsilon^8}\: \vartheta_0^8 \:\big( 1+ \O(\vartheta_0^2) \big) \notag \\
&= \frac{T \ell^3}{\varepsilon^8}\: \Big( \frac{\dim \H^\sp}{\dim \H^\he} \Big)^4 \:\big( 1+ \O(\vartheta_0^2) \big) 
\overset{\eqref{Idim}}{=} \frac{T \ell^3}{\varepsilon^8}\: \big(\dim \H^\sp \big)^4\:
\Big( \frac{\varepsilon}{\ell} \Big)^{12}
\:\big( 1+ \O(\vartheta_0^2) \big) \notag \\
&= T\: \frac{\varepsilon^4}{\ell^9}\: \big(\dim \H^\sp \big)^4\:\big( 1+ \O(\vartheta_0^2) \big) \:. \label{Tfinal}
\end{align}

After these approximations and computations, we are in a setting which is very similar
to the one-dimensional example~\eqref{groupintegral}. In order to see the similarity,
we first make the simplifying assumption that there is a basis of~$\H^\sp$ consisting of wave packets
having mutually disjoint supports in the spacetime region~$V$. Identifying each wave packet with
a point~$x \in X$ in~\eqref{Xdef}, we are precisely in the setting of the example~\eqref{groupintegral} with
\[ q = \dim \H^\sp \qquad \text{and} \qquad \beta = \alpha \,T\, \frac{\varepsilon^4}{\ell^9}\: q^3 \]
(compare~\eqref{groupintegral} and the linear scaling in~$q$ of~\eqref{fform} with~\eqref{Zreflocfinal} 
and the quartic scaling in~$q$ in~\eqref{Tfinal}).
Clearly, due to dispersion effects and the fact that we only have solutions of negative energies,
there is no basis consisting of wave packets with mutually disjoint supports.
Nevertheless, we can apply the result of the model example studied in Section~\ref{secquartic}
in various situations, as we now explain.
The simplest method is to choose the width~$T$ of the spacetime region~$V$
so small that wave packets do not get dispersed in time.
Then the remaining error terms are of higher order in~$\varepsilon\ell$, $\varepsilon m$
and~$(T \vartheta_0/\varepsilon)^2$ (these error terms will be explained
for specific wave packets in Section~\ref{secphasefree} below).
In this setting, we can apply Theorem~\ref{thmsaddle}.
Noting that saddle points exist
only if~$\beta$ is larger than four, it is most convenient to write~$\beta$ in Theorem~\ref{thmsaddle} as
\[ \beta = 4\: \frac{q^3}{q_{\min}^3} \:, \]
which means that the prefactor~$\alpha$ in~\eqref{Zreflocfinal} is chosen as
\beq \label{alphachoice}
\alpha = \frac{\ell^9}{T\:\varepsilon^4}\: \frac{4}{q_{\min}^3} \:.
\eeq
Note that~$\alpha$ is a constant independent of~$q$ of length dimension four.
We thus obtain the following result.
\begin{Prp} \label{prpsaddle}
We consider the functional~$\T^t_V$ for~$V$ a
four-dimensional cu\-boid~\eqref{Vbox} inside a spacetime cylinder~\eqref{Mcylinder} with
\beq \label{Tsmall}
T \lesssim \frac{\varepsilon}{\vartheta_0^\frac{5}{3}} \:.
\eeq
Then the subsystem formed by the wave functions~$\H^\sp$ inside a cone of opening angle~$\vartheta_0$
(see~\eqref{Ccone}) gives rise to the saddle point contribution
\begin{align}
\fint_\G &d\mu_\G(\scrU_>) \fint_\G d\mu_\G(\scrU_<) \:e^{\alpha N \T^t_V(\rho, \rho)}
\asymp  (2 \pi N)^q\: \big( 1-c^4 \big)^{-q^2} \: \bigg( 
\frac{1+c^2}{1-c^2} \:\frac{c^2}{\sqrt{2c^2-1}} \bigg)^{q} \notag \\
&\times \exp \Big\{ N \,q\, \Big( \frac{c^2}{1-c^2} + 2 \log \big(1-c^2 \big) \Big) \Big\}\:
 \bigg(1 + \O \Big( \frac{1}{N} \Big) +  \O \Big( \frac{T^2}{\varepsilon^2}\, \vartheta_0^\frac{10}{3} \Big) \bigg) \:, \label{Msaddle}
\end{align}
provided that~$q>q_{\min}$. Here~$\alpha$ is chosen according to~\eqref{alphachoice} and
\begin{align}
N &= \dim \H \sim \Big( \frac{L}{\varepsilon} \Big)^3 \label{Nscale} \\
q &= \dim \H^\sp \sim \vartheta_0^2 \:\Big( \frac{\ell}{\varepsilon} \Big)^3 \label{qscale} \\
c(q)^2 &= \frac{1}{2} + \frac{1}{2} \:\sqrt{1 - \frac{q_{\min}^3}{q^3}} \:. \label{calpha}
\end{align}
\end{Prp}
\Proof The previous computation gives~\eqref{Msaddle}, except for the last error term.
In order to determine the scaling of this error term, 
we form~$q$ wave packets localized evenly in the box of size~$\ell$. According to~\eqref{qscale},
the Euclidean distance~$d$ between the wave packets scales
like~$d \simeq \varepsilon\, \vartheta_0^{-2/3}$. The wave packets disperse with
an opening angle~$\vartheta_0$ (see Figure~\ref{figIcone}).
In order for this effect to be negligible, we need to assume that~$d \gtrsim T \sin \vartheta_0 \simeq T \vartheta_0$, giving~\eqref{Tsmall}.
The resulting overlap of the wave packets is integrated over a set of dimension at least two.
Therefore, we may square the corresponding error term. This gives the result.
\QED

We point out that the last error term in~\eqref{Msaddle} makes it necessary to choose~$T$
very small, as quantified by~\eqref{Tsmall}. This does not cause any problems, because at this stage we
are free to choose~$T$ arbitrarily. Nevertheless, in order to model a realistic measurement device,
it may be preferable to choose~$T$ on the Compton scale or even larger.
Before explaining how to treat this situation, we briefly discuss what the result of Proposition~\ref{prpsaddle}
means for the refined localized state~\eqref{Zreflocfinal}.
In this proposition we computed the contribution of one saddle point corresponding to
the subsystem~$\H^\sp$. Clearly, there are many possible choices for~$\H^\sp$, giving rise to many
saddle points. We do not need to work out the combinatorics of these saddle points,
because the relative size of the different contributions can be determined with the following scaling argument.
We consider the asymptotics for small~$\varepsilon$ in the limit when~$N$ tends to infinity.
The combinatorics of the saddle points depend on~$\ell$ and~$\varepsilon$, but it is independent
of the size~$L$ of the spacetime cylinder. Therefore, the asymptotics for large~$N$
can be studied by taking the limit~$L \rightarrow \infty$ (see~\eqref{Nscale}).
Thus the leading contributions to the refined localized state are obtained by
those subsystems for which the function in the exponent
\beq \label{hdef}
h(q) := q\, \Big( \frac{c(q)^2}{1-c(q)^2} + 2 \log \big(1-c(q)^2 \big) \Big)
\eeq
with~$c(q)$ according to~\eqref{calpha} is largest.

\begin{Lemma} \label{lemmamonotone}
The function~$h(q)$ is strictly monotone increasing for~$q \geq q_{\min}$.
\end{Lemma}
\Proof A direct computation yields
\begin{align*}
h'\big( q_{\min} \big) &= 4 - 2\, \log 2 > 0 \\
h''(q) &= \frac{3\,\big( 2x - 2 \,\sqrt{x}\: \sqrt{1+x} + 3 \big)}{q_{\min}\: \sqrt{x} \:(1+x)^\frac{1}{3}\: \big(\sqrt{1+x} -\sqrt{x} \big)^3 } \quad \text{for~$q>q_{\min}$} \:,
\end{align*}
where we parametrized~$q$ in terms of~$x > 0$ by
\[ q = q_{\min}\: (1+x)^\frac{1}{3}\:. \]
Clearly, the denominator of our formula for~$h''(q)$ is positive. The same is true for the numerator
in view of the Cauchy-Schwarz inequality
\[ 2 \,\sqrt{x}\: \sqrt{1+x} \leq \big( \sqrt{x} \,\big)^2 + \big( \sqrt{1+x} \,\big)^2 = 2x+2 < 2x+3 \:. \]
Therefore, the function~$h$ is convex on~$(q_{\min}, \infty)$.
Using that~$h'(q_{\min})$ is strictly positive, we conclude that~$h$ is strictly monotone on~$[q_{\min}, \infty)$.
\QED

This lemma shows that in the considered limiting case~$N \rightarrow \infty$, the
main contribution to the dominant saddle points come from the large systems.
The largest possible system is that formed of all wave functions in~$\H^\he$, i.e.\
\[ \H^\sp = \H^\he \qquad \text{and} \qquad \vartheta_0 = 2 \pi \:. \]

We finally explain how one can relax the condition~\eqref{Tsmall}. If this condition is violated, we cannot
choose a basis of~$\H^\sp$ consisting of wave packets whose supports are mutually disjoint
in~$V$ up to small error terms. Instead, these wave packets will have substantial overlaps
due to {\em{dispersion}}.  The main observation is that this has no effect on the exponential factor in~\eqref{Msaddle},
but it may change the prefactors.
\begin{Corollary}
Consider the setting of Proposition~\ref{prpsaddle}, but now without assuming~\eqref{Tsmall}.
Then the subsystem formed by the wave functions~$\H^\sp$ inside a cone of opening angle~$\vartheta_0$
(see~\eqref{Ccone}) gives rise to the saddle point contribution
\[ \fint_\G d\mu_\G(\scrU_>) \fint_\G d\mu_\G(\scrU_<) \:e^{\alpha N \T^t_V(\rho, \rho)}
\asymp \exp \Big\{ N \,h(q) + \O \big( \log N \big) \Big\} \:, \]
provided that~$q>q_{\min}$, where~$\alpha$, $q$, $c$ and~$h(q)$ are again
given by~\eqref{alphachoice}, \eqref{Nscale}--\eqref{calpha} and~\eqref{hdef}.
\end{Corollary}
\Proof We denote the localized wave packets by~$e_x$ with~$x \in X:=\{1,\ldots, q\}$.
Compared to the model example~\eqref{gammaex}, we need to take into account that for every~$x$,
also other waves~$e_y$ with~$y \neq x$ contribute. 
In order to describe this situation in the most general setting, we consider a functional~$\T^\text{dis}$
of the general form
\beq \label{gammexdis}
\T^\text{dis}(A_<, A_>) = \sum_{x \in X} \sum_{\alpha =1}^{p_x}
\Big| \tr_I \big( \scrU_> \,\Psi(x)^* \,\Psi(x) \,\scrU_<^{-1}\: P_{x,\alpha}  \big) \Big|^2
\eeq
with parameters~$p_1, \ldots, p_q \geq 0$ and
positive operators~$P_{x, 1}, \ldots, P_{x,p_x}$ on~$I$. These operators can be thought of as
being non-zero on all wave packets which have a substantial overlap with~$e_x$.
Since the wave function~$e_x$ clearly has an overlap, we may assume that the new functional
bounds the original functional from below, i.e.\
\beq \label{lowerbound}
\T(A_<, A_>) \leq \T^\text{dis}(A_<, A_>) \qquad \text{for all~$A_<, A_> \in \scrA_q$}
\eeq
(and~$\scrA_q$ as introduced in~\eqref{Aqdef}).

Similar to~\eqref{fform}, we write~\eqref{gammexdis} as
\[ 
\T^\text{dis}(A_<, A_>) 
= \sum_{x \in X} \sum_{\alpha =1}^{p_x}
\Big| \tr_I \big( A_> \,\pi_x \,A_<^*\: P_{x,\alpha}  \big) \Big|^2 \]
(where~$\pi_x$ is the orthogonal projection to the subspace spanned by~$e_x$).
Given~$\scrU_<$ and~$\scrU_>$, we now consider the integral over all matrices
obtained by multiplying from the left by unitary operators acting on~$\U(I)$,
\beq \label{intV}
\fint_{\U(I)} d\mu_{\U(I)}(V_<) \fint_{\U(I)} d\mu_{\U(I)}(V_<) \: e^{\alpha N \T^\text{dis}(V_< A_<, V_> A_>)}
\eeq
(here~$d\mu_{\U(I)}$ denotes again the normalized Haar measure).
Integrating subsequently over the whole group~$\G \times \G$ gives the partition function.

In order to estimate~\eqref{intV}, we first write the functional~$\T^\text{dis}(V_< A_<, V_> A_>)$ as
\begin{align*}
\T^\text{dis}(V_< A_<, V_> A_>) &=
\sum_{x \in X} \sum_{\alpha =1}^{p_x}
\Big| \tr_I \big( V_>  A_> \,\pi_x \,A_<^* V_<^* \: P_{x,\alpha}  \big) \Big|^2 \\
&= \sum_{x \in X} \sum_{\alpha =1}^{p_x}
\big| \la e_x \,|\, A_<^* V_<^* \: P_{x,\alpha} V_>  A_> \, e_x \ra \big|^2 \:.
\end{align*}
For any~$x$ and~$\alpha$, we let~$u$ be a normalized eigenvector of~$P_{x,\alpha}$
corresponding to the largest eigenvalue. We choose~$V_>$ such that it maps the
vector~$A_> \, e_x$ to a multiple of~$u$. Likewise, we choose~$V_<$ such that it maps the
vector~$A_< \, e_x$ to a multiple of~$u$. Then
\[ \big| \la e_x \,|\, A_<^* V_<^* \: P_{x,\alpha} V_>  A_> \, e_x \ra \big| = \| A_> \,e_x \| \: \|A_<^* \,e_x\| \: \|P_{x,\alpha} \| \:. \]
Using continuity and the Schwarz inequality, we conclude that the double integral~\eqref{intV}
can be estimated from above by the corresponding integral with~$\T^\text{dis}$
replaced by our original functional~$\T$ in~\eqref{fform}.
The point is that the constant involved in this estimate depends only on~$q$, but not on~$N$
(this could be worked out in more detail with a covering argument, using that the set~$\scrA_q$
in~\eqref{Aqdef} is relatively compact).
Now we can compute the saddle point contribution for the upper bound.
Moreover, \eqref{lowerbound} gives a lower bound for the saddle point contribution.
Clearly, the above estimates change the $q$-dependent prefactors in~\eqref{Msaddle}.
The exponential factor in~\eqref{Msaddle}, however, remains unchanged, because it
is determined solely by the value of the integrand at the saddle point (see~\eqref{Tfinal}) and the Haar measure
on~$\G$.
\QED

\subsection{Introducing a Weight for the Size of the Subsystems} \label{secweightstate}
\begin{figure}
\psscalebox{1.0 1.0} 
{
\begin{pspicture}(0,28.18779)(6.9497313,31.391672)
\psline[linecolor=black, linewidth=0.04, arrowsize=0.05291667cm 2.0,arrowlength=1.4,arrowinset=0.0]{->}(0.61,28.20194)(0.61,31.35194)
\psline[linecolor=black, linewidth=0.04, arrowsize=0.05291667cm 2.0,arrowlength=1.4,arrowinset=0.0]{->}(0.01,30.80194)(6.91,30.80194)
\psbezier[linecolor=black, linewidth=0.03](1.025,28.21194)(1.565,28.401941)(2.505,29.52194)(2.825,30.51194091796875)
\psline[linecolor=black, linewidth=0.03](2.81,30.501942)(6.81,29.911942)
\psline[linecolor=black, linewidth=0.04](1.01,30.901941)(1.01,30.70194)
\psline[linecolor=black, linewidth=0.04](2.81,30.901941)(2.81,30.70194)
\rput[bl](0.85,31.05){$q_{\min}$}
\rput[bl](2.7,30.97){$\underline{q}$}
\rput[bl](-0.2,31.2){$h(q)$}
\rput[bl](6.7,30.4){$q$}
\end{pspicture}
}
\caption{The function~$h(q)$ for $\eta$ a sharp cutoff function.}
\label{figh}
\end{figure}
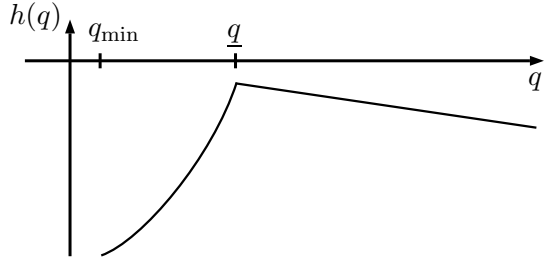

We now explain a method which allows us to introduce a weight for the size of the subsystems.
In particular, this makes it possible to get finer information on the smaller subsystems formed
by wave functions pointing in different spatial directions. To this end, we choose a bounded cutoff function~$\eta$
supported in a neighborhood of the origin, i.e.\
\[ \eta \in L^\infty_0([0,1], \R^+) \qquad \text{with} \qquad \eta|_{[0, \delta)} \equiv 1 \]
for some~$\delta>0$.
We now modify the functional~$\T^t_V$ in~\eqref{TMink}
by inserting the cutoff function
\beq \label{etacutoff}
\eta \bigg( \frac{\|xy\|}{\|x\| \|y\|} \bigg)
\eeq
into the integrand (here~$\|.\|$ is the operator norm on~$\Lin(\H)$),
\begin{align}
\T^t_V( \tilde{\rho}, T_{\scrU_<, \scrU_>} \rho) \notag 
&= \bigg( \int_{\Omega \cap V}\!\! d^4x \int_{\Omega \cap V} \!\!d^4y
+ \int_{(M \setminus \Omega) \cap V} \!\!\!\!\!d^4x
\int_{(M \setminus \Omega) \cap V} \!\!\!\!\!d^4y \bigg) \\
&\qquad \times 
\eta \bigg( \frac{\|xy\|}{\|x\| \|y\|} \bigg)\:\big| F^\varepsilon(x)\, \scrU_> \,F^\varepsilon(y)\, \scrU_<^{-1} \big|^2 \:.
\label{TMinketa}
\end{align}
As a result, the contribution for small~$\xi$ in Figure~\ref{figsuppA} is suppressed.
In order to clarify the effect of the cutoff function on the scaling behavior, we consider
the particularly simple choice of a Heaviside function, i.e.\
\beq \label{tau0def}
\eta(\tau) = \Theta(\tau_0-\tau)
\eeq
with a dimensionless constant~$\tau_0>0$.
As a consequence, in~\eqref{Trange} we integrate only over the region
\beq \label{xi0cutoff}
|\xi^0| > \underline{t} \:,
\eeq
where~$\underline{t}$ is a function of~$\tau_0$ with the scaling
\beq \label{utdef}
\underline{t} \simeq \frac{\varepsilon}{\tau_0} \:.
\eeq
Therefore, we can choose~$\underline{t}$ arbitrary by a suitable choice of the cutoff function.
This changes the computation of~$\T^t_V$ after~\eqref{Trange} to
\begin{align*}
\T^t_V(\rho, \rho) &\simeq T \ell^3 \int_{\R^4} |\lambda^{xy}_i|^2\: d^4y \\
&\simeq T \ell^3 \int_{\underline{t}}^\infty r^2\:dr\:
\frac{1}{\varepsilon^4 r^6} \:\frac{1}{t_0^6}\:\big( 1+ \O(\varepsilon/r) \big) 
\times \left\{ \begin{array}{cl} t_0^5 & \text{if~$r \geq t_0$} \\[0.2em]
r^5 & \text{if~$r < t_0$} \end{array} \right. \\
&\simeq \frac{T \ell^3}{\varepsilon^4}\:  \frac{1}{t_0\: \max(t_0, \underline{t})^3}\:
\bigg( 1+ \O\Big( \frac{\varepsilon}{\max(t_0, \underline{t})} \Big) \bigg) \\
&= \frac{T \ell^3}{\varepsilon^8}\: \vartheta_0^2\, \min \big( \vartheta_0^2, \underline{\vartheta}^2 \big)^3\:
\bigg( 1+ \O\Big( \min \big( \vartheta_0^2, \underline{\vartheta}^2 \big) \Big) \bigg) \\
&= T\: \frac{\varepsilon^4}{\ell^9}\: q\: \min(q, \underline{q})^3\:
\bigg( 1+ \O\Big( \min \big( \vartheta_0^2, \underline{\vartheta}^2 \big) \Big) \bigg) \:,
\end{align*}
where we used the notation
\beq \label{uthetadef}
\underline{\vartheta}^2 := \frac{\varepsilon}{\underline{t}} \qquad \text{and} \qquad
\underline{q} := \underline{\vartheta}^2 \:\Big( \frac{\ell}{\varepsilon} \Big)^3
= \frac{\varepsilon}{\underline{t}}\: \Big( \frac{\ell}{\varepsilon} \Big)^3\:.
\eeq
Thus for~$q>\underline{q}$, the functional~$\T^t_V$
scales only linearly in~$q$. This can be taken into account simply by modifying the function~$c(q)$
in~\eqref{calpha} to
\beq \label{calphamod}
c(q)^2 = \frac{1}{2} + \frac{1}{2} \:\sqrt{1 - \frac{q_{\min}^3}{\min(q, \underline{q})^3}} \:.
\eeq
On the interval~$[q_{\min}, \underline{q}]$, the function~$h(q)$ is unchanged.
In particular, we know from~\eqref{lemmamonotone} that this function is strictly monotone increasing.
For~$q>\underline{q}$, however, $h(q)$ simply is a linear function. Its monotonicity properties
are determined by the slope
\[ \frac{c(\underline{q})^2}{1-c(\underline{q})^2} + 2 \log \big(1-c(\underline{q})^2 \big) \:. \]
If this slope is arranged to be negative, then the function~$h(q)$ has a maximum at~$\underline{q}$
(see Figure~\ref{figh}).
If this is done, the systems with~$q=\underline{q}$ give the main contribution to the saddle points.

More generally, with the cutoff function~$\eta$ in~\eqref{etacutoff} one can
introduce an arbitrary weight function which determines to what extent
subsystems of a certain size contribute to the quantum state.
We finally give a possible interpretation for the parameter~$\underline{t}$.

\begin{Remark} {\bf{(potential significance of~$\underline{t}$)}} \label{remtbar} {\em{
In the above construction, the parameter~$\tau_0$ in~\eqref{tau0def}, and consequently
also the parameter~$\underline{t}$ in~\eqref{utdef}, were introduced ad-hoc as a free parameter
which can be used in order to introduce a weight for the size of the subsystems forming the saddle points.
Ultimately, it might be preferable conceptually to replace the ad-hoc functional~\eqref{TMinketa}
by another functional which arises more naturally in the theory of causal fermion systems.
The most obvious candidate would be the functional~\eqref{Tgen2} involving the Lagrangian mentioned
in Section~\ref{seclocdetail}. Indeed, the Lagrangian involves a natural cutoff for
small distances~\eqref{xi0cutoff} which comes from the fact that the formalism of the continuum
limit applies only away from the origin (i.e.\ if~$|\xi^0|$ is much larger than the Planck scale;
for details see~\cite[Chapter~4]{pfp} or~\cite[Section~2.4]{cfs}).
It seems a good idea to identify the corresponding length scale with~$\underline{t}$.
Here we shall not enter the details because, as mentioned in Section~\ref{seclocdetail},
the computation of the Lagrangian is more subtle due to cancellations of summands in~\eqref{Lagrange}.
}} \QEDrem
\end{Remark}

\subsection{The Phase Freedom of the Saddle Points} \label{secphasefree}
For the model example in Section~\ref{secquartic}, the saddle point is unique only up to the
phase freedom described by the functions~$\phi^1_<, \ldots, \phi^q_<$ and~$\phi^1_>, \ldots, \phi^q_>$
in~\eqref{Aphase}. In Section~\ref{sechighenergy} we saw that, in a suitable approximation
where we took into account only the leading order in~$\varepsilon/\ell_{\text{macro}}$,
the high-energy wave functions in Minkowski space could be described by this model example.
It remains to clarify how the phase freedom in~\eqref{Aphase} translates to the saddle point in
Minkowski space. As we will now explain, this saddle point has indeed a phase freedom,
but only if we restrict attention to the leading order in~$\varepsilon/\ell_{\text{macro}}$.
This phase freedom is in general broken if
\[ \text{higher orders in} \quad \frac{\varepsilon}{\ell} \quad \text{and} \quad m \varepsilon \]
are taken into account.
This means more concretely that all the computations of Sections~\ref{sechighenergy}
and~\ref{secweightstate} do involve the phase freedom, but the error terms do not.

\begin{figure}
\psscalebox{1.0 1.0} 
{
\begin{pspicture}(0,27.072796)(8.130381,31.846668)
\definecolor{colour0}{rgb}{0.9019608,0.9019608,0.9019608}
\definecolor{colour1}{rgb}{0.5019608,0.5019608,0.5019608}
\definecolor{colour2}{rgb}{0.7019608,0.7019608,0.7019608}
\pspolygon[linecolor=black, linewidth=0.01, fillstyle=solid,fillcolor=colour0](1.5906494,27.586937)(0.09064941,29.041937)(0.08814941,31.571936)(5.5881495,31.586937)
\pspolygon[linecolor=colour1, linewidth=0.01, fillstyle=solid,fillcolor=colour1](5.5906496,31.586937)(5.6906495,31.586937)(1.7031494,27.601936)(1.5906494,27.586937)
\pspolygon[linecolor=colour1, linewidth=0.01, fillstyle=solid,fillcolor=colour1](7.2981496,31.589436)(7.3906493,31.586937)(3.3906493,27.586937)(3.3181493,27.601936)
\pspolygon[linecolor=colour2, linewidth=0.01, fillstyle=solid,fillcolor=colour2](1.6906494,27.586937)(3.2906494,27.586937)(7.2906494,31.586937)(5.6906495,31.586937)
\psline[linecolor=black, linewidth=0.04, arrowsize=0.05291667cm 2.0,arrowlength=1.4,arrowinset=0.0]{->}(0.09064941,27.586937)(8.09065,27.586937)
\psline[linecolor=black, linewidth=0.04, arrowsize=0.05291667cm 2.0,arrowlength=1.4,arrowinset=0.0]{->}(0.09064941,27.586937)(0.09064941,31.806936)
\psline[linecolor=black, linewidth=0.02](0.09064941,27.586937)(4.0906496,31.586937)
\psline[linecolor=black, linewidth=0.02](1.6906494,27.586937)(5.6906495,31.586937)
\psline[linecolor=black, linewidth=0.02](3.2906494,27.586937)(7.2906494,31.586937)
\psline[linecolor=black, linewidth=0.02](4.8906493,27.586937)(8.09065,30.786936)
\psline[linecolor=black, linewidth=0.02](6.490649,27.586937)(8.09065,29.186935)
\psline[linecolor=black, linewidth=0.02](0.09064941,29.186935)(2.4906495,31.586937)
\psline[linecolor=black, linewidth=0.02](0.09064941,30.786936)(0.89064944,31.586937)
\psline[linecolor=black, linewidth=0.02](1.5656494,27.586937)(5.5656495,31.586937)
\psline[linecolor=black, linewidth=0.02](3.4156494,27.586937)(7.3906493,31.586937)
\psbezier[linecolor=black, linewidth=0.02, arrowsize=0.05291667cm 2.0,arrowlength=1.4,arrowinset=0.0]{<-}(6.9006495,31.106936)(7.3306494,30.946936)(7.5006495,31.126936)(7.5206494,31.11693603515625)
\psline[linecolor=black, linewidth=0.02](1.6906494,27.496937)(1.8906494,27.296936)(2.2906494,27.296936)(2.4906495,27.096935)(2.6906495,27.296936)(3.0906494,27.296936)(3.2906494,27.496937)
\psbezier[linecolor=black, linewidth=0.02, arrowsize=0.05291667cm 2.0,arrowlength=1.4,arrowinset=0.0]{<-}(3.3906493,27.506935)(3.5206494,27.226936)(3.6306493,27.236937)(3.7906494,27.19693603515625)
\rput[bl](4.5,29.65){$S$}
\rput[bl](7.65,30.98){(I)}
\rput[bl](1.97,30.35){(II)}
\rput[bl](2.4,26.75){$d$}
\rput[bl](3.9,27.1){$\sim \varepsilon$}
\rput[bl](8,27.25){$x$}
\rput[bl](-0.2,31.5){$t$}
\rput[bl](0.4,30.3){$e^{i \varphi^0}$}
\rput[bl](1.2,29.5){$e^{i \varphi^1}$}
\rput[bl](1.9,28.8){$e^{i \varphi^2}$}
\rput[bl](2.7,28){$e^{i \varphi^3}$}
\rput[bl](4.4,28){$e^{i \varphi^4}$}
\rput[bl](6.1,28){$e^{i \varphi^5}$}
\rput[bl](7.6,28){$\cdots$}
\end{pspicture}
}
\caption{Phase freedom of the saddle points.}
\label{figphase}
\end{figure}
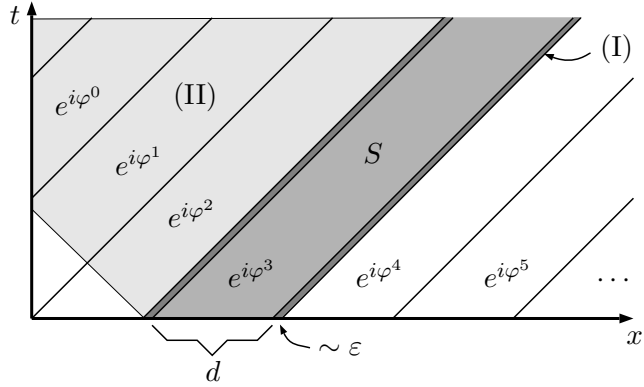

In order to derive this result, we decompose the system~$[-\ell, \ell]^3 \subset \R^3$
describing our laboratory into strips in $x$-direction of width~$d$ with the scaling
\[ \varepsilon \ll d \lesssim \ell \:. \]
We consider solutions of the scalar wave equation~$\Box \phi =0$ which do not depend on
the coordinates~$y$ and~$z$. Then~$\phi$ is a solution of the two-dimensional wave equation
\[ \big( \partial_t^2 - \partial_x^2)\, \phi = 0 \:. \]
These solutions can be decomposed into the left- and right-moving solutions.
We consider for example the right-moving solutions~$\phi(t-x)$.
We multiply the initial data by phase factors~$e^{i \varphi_k}$ and solve the Cauchy problem.
In this way, the solution is multiplied in each right-moving strip by a corresponding phase factor;
see Figure~\ref{figphase}.
Clearly, we are interested in solutions of the Dirac equation. They are obtained by acting with the 
operator~$i \Pdd$, without changing the picture of the right-moving waves propagating independently
in the right-moving time strips. But we obtain error terms for the following reasons:
\bitem
\item[(I)] The waves must be formed of wave functions contained in~$\H^\sp$. To this end, we choose a
smooth cutoff function~$\eta(k)$ which vanishes outside~$\H^\sp$
(i.e.\ the space of high-energy wave functions inside the 
cone with opening angle~$\vartheta_0$; see Figure~\ref{figIcone}). This cutoff function varies on the
scale~$\varepsilon^{-1}$. Therefore, multiplying by this cutoff function in momentum spaces
gives rise to a convolution in position space with a kernel which decays on the regularization scale~$\varepsilon$,
as is indicated in Figure~\ref{figphase} for the solutions in the time strip with phase~$e^{i \varphi_3}$
by the dark shaded regions near the boundary of the strip.
\item[(II)] The non-zero mass~$m$ gives rise to contributions which violate the strong Huygens principle
and propagate into other time strips (see the light gray region in Figure~\ref{figphase}).
\eitem
In order to describe the waves in the different strips, we introduce
symmetric linear operators~$E_k$ acting on~$\H^\lab$. Disregarding
the convolution in position space~(I), these operators could be chosen as orthogonal projection operators.
Taking these convolutions into account, we can nevertheless choose the~$E_k$
as orthogonal projection operators
outside the transition regions of size~$\sim \varepsilon$. Next, in analogy to~\eqref{Aphase}
we introduce the unitary operators
\[ A_< = c\, \exp \Big( i \sum_k \varphi^k_< E_k \Big) \qquad \text{and} \qquad
A_> = c\, \exp \Big( i \sum_k \varphi^k_> E_k \Big) \]
(with~$c$ as in~\eqref{calpha} or~\eqref{calphamod}).

Let us consider how the phases~$\varphi^k_<$
and~$\varphi^k_>$ enter the functional~$\T^t_V( \tilde{\rho}, T_{\scrU_<, \scrU_>} \rho)$. Clearly, if both~$x$ and~$y$ lie inside the same time strip, then the phases drop out.
Therefore, we only need to consider the case that~$x$ and~$y$ are in different time strips.
The contribution of the error terms~(I) from the convolution by the cutoff function~$\eta$ to the integrand in~\eqref{TMink} are not small. But they vanish unless~$y$ lies in a strip of width~$\sim \varepsilon$. Therefore,
after integrating over~$y$, the resulting
contribution to~$\T^t_V( \tilde{\rho}, T_{\scrU_<, \scrU_>} \rho)$ is of higher order in~$\varepsilon/\ell$.
The error terms~(II) due to the rest mass, on the other hand, do not affect the leading degree
of the singularity on the light cone (as defined before~\eqref{cont0}). Therefore, their contribution to~$\T^t_V( \tilde{\rho}, T_{\scrU_<, \scrU_>} \rho)$ involves a scaling factor~$m \varepsilon$. This proves the claim.

\subsection{Saddle Points for the Low-Energy Wave Functions} \label{seclowenergy}
In the previous sections (Sections~\ref{sechighenergy} and~\eqref{secweightstate})
we saw that the high-energy wave functions give rise to saddle points. If~$\scrU_<$ and~$\scrU_>$
are near these saddle points, the surface layer integrals can be computed using the formalism of the
continuum limit; in particular, they have a well-defined scaling in negative powers of~$\varepsilon$.
By introducing a parameter~$\underline{t}$ with the dimension of length into the functional~$\T^t_V$,
we could prescribe the number of wave functions forming the saddle points.
This procedure leads to many saddle points, which come with a certain combinatorics.
Before entering this combinatorics (see Section~\ref{secsaddlecombi} below), we now
consider a single saddle point and analyze how
it is modified by the low-energy wave functions.

In the following consideration, we may disregard the space~$(\H^\sp)^\perp \cap \H^\he$ formed of
high-energy wave functions which are not part of the saddle point (because if they had an effect, we would
have included them in~$\H^\sp$). Therefore, it remains to consider the subspace
\beq \label{Hspdecomp}
\H^\sp \oplus \H^\le \subset \H^\lab \:.
\eeq
We denote the restriction of a unitary operator~$\scrU$ to this subspace by~$\hat{A}$.
Using a block matrix notation in the decomposition~\eqref{Hspdecomp}, this matrix has the form
\beq \label{hatA}
\hat{A} = \begin{pmatrix} A & * \\ * & * \end{pmatrix} \:,
\eeq
where~$A : \H^\sp \rightarrow \H^\sp$ is the matrix used in the computations of Sections~\ref{sechighenergy}
and~\ref{secweightstate}, and the stars denote the other (still undetermined) matrix entries.
The saddle point of the high-energy wave functions is again characterized by~\eqref{Aid}.
Being close to this saddle point means that the functional~$\T^t_V$ can be computed
using the formalism of the continuum limit, exactly as explained in Section~\ref{sechighenergy}.
Therefore, we can expand this functional in powers of individual matrix entries in~\eqref{hatA}.
Clearly, expanding in matrix elements of~$A$ gives us back the expansion of Lemma~\ref{lemmaexpand}.
But we can also expand in the matrix elements of the ``starred'' block matrices in~\eqref{hatA}.
Similar to~\eqref{delLdef}, the first variation of the integrand in~\eqref{Tgen} can be expressed
with a trace involving the variation of the kernel of the fermionic projector, i.e.\
\beq \label{deltint}
\delta \Big( \big| x\, \scrU_> \,y\, \scrU_<^{-1} \big|^2 \Big)
= 2 \re \Tr_{S_y} \!\Big( R_<(y,x)\, \delta P_<(x,y) + R_>(y,x)\, \delta P_>(x,y) \Big)
\eeq
with integral kernels~$R_<(y,x)$ and~$R_>(y,x)$. Here the kernel of the fermionic projector with index~$>$ and~$<$ is
defined in analogy of~\eqref{Pxydef} by
\[ P_>(x,y) = -\Psi(x)\, \scrU_>\, \Psi(y)^* \:,\qquad P_<(x,y) = -\Psi(x)\, \scrU_<\, \Psi(y)^* \:. \]
The kernels~$R_<(y,x)$ and~$R_>(y,x)$ have singularities on the light cone
which are generated by the wave functions in~$\H^\sp$ and thus depend on~$A$ alone.
The variations~$\delta P$, on the other hand,
are formed of individual wave functions which are bounded in spacetime.
Therefore, an expansion in powers of~$\delta P$ amounts to an expansion in
orders on the light cone (see the preliminaries at the end of Section~\ref{seccfsmink}).
After carrying out the integrals in~\eqref{Tgen}, we thus obtain an expansion
of the form
\beq \label{Tpert}
\Delta \T^t_V( \tilde{\rho}, T_{\scrU_<, \scrU_>} \rho) = \re \Phi_< \big( \delta P_< \big)
+ \re \Phi_> \big( \delta P_> \big) + 
\text{(higher orders in~$\varepsilon/\ell_{\text{macro}}$)} \:,
\eeq
where~$\Phi_<$ and~$\Phi_>$ are complex-linear functionals, and~$\ell_{\text{macro}}$ denotes
the length scale characterizing the low-energy wave functions.
Note that~$\Phi_<$ and~$\Phi_>$ are determined by the saddle point computation in
Section~\ref{sechighenergy}.

Varying more specifically the unitary operators on~$\H^\sp \oplus \H^\le$,
we obtain the formula
\beq \label{Ttlin}
\begin{split}
\T^t_V( \tilde{\rho}, T_{\scrU_<, \scrU_>} \rho)
&= \T^t_V(c \1) + \re \Tr(\delta \hat{A}_< \,Y_<) + \re \Tr(\delta \hat{A}_> \,Y_>)  \\
&\quad\: + \text{(higher orders in~$\varepsilon/\ell_{\text{macro}}$)} \:,
\end{split}
\eeq
for suitable symmetric matrices~$Y_<$ and~$Y_>$,
where~$\T^t_V(c \1)$ denotes the functional at the saddle point.
Since first variations of~$A_<$ and~$A_>$ 
(where we again use the block matrix form~\eqref{hatA})
give us back the results of Lemma~\ref{lemmaexpand},
we know that~$Y_<$ and~$Y_>$ have the block matrix form
\[ 
Y_{<,>} = \begin{pmatrix} d_{<,>} \1 & * \\ * & * \end{pmatrix} \:, \]
where the parameters~$d_{<,>}$ describe the first variation of~$\T$ as computed in~\eqref{dT1}.

Next, we diagonalize~$Y_<$ and~$Y_>$ by a unitary transformation. 
This unitary transformation
can be regarded as a redefinition of the subspaces~$\H^\le$ and~$\H^\sp$.
With this in mind, for ease in notation we do not write out the unitary transformation,
but simply replace~$Y_<$ and~$Y_>$ by the unitarily transformed matrices.
Using that
the dimension of~$\H^\le$ is independent of~$\varepsilon$, whereas the dimension of~$\H^\sp$
is~$\sim \varepsilon^{-3}$, we conclude that, to leading order in~$\varepsilon/\ell_\text{macro}$,
this diagonalization does not change the upper left block matrix entries, i.e.\
\beq \label{Yerror}
Y_{<,>} = \begin{pmatrix} d_{<,>} \1 & 0 \\ 0 & Z_{<,>} \end{pmatrix} 
+ \text{(higher orders in~$\varepsilon/\ell_{\text{macro}}$)}
\eeq
with~$Z$ a symmetric operator on~$\H^\le$. Here the error term can be understood in more
detail if one proceeds in the following two steps.
In the first step, one chooses an orthonormal basis of~$\H^\sp$ such that the
matrices~$Y_<$ and~$Y_>$ take the form
\[ 
Y_{<,>} = \begin{pmatrix} d_{<,>} \1 & 0 & 0 \\ 0 & d_{<,>} \1 & * \\ 0 & * & * \end{pmatrix} \:, \]
where we decomposed the first block component into two blocks, where
the second block refers to a subspace of~$\H^\sp$ of dimension~$d:= \dim \H^\le$.
In the second step we diagonalize the lower $(2d \times 2d)$-block.
This diagonalization redefines~$\H^\sp$. Consequently, it also changes the computation
of the saddle point of the high-energy wave functions. More precisely, we change at most $2d$ physical
wave functions. Since each change of physical wave function can be described
perturbatively as explained after~\eqref{Tpert} above, one sees that it only affects the
higher orders in~$\varepsilon/\ell_{\text{macro}}$. Using that the dimension~$d$
is independent of~$\varepsilon$, we obtain the desired error term in~\eqref{Yerror}.

After these preparations, the exponential in~\eqref{Zreflocfinal} can be written as a product
of a function depending only on~$A_<$ and~$A_>$, a function of~$\pi_{\H^\le} \hat{A}_< \pi_{\H^\le}$
and a function of~$\pi_{\H^\le} \hat{A}_> \pi_{\H^\le}$.
Consequently, the $\scrU_<$-integral can be analyzed with the help of Proposition~\ref{prpfact2}
if we set
\[ A = A_< \qquad \text{and} \qquad D = \pi_{\H^\le} \,\hat{A}_<\, \pi_{\H^\le} \:. \]
Now we argue as follows. Using that
\[  A^* A \otimes D^* D \leq A^* A \otimes \1_{\C^p} \:, \]
the factor in the denominator in~\eqref{fgform} can be estimated by
\[ \det ( \1 - A^* A)^p \leq \det \big( \1 - A^* A \otimes D^* D \big) \leq 1 \:. \]
This shows that we can again apply the saddle point computation of Section~\ref{secquartic}.
An additional factor~$\det ( \1 - A^* A)^{-p}$ in the integrand
changes the formula~\eqref{formulathmsaddle} in Theorem~\ref{thmsaddle} only by a prefactor~$(1-c^2)^{-p}$,
which is independent of~$N$. Therefore, independent of the choice of~$D$,
we get a saddle point in the high-energy region, exactly as explained in Sections~\ref{sechighenergy}
and~\ref{secweightstate}. At the saddle point, we know that
\[ \det \big( \1 - A^* A \otimes D^* D \big) = \det (\1 - c^2 D^* D)^q \]
with the parameter~$c$ given by~\eqref{cval}, being bounded
below by~$1/2$ and bounded from above away from one.
Therefore, we can again apply Proposition~\ref{prpfact2} to conclude that
the saddle point for the low-energy wave functions can be described
by applying Proposition~\ref{prpsaddlelin}.
The denominator in~\eqref{fgform} simply gives
rise to the prefactor
\[ \frac{1}{\det (\1 - c^2 A_0^2)^q} \]
with~$A_0$ as in~\eqref{A0def} and~$Y=Z_{<,>}$.

This consideration shows that, taking the low-energy wave functions into account,
the saddle points can be computed as follows: We first compute the saddle points of the
high-energy wave functions as described in Sections~\ref{sechighenergy} and~\eqref{secweightstate}.
For each of the resulting high-energy saddle points, we obtain a unique corresponding saddle point
for the low-energy wave functions, which can be computed with the help of Proposition~\ref{prpsaddlelin}.
Due to the linear expansion in~\eqref{deltint}, the dependence on the low-energy wave functions
decouples into a product, making it possible to integrate separately
over~$\pi_{\H^\le} \,\scrU_<\, \pi_{\H^\le}$ and~$\pi_{\H^\le} \,\scrU_>\, \pi_{\H^\le}$.

\subsection{The Structure of the Saddle Points} \label{secsaddlecombi}
In the previous sections, we located and computed various saddle points.
We now analyze the saddle points systematically and work out the combinatorics.

The saddle points for the high-energy wave functions were computed starting from a 
subspace~$\H^\sp \subset \H^\lab$. Clearly, there are many possible choices for this subspace, which can be
described by a combinatorics depending on the dimensions of~$\H^\lab$ and~$\H^\sp$.
We point out that this combinatorics is independent of~$N$. This observation is very helpful because,
as computed in Proposition~\ref{prpsaddle}, the contribution by each saddle point involves the exponential factor
\beq \label{Nexp}
\exp \big\{ N \,h(q) \big\}
\eeq
with~$h(q)$ defined by~\eqref{hdef}.
This shows that, asymptotically for large~$N$, those saddle points will be dominant for which
the function~$h$ is maximal.
The combinatorics describing the possible choices of the subspace~$\H^\sp \subset \H^\lab$
only gives prefactors which are independent of~$N$ and will therefore be dominated by the exponential~\eqref{Nexp}.
In the setting of Section~\ref{sechighenergy}, we saw in Lemma~\ref{lemmamonotone}
that the function~$h(q)$ is monotone increasing. Therefore, as already explained
after the statement of Proposition~\ref{prpsaddle}, in this setting
it suffices to take into account one saddle point
corresponding to choosing~$\H^\sp = \H^\lab$ formed by all high-energy wave functions.
In the setting of Section~\ref{secweightstate}, on the other hand, we arranged that the function~$h(q)$
has one maximum at~$q=\underline{q}$ (see Figure~\ref{figh}).
Consequently, we need to take into account all the saddle points involving approximately~$\underline{q}$
wave functions. The combinatorics of these saddle points simply corresponds to the possible choices
of cones of opening angle~$\underline{\vartheta}$ defined by~\eqref{uthetadef}.
For our purposes, it is unnecessary to work out the detailed combinatorics of these saddle points.
Instead, we simply label them by an index~$a \in {\mathfrak{S}}$, where~${\mathfrak{S}}$
is an abstract index set. For notational simplicity, we treat~${\mathfrak{S}}$ as a discrete set and sum
over~$a \in {\mathfrak{S}}$ with weights~$c_a \geq 0$.

An important point for what follows is that the saddle point of the model example in Section~\ref{secquartic}
is not unique, but involves the freedom in choosing the phases~$\phi^1_<, \ldots, \phi^q_<$
and~$\phi^1_>, \ldots, \phi^q_>$ in~\eqref{Aphase}. In this way, every~$a$ does not stand
for a single saddle point, but instead for a whole family of saddle points. For a convenient notation, we denote
the phases by~$\phi^<_\alpha$ (for {\em{bra}}) and~$\phi^>_\beta$ (for {\em{ket}}), i.e.\
\[ e^{i \varphi^<_\alpha} = \phi^\alpha_< \quad \text{and} \quad
e^{i \varphi^>_\beta} = \phi^\beta_< \qquad \text{with} \qquad \alpha, \beta \in \{1, \ldots, q\} \:. \]
As explained in detail in Section~\ref{secphasefree}, there is also a phase freedom
for the saddle points formed of the high-energy wave functions in Minkowski space, which can be expressed as the freedom in performing transformations of the form
\beq \label{saddlephase}
P_<(x,y) \rightarrow e^{i \varphi^<_\alpha(x,y)}\: P_<(x,y) \qquad \text{and} \qquad
P_>(x,y) \rightarrow e^{i \varphi^>_\beta(x,y)}\: P_>(x,y)
\eeq
for suitable functions~$\varphi^<_\alpha$ and~$\varphi^>_\beta$.
As the details are not needed later on, we simply work with
these phases abstractly and label them by indices~$\alpha, \beta \in {\mathfrak{T}}_a$, where~${\mathfrak{T}}_a$ is any index set which may depend on~$a \in {\mathfrak{S}}$.
For notational simplicity, we again treat~${\mathfrak{T}}$ as a discrete set and sum over~$\alpha, \beta \in {\mathfrak{T}}_a$. In cases when~${\mathfrak{T}}_a$ is continuous, the sum must be replaced by corresponding integrals
over phases.

We now come to the saddle points for the low-energy wave functions.
We saw in Section~\ref{seclowenergy} that these saddle points can be described with the help of
Proposition~\ref{prpsaddlelin}, where~$A_0$ is determined by the saddle point~$(a, \alpha, \beta)$
(with~$a \in {\mathfrak{S}}$ and~$\alpha, \beta \in {\mathfrak{T}}_a$)
of the high-energy wave functions. Proposition~\ref{prpsaddlelin} gives a unique saddle point.
Moreover, from~\eqref{linstrict} we know that the contribution by the saddle point in~\eqref{saddlelin}
grows exponentially in~$N$. Therefore, for large~$N$, it suffices to consider the contribution at this unique
saddle point. The interesting issue is the dependence on~$\alpha$ and~$\beta$.
Clearly, the phase factors in~\eqref{saddlephase} drop out of the integrand in~\eqref{Tgen}
when taking the spectral weight. However, being created by unitary transformations
acting on the high-energy wave functions, these phases do not appear in the variations of the
low-energy wave functions, i.e.\
\[ \delta P_<(x,y) \rightarrow \delta P_<(x,y) \qquad \text{and} \qquad
\delta P_>(x,y) \rightarrow \delta P_>(x,y) \:, \]
where we vary~$A_>$ and~$A_<$. As a consequence, the first variation of the integrand in~\eqref{Tgen}
does involve the inverse phases. More precisely, \eqref{deltint} transforms to
\beq \label{deltint2}
2 \re \Tr_{S_x} \!\Big( R_<(y,x)\, e^{-i \varphi^<_\alpha(x,y)}\:
\delta P_<(x,y) + R_>(y,x)\, e^{-i \varphi^>_\alpha(x,y)}\: \delta P_>(x,y) \Big) \:.
\eeq
Integrating over~$x$ and~$y$, in~\eqref{Tpert} we obtain a dependence on~$\alpha$ and~$\beta$,
which can be described most conveniently by adding indices to the matrices~$Y_<$ and~$Y_>$
in~\eqref{Ttlin},
\begin{align*} 
\T^t_V( \tilde{\rho}, T_{\scrU_<, \scrU_>} \rho)
&= \T^t_V(c \1) + \re \Tr(\delta \hat{A}_< \,Y^{a,\alpha}_<) + \re \Tr(\delta \hat{A}_> \,Y^{a,\beta}_>)  \\
&\quad\: + \text{(higher orders in~$\varepsilon/\ell_{\text{macro}}$)} \:,
\end{align*}
We point out that~$Y_<$ depends only on~$\alpha$, whereas~$Y_>$ depends only on~$\beta$.
This is a consequence of how the phase factors show up in~\eqref{deltint2}.
The index~$a$ clarifies that all the terms also depend on the
saddle point~$a \in {\mathfrak{S}}$ of the high-energy wave functions under consideration.

\subsection{Construction of the Insertions} \label{secinsert}
Our next task is to construct the insertions to be placed into the integrand of the 
localized refined pre-state.
Given~$\scrU_<$ and~$\scrU_>$, we form the {\em{bosonic insertions}} exactly as explained
in~\cite[Section~5.3]{fockfermionic} as variational derivatives of the nonlinear surface layer integral.
More precisely, given holomorphic linearized solution~$z'_1, \ldots, z'_p$
and~$z_1, \ldots, z_q$, we work with the bosonic insertion
\[ D^<_{z'_1} \gamma^t(\tilde{\rho}, T_{\scrU_<, \scrU_>} \rho) \cdots D^<_{z'_p} \gamma^t(\tilde{\rho}, T_{\scrU_<, \scrU_>} \rho)\;
D^>_{\overline{z}_1} \gamma^t(\tilde{\rho}, T_{\scrU_<, \scrU_>} \rho) \cdots D^>_{\overline{z}_q} \gamma^t(\tilde{\rho}, T_{\scrU_<, \scrU_>} \rho) \:. \]

For the {\em{fermionic insertions}}, we improve the construction given in~\cite{fockfermionic}.
We first give the improved construction and explain the differences to the earlier construction in~\cite{fockfermionic}
afterward (see Remark~\ref{remcompare} below).
Given unitary operators~$\scrU_<, \scrU_>$ as well as a vector~$e_\ell \in \H^\fermi$, we consider the
anti-linear mapping
\[ b^>_\ell : \H^\fermi_\rho \rightarrow \C \:,\qquad b^>_\ell(\phi) = 
D^>_{|e_\ell \ra \, \overline{\phi}} \,\gamma^t \big(\tilde{\rho}, T_{\scrU_<, \scrU_>}  \rho \big) \:. \]
We represent this anti-linear mapping by a
vector~$\psi^>_\ell(\scrU_<, \scrU_>)$ in the Hilbert space~$\H^\fermi_\rho$, i.e.\
\[ b^>_\ell(\phi) = \la \phi \,|\, \psi^>_\ell \ra_\rho \qquad \text{for all~$\phi \in \H^\fermi_\rho$}\:. \]
Intuitively speaking, the wave function~$\psi_\ell^>(\scrU_<, \scrU_>)$ tells us how the physical wave
corresponding to~$e_\ell$ of the {\em{interacting}} spacetime looks like for an observer in the vacuum.
Next, we let~$(e_\ell)_{\ell =1,\ldots, f_\fermi}$ be an orthonormal basis of~$\H^\fermi_\rho$
and form the Hartree-Fock state
\[ \Phi^>(\scrU_<, \scrU_>) := \psi^>_1(\scrU_<, \scrU_>) \wedge \cdots \wedge \psi^>_{f_\fermi}(\scrU_<, \scrU_>) 
\:\in\: \Fock^\fermi_{\rho, f_\fermi} \:. \]
Similarly, we define the Hartree-Fock state~$\Phi^<$ by representing the linear mapping
\[ b^<_\ell : \H^\fermi_\rho \rightarrow \C \:,\qquad b^<_\ell(\phi) = 
D^<_{\phi \, \la e_\ell|} \,\gamma^t \big(\tilde{\rho}, T_{\scrU_<, \scrU_>}  \rho \big) \]
with a vector~$\psi^<_\ell(\scrU_<, \scrU_>)$,
\[ b^<_\ell(\phi) = \la \psi^<_\ell \,|\, \phi \ra_\rho \qquad \text{for all~$\phi \in \H^\fermi_\rho$}\:, \]
and by taking the resulting Hartree-Fock state,
\[ \Phi^<(\scrU_<, \scrU_>) := \psi^<_1(\scrU_<, \scrU_>) \wedge \cdots \wedge \psi^<_{f_\fermi}(\scrU_<, \scrU_>) 
\:\in\: \Fock^\fermi_{\rho, f_\fermi} \:. \]
The fermionic insertion is introduced as the expectation value of these Hartree-Fock states,
\beq \label{fermiinsert}
\big\la \Phi^<(\scrU_<, \scrU_>)
\,\big|\, \Psi^\dagger(\phi'_1) \cdots \Psi^\dagger(\phi'_{r'}) \;\Psi(\overline{\phi_1}) \cdots \Psi(\overline{\phi_r})
\,\big|\, \Phi^>(\scrU_<, \scrU_>) \big\ra_{\Fock^\fermi_{\rho, f_\fermi}} \:.
\eeq
This leads us to the following definition.

\begin{Def} \label{defstateref}
The {\bf{localized refined pre-state}}~$\omega^t_V$ at time~$t$ is defined by
\begin{align*}
&\omega^t_V\Big( a^\dagger(z'_1) \cdots a^\dagger(z'_p)\:\Psi^\dagger(\phi'_1) \cdots \Psi^\dagger(\phi'_{r'}) \;
a(\overline{z_1}) \cdots a(\overline{z_q}) \: \Psi(\overline{\phi_1}) \cdots \Psi(\overline{\phi_r}) \Big) \\
&:= \frac{1}{Z^t_V\big( \beta, \tilde{\rho} \big)} 
\fint_\G d\mu_\G \big(\scrU_< \big)  \fint_\G d\mu_\G \big( \scrU_> \big)\: 
e^{\alpha N \T^t_V \big(\tilde{\rho}, T_{\scrU_<, \scrU_>}  \rho \big)} \\
&\qquad\: \times
\big\la \Phi^<(\scrU_<, \scrU_>)
\,\big|\, \Psi^\dagger(\phi'_1) \cdots \Psi^\dagger(\phi'_{r'}) \;\Psi(\overline{\phi_1}) \cdots \Psi(\overline{\phi_r})
\,\big|\, \Phi^>(\scrU_<, \scrU_>) \big\ra_{\Fock^\fermi_{\rho, f_\fermi}}  \\
&\qquad\: \times 
D^<_{z'_1} \gamma^t(\tilde{\rho}, T_{\scrU_<, \scrU_>} \rho) \cdots D^<_{z'_p} \gamma^t(\tilde{\rho}, T_{\scrU_<, \scrU_>} \rho)\;
D^>_{\overline{z}_1} \gamma^t(\tilde{\rho}, T_{\scrU_<, \scrU_>} \rho) \cdots D^>_{\overline{z}_q} \gamma^t(\tilde{\rho}, \scrU \rho)
\end{align*}
with the normalization constant given by
\begin{align*}
Z^t_V
&:= \fint_\G d\mu_\G \big(\scrU_< \big)  \fint_\G d\mu_\G \big( \scrU_> \big)\: 
e^{\alpha N \T^t_V \big(\tilde{\rho}, T_{\scrU_<, \scrU_>}  \rho \big)}\: \big\la \Phi^<(\scrU_<, \scrU_>)
\,\big|\, \Phi^>(\scrU_<, \scrU_>) \big\ra_{\Fock^\fermi_{\rho, f_\fermi}} \:.
\end{align*}
\end{Def}

\begin{Remark} {\bf{(discussion of fermionic insertions)}} \label{remcompare} {\em{
We now compare the insertion~\eqref{fermiinsert} with the construction used in~\cite{fockfermionic}.
For clarity, we begin with the case~$\scrU_< = \scrU_> = \scrU$ without refinement.
In this case, we can leave out all indices~$<$ and~$>$, so that~\eqref{fermiinsert} simplifies to
\[ \big\la \Phi(\scrU)
\,\big|\, \Psi^\dagger(\phi'_1) \cdots \Psi^\dagger(\phi'_{r'}) \;\Psi(\overline{\phi_1}) \cdots \Psi(\overline{\phi_r})
\,\big|\, \Phi(\scrU) \big\ra_{\Fock^\fermi_{\rho, f_\fermi}} \:. \]
This expectation value can be computed directly with the Wick rules.
The result becomes particularly simple in the case that the effective one-particle wave functions~$\psi_\ell$
are orthonormal
\[ 
\la \psi_\ell \,|\, \psi_{\ell'} \ra_\rho = \delta_{\ell, \ell'}\:. \]
In this case, the Hartree-Fock wave function~$\Phi(\scrU)$ is normalized. Moreover, the
projection operator~$\pi^{\fermi, t}$ to the image of the effective one-particle wave functions~$\psi_\ell$ 
can be written as
\[ \pi^{\fermi, t} := \sum_{\ell=1}^{f_\fermi} | \psi_\ell \ra_\rho \la \psi_\ell | \:. \]
A straightforward computation yields
\beq \label{HFcorres}
\begin{split}
&\big\la \Phi(\scrU)
\,\big|\, \Psi^\dagger(\phi'_1) \cdots \Psi^\dagger(\phi'_{r'}) \;\Psi(\overline{\phi_1}) \cdots \Psi(\overline{\phi_r})
\,\big|\, \Phi(\scrU) \big\ra_{\Fock^\fermi_{\rho, f_\fermi}} \\
&= \delta_{r' r}\:\frac{1}{r!} \sum_{\sigma, \sigma' \in S_{r}}
(-1)^{\sign(\sigma)+\sign(\sigma')} \:\la \phi_{\sigma(1)} \,|\, \pi^{\fermi, t}\, \phi'_{\sigma'(1)} \ra^t_\rho
\cdots \la \phi_{\sigma(r)} \,|\, \pi^{\fermi, t} \, \phi'_{\sigma'(r)} \ra^t_\rho \:,
\end{split}
\eeq
giving agreement with the fermionic insertions in~\cite[Definition~4.1]{fockfermionic}.

Clearly, in typical situations the effective one-particle wave functions will {\em{not}} be orthonormal.
The construction in~\cite[Section~4.3]{fockfermionic} amounts to making these wave functions orthonormal
with a Gram-Schmidt procedure. After this has been done, the relation~\eqref{HFcorres} again holds.
This consideration shows that the construction in~\cite{fockfermionic} differs from
Definition~\ref{defstateref} precisely by the orthonormalization of the effective one-particle wave functions.
This raises the question whether such an orthonormalization is desirable or physically preferable.
The only advantage of the orthonormalization is that a $p$-particle measurement only involves~$p$
one-particle wave functions (and not all wave functions via the normalization factors coming up in the
expectation value). On the other hand, this advantage disappears if one keeps in mind that
through the Gram-Schmidt procedure, these~$p$ wave functions depend on all the other
wave functions as well.
More importantly, the orthonormalization seems unnatural, because if the one-particle wave functions
are small, then they should also give a small contribution to the expectation value.
Finally, Definition~\ref{defstateref} has the advantage that the positivity properties are more apparent,
as will be explained in Section~\ref{secpositive}.

These considerations apply analogously to the refined state. In particular, the
suggestion for the fermionic insertions made for the refined state in~\cite[Section~5]{fockfermionic}
seems superseded by Definition~\ref{defstateref}.
}} \QEDrem
\end{Remark}

\subsection{Positivity of the Localized Refined Pre-State} \label{secpositive}
Having defined the localized refined pre-state (see Definition~\ref{defstateref}), the next step is to
compute it in more detail and to prove positivity.
As explained in Section~\ref{secsaddlecombi}, to leading order in~$1/N$ it suffices to
evaluate the integrand at the saddle points.
Moreover, we may expand the integrand in powers of the operators~$A_<$ and~$A_>$.
To leading order on the light cone, the insertions are linear in these operators
and thus depend only on one of these operators. We take these dependencies into account
in our notation by evaluating at the saddle points using the notation
\[ \Phi(a, \alpha) := \Phi^<(\scrU_<, \scrU_>) \:,\qquad \Phi(a, \beta) := \Phi^>(\scrU_<, \scrU_>) \:, \]
and similarly for the bosonic insertions. Evaluating the group integrals at the saddle points as explained
 in Section~\ref{secsaddlecombi} gives the following result.
\begin{Thm} \label{thmstatesaddle} To leading order in~$1/N$ and~$\varepsilon/\ell_{\text{macro}}$,
the localized refined pre-state~$\omega^t_V$ takes the form
\begin{align*}
&\omega^t_V\Big( a^\dagger(z'_1) \cdots a^\dagger(z'_p)\:\Psi^\dagger(\phi'_1) \cdots \Psi^\dagger(\phi'_{r'}) \;
a(\overline{z_1}) \cdots a(\overline{z_q}) \: \Psi(\overline{\phi_1}) \cdots \Psi(\overline{\phi_r}) \Big) \\
&= \sum_{a \in {\mathfrak{S}}} c_a \!\!\sum_{\alpha, \beta \in {\mathfrak{T}}_a}
\big\la \Phi(a, \alpha)
\,\big|\, \Psi^\dagger(\phi'_1) \cdots \Psi^\dagger(\phi'_{r'}) \;\Psi(\overline{\phi_1}) \cdots \Psi(\overline{\phi_r})
\,\big|\, \Phi(a, \beta) \big\ra_{\Fock^\fermi_{\rho, f_\fermi}}  \notag \\
&\qquad\qquad\qquad \: \times 
D^<_{z'_1} \gamma^t(a, \alpha) \cdots D^<_{z'_p} \gamma^t(a, \alpha)\;
D^>_{\overline{z}_1} \gamma^t(a, \beta) \cdots D^>_{\overline{z}_q} \gamma^t(a, \beta) \\
&\quad\: + \text{higher orders in~$1/N$ and~$\varepsilon/\ell_{\text{macro}}$}
\end{align*}
with non-negative weights~$c_a$.
\end{Thm} \noindent
To avoid confusion, we point out that, for notational convenience, 
the factor~$1/Z^t_V\big( \beta, \tilde{\rho} \big)$ was absorbed into the weights~$c_a$.

\begin{Thm} \label{thmpositive} To leading order in~$1/N$ and~$\varepsilon/\ell_{\text{macro}}$,
the localized refined pre-state~$\omega^t_V$ is positive, thereby defining
a quantum state~\eqref{state}.
\end{Thm} 
\Proof Before beginning, we note that, using linearity together with
the canonical anti-commutation relations, one sees that
the formula in Proposition~\ref{thmstatesaddle} holds more generally for any
combination~$B_\text{fermi}$ of fermionic field operators, i.e.\
\begin{align}
&\omega^t_V\Big( a^\dagger(z'_1) \cdots a^\dagger(z'_p)\:
a(\overline{z_1}) \cdots a(\overline{z_q}) \:B_\text{fermi} \Big) \label{omegaorder} \\
&= \sum_{a \in {\mathfrak{S}}} c_a \!\!\sum_{\alpha, \beta \in {\mathfrak{T}}_a}
\big\la \Phi(a, \alpha)
\,\big|\, B_\text{fermi}
\,\big|\, \Phi(a, \beta) \big\ra_{\Fock^\fermi_{\rho, f_\fermi}}  \notag \\
&\qquad\qquad\qquad \: \times 
D^<_{z'_1} \gamma^t(a, \alpha) \cdots D^<_{z'_p} \gamma^t(a, \alpha)\;
D^>_{\overline{z}_1} \gamma^t(a, \beta) \cdots D^>_{\overline{z}_q} \gamma^t(a, \beta) \\
&\quad\: + \text{higher orders in~$1/N$ and~$\varepsilon/\ell_{\text{macro}}$} \:. \notag
\end{align}
Given~$A \in \A$, our task is to show that~$\omega^t_V(A^* A) \geq 0$.
We first need to order the bosonic field operators in~$A^* A$
as in~\eqref{omegaorder}.
To this end, we use the canonical commutation relations~\eqref{CCR} to bring all
the creation operators to the left and all the annihilation operators to the right.
This gives rise to pairings according to
\[ a(\overline{z})\: a^\dagger(z') \;\text{ gives }\; (z | z')^t_\rho \:, \]
where always one annihilation operator in~$A^*$ is combined with one creation operator in~$A$.
After having performed all these commutations, we end up with a product of bosonic field
operators of the form as in Definition~\ref{defstateref}, where each creation and annihilation operator
gives rise to an insertion~$D_{z} \gamma^t(a,\alpha)$ and~$D_{\overline{z}} \gamma^t(a,\beta)$, respectively.
After these transformations, the expectation value can be written as follows,
\begin{align*}
\omega_V^t(A^* A) &= 
\sum_{a \in {\mathfrak{S}}} c_a \sum_{r=0}^\infty \sum_{\alpha, \beta \in {\mathfrak{T}}_a}
\sum_{k_1, \ldots, k_r} \sum_{n_1,\ldots, n_r}
\overline{T_{k_1 \cdots k_r}}(a, \alpha) \;
T_{n_1 \cdots n_r}(a, \beta) \\
&\quad\: \times(z_{k_1} | z_{n_1})^t_\rho \cdots (z_{k_r} | z_{n_r})^t_\rho\: \big\la \Phi(a,\alpha) \:\big|\:  A_\text{fermi}^*(r, \alpha) \:A_\text{fermi}(r, \beta) \:\big|\: \Phi(a,\beta) \big \ra_{\Fock^\fermi_{\rho, f_\fermi}} \:.
\end{align*}
In order to clarify the structure of this formula, we introduce the bosonic Fock vector
\[ Z_r(T) := \sum_{k_1, \ldots, k_r} T_{k_1 \cdots k_r}\: z_{k_1} \otimes \cdots \otimes z_{k_r} \:. \]
We thus obtain
\begin{align*}
\omega_V^t(A^* A) &= \sum_{a \in {\mathfrak{S}}} c_a \sum_{r=0}^\infty \sum_{\alpha, \beta \in {\mathfrak{T}}_a}
\big\la Z_r \big( T(a, \alpha)\big) \otimes \Phi(a,\alpha) \:\big|\:
 A_\text{fermi}^*(r, \alpha) \\
& \qquad\qquad\qquad\qquad\qquad\qquad\;\, \times A_\text{fermi}(r, \beta) \:\big|\: 
Z_r \big( T(a, \beta)\big) \otimes \Phi(a,\beta) \big \ra_{\Fock_\rho} \:.
\end{align*}
Bringing the operator~$A_\text{fermi}^*(r, \alpha)$ to the left and introducing the vector
\[ \Psi(a,r) = \sum_{\alpha\in {\mathfrak{T}}_a} Z_r \big( T(a, \alpha)\big) \otimes A_\text{fermi}(r, \alpha)
\, \Phi(a,\alpha) \:, \]
we conclude that
\[ \omega_V^t(A^* A) = \sum_{a \in {\mathfrak{S}}} c_a \sum_{r=0}^\infty 
\big\la \Psi(a,r) \big| \Psi(a,r) \big\ra_{\Fock_\rho} \:. \]
This is obviously non-negative, concluding the proof.
\QED

\section{Description of Entanglement} \label{secentangle}
The formula derived in Theorem~\ref{thmstatesaddle} reveals that the refined state
makes it possible to describe entanglement. Indeed, the state~$\omega^t_V$
can be written as the expectation value of a density operator~$\sigma^t$ 
\beq \label{fockrep}
\omega^t_V(A) = \tr_{\Fock_\rho}(\sigma^t A) \qquad \text{for all~$A \in \A$} \:.
\eeq
This density operator can be written in {\em{bra/ket}} notation as
\beq \label{sigmaform}
\sigma^t = \sum_{a \in {\mathfrak{S}}} c_a \;\Big| \sum_{\alpha \in {\mathfrak{T}}_a} \Psi^\Fock_{a \alpha} \Big\ra
\Big\la \sum_{\beta \in {\mathfrak{T}}_a} \Psi^\Fock_{a \beta}  \Big| \:,
\eeq
where the Fock vector~$\Psi^\Fock_{a \alpha}$ is given as the tensor product of the fermionic
Fock vector~$\Phi(a, \beta)$ in the statement of Theorem~\ref{thmstatesaddle}
with a bosonic Fock vector which realizes the bosonic expectation values.
From~\eqref{sigmaform} one sees that, in the case of only one saddle point~$a$,
one gets a pure state, and the trace in~\eqref{fockrep} reduces to the usual expectation value,
\[ \omega^t_V(A) = 
\Big\la \sum_{\beta \in {\mathfrak{T}}} \Psi^\Fock_{\beta}  \Big| A \Big| \sum_{\alpha \in {\mathfrak{T}}} \Psi^\Fock_{\alpha} \Big\ra_{\Fock_\rho} \:.   \]
If we have more than one saddle point, we can describe a mixed state.
The sums over~$\alpha, \beta \in {\mathfrak{T}}_a$ run over the phase freedom of the
high-energy saddle point~$a$. In this way,
the state~\eqref{sigmaform} allows for the description of general entangled states.

Based on these results, we can make the qualitative picture described in Section~\ref{secnoentangle}
precise. The index~$\alpha$ in~\eqref{Phiansatz} and in the insertions~\eqref{sumdecomp}
describes the phase freedom of our saddle points. This phase freedom appears separately
for {\em{bra}} and {\em{ket}}, giving the synchronization aimed for in~\eqref{stateapprox}.
In this way, we have bypassed the counter argument of Lemma~\ref{lemmacounter},
making it possible to describe entanglement.

These constructions and considerations even give some insight into the structure of the
{\em{long-range correlations}} and {\em{dephasing effects}} mentioned in the paragraph 
after~\eqref{omegareflocfinal} in the introduction. Indeed, the saddle points are formed of
wave functions in a cone in momentum space (see Figure~\ref{figIcone}). These wave functions
are ``in phase'' even for large distances. Moreover, the phase freedom of each saddle point propagates
in lightlike directions (see Figure~\ref{figphase}).
This phase freedom has the effect that every saddle point is formed of many components having different
relative phases.
Each component encodes information on a specific Fock component,
including both the fermionic and bosonic tensor factors.
When integrating over the unitary group, dephasing effects (destructive interference) make it possible
to detect and distinguish the different components.
These wave components give rise to the summands~$\alpha$ in the decomposition of a Fock vector
into product states as in~\eqref{sigmaform} and~\eqref{Phiansatz}.
In summary, we thus obtain nonlocal effects and long-range correlations,
giving a natural explanation for Einstein's ``spooky action at a distance.''

\appendix
\section{Diagrammatic Derivation of the Gaussian Asymptotics} \label{appA}
\begin{figure}
\psscalebox{1.0 1.0} 
{
\begin{pspicture}(0,28.199356)(11.055777,30.220644)
\definecolor{colour1}{rgb}{0.7019608,0.7019608,0.7019608}
\psline[linecolor=black, linewidth=0.04](0.22788818,28.210644)(0.22788818,30.210644)
\psline[linecolor=black, linewidth=0.04](0.027888184,29.010643)(0.22788818,29.410643)(0.42788818,29.010643)
\psline[linecolor=black, linewidth=0.04](0.8278882,28.210644)(0.8278882,30.210644)
\psline[linecolor=black, linewidth=0.04](1.0278882,29.410643)(0.8278882,29.010643)(0.6278882,29.410643)
\psline[linecolor=black, linewidth=0.04](1.8278881,28.210644)(1.8278881,30.210644)
\psline[linecolor=black, linewidth=0.04](2.4278882,28.210644)(2.4278882,30.210644)
\psline[linecolor=black, linewidth=0.04](4.4278884,28.210644)(4.4278884,30.210644)
\psline[linecolor=black, linewidth=0.04](3.8278883,28.210644)(3.8278883,30.210644)
\psline[linecolor=colour1, linewidth=0.04](6.627888,28.210644)(6.627888,30.210644)
\psline[linecolor=colour1, linewidth=0.04](6.4278884,29.010643)(6.627888,29.410643)(6.827888,29.010643)
\psline[linecolor=colour1, linewidth=0.04](7.227888,28.210644)(7.227888,30.210644)
\psline[linecolor=colour1, linewidth=0.04](7.4278884,29.410643)(7.227888,29.010643)(7.0278883,29.410643)
\psline[linecolor=colour1, linewidth=0.04](8.227888,28.210644)(8.227888,30.210644)
\psline[linecolor=colour1, linewidth=0.04](8.8278885,28.210644)(8.8278885,30.210644)
\psline[linecolor=colour1, linewidth=0.04](10.8278885,28.210644)(10.8278885,30.210644)
\psline[linecolor=colour1, linewidth=0.04](10.227888,28.210644)(10.227888,30.210644)
\psline[linecolor=black, linewidth=0.04](3.6278882,29.010643)(3.8278883,29.410643)(4.0278883,29.010643)
\psline[linecolor=black, linewidth=0.04](1.6278882,29.010643)(1.8278881,29.410643)(2.0278883,29.010643)
\psline[linecolor=colour1, linewidth=0.04](8.027888,29.010643)(8.227888,29.410643)(8.427888,29.010643)
\psline[linecolor=colour1, linewidth=0.04](10.027888,29.010643)(10.227888,29.410643)(10.427888,29.010643)
\psline[linecolor=colour1, linewidth=0.04](11.027888,29.410643)(10.8278885,29.010643)(10.627888,29.410643)
\psline[linecolor=colour1, linewidth=0.04](9.027888,29.410643)(8.8278885,29.010643)(8.627888,29.410643)
\psline[linecolor=black, linewidth=0.04](2.6278882,29.410643)(2.4278882,29.010643)(2.227888,29.410643)
\psline[linecolor=black, linewidth=0.04](4.627888,29.410643)(4.4278884,29.010643)(4.227888,29.410643)
\psbezier[linecolor=black, linewidth=0.04](7.227888,30.200644)(7.2223153,29.983429)(7.0678883,29.930643)(6.917888,29.930643310546873)(6.767888,29.930643)(6.6507816,29.983536)(6.627888,30.200644)
\psbezier[linecolor=black, linewidth=0.04](8.826777,30.1862)(8.821204,29.968985)(8.666777,29.916199)(8.516777,29.91619886610243)(8.366777,29.916199)(8.24967,29.969091)(8.226777,30.1862)
\psbezier[linecolor=black, linewidth=0.04](10.8278885,30.205088)(10.822315,29.987875)(10.667889,29.935087)(10.517888,29.935087754991383)(10.367888,29.935087)(10.250781,29.987982)(10.227888,30.205088)
\psbezier[linecolor=black, linewidth=0.04](8.823443,28.213976)(8.802127,28.520205)(8.39011,28.51731)(7.73011,28.517309977213518)(7.0701103,28.51731)(6.587189,28.546478)(6.6234436,28.213976)
\psbezier[linecolor=black, linewidth=0.04](8.227888,28.212866)(8.194299,28.532341)(8.832333,28.585087)(9.487888,28.58064331054685)(10.143444,28.576199)(10.897871,28.605635)(10.8278885,28.212866)
\psbezier[linecolor=black, linewidth=0.04](10.223444,28.215088)(10.218982,28.93119)(9.734555,28.80842)(8.798999,28.81064331054682)(7.863444,28.812866)(7.159645,28.890116)(7.2234435,28.215088)
\rput[bl](2.9,29.2){$\cdots$}
\rput[bl](9.3,29.2){$\cdots$}
\rput[bl](0.12,27.75){$j_1$}
\rput[bl](0.12,30.35){$i_1$}
\rput[bl](0.72,27.75){$l_1$}
\rput[bl](0.72,30.35){$k_1$}
\rput[bl](1.7,27.75){$j_2$}
\rput[bl](1.7,30.35){$i_2$}
\rput[bl](2.3,27.75){$l_2$}
\rput[bl](2.3,30.35){$k_2$}
\rput[bl](3.7,27.75){$j_p$}
\rput[bl](3.7,30.35){$i_p$}
\rput[bl](4.3,27.75){$l_p$}
\rput[bl](4.3,30.35){$k_p$}
\rput[bl](6.83,30.1){$1$}
\rput[bl](8.43,30.1){$2$}
\rput[bl](10.45,30.1){$p$}
\end{pspicture}
}
\caption{Contractions among unitary factors.}%
\label{figU1}
\end{figure}
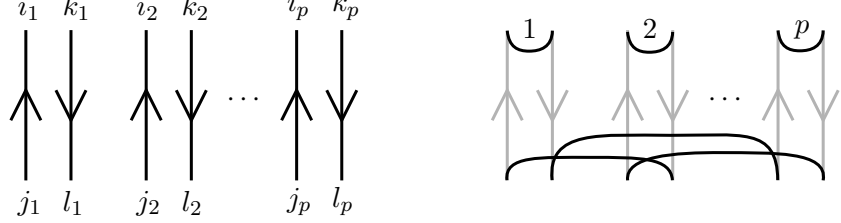

In this appendix we given the proof of Lemma~\ref{lemmagroup}.
Before beginning, we note that, since the group integral in~\eqref{groupint} involves
as many factors~$\scrU$ as~$\scrU^{-1}$, phase factors in~$\scrU$ cancel out.
Therefore, instead of considering the group~$\U(N)$,
we can just as well work with the special unitary group~$\SU(N)$.
We closely follow the procedure in~\cite{creutz}, also using a similar graphical notation.
We write each factor~$\scrU$ as an up arrow and each factor~$\scrU^{-1}$ as a down arrow
(see the left of Figure~\ref{figU1}).
Next, we use Cramer's rule
\[ (\scrU^{-1})^{ij} = \frac{1}{(N-1)!} \:\epsilon^{i\, i_1\, \cdots\, i_{N-1}} \:
\epsilon^{j\, j_1\, \cdots\, j_{N-1}} \:\scrU_{i_1\, j_1} \cdots \scrU_{i_{N-1}\, j_{N-1}} \]
in order to rewrite the factors~$\scrU^{-1}$ in terms of~$\scrU$
(here~$\epsilon$ is the totally anti-symmetric Levi-Civita symbol; moreover we use the
Einstein summation convention).
The resulting expression involves~$Np$ factors~$\scrU$. Its group integral
gives pairings of the indices with the Levi-Civita symbol (for details see~\cite[Figure~6]{creutz}).
Pairs of Levi-Civita symbols whose indices are contracted with each other can be rewritten
as sums of products of Kronecker deltas (for details see~\cite[eq.~(18)]{creutz}).
After these transformations, the up and down arrows on the left of Figure~\ref{figU1} are
connected both at the top and the bottom. In order to simplify the graphical notation, we
permute the down arrows in such a way that every up arrow is connected at the top to
the down arrow at its right (these permutations must be taken into account in our
end formula by a simultaneous symmetrization in the indices~$k_1, \ldots, k_p$ and~$l_1, \ldots, l_p$). At the bottom,
the arrows are connected as shown on the right of Figure~\ref{figU1}, with a certain combinatorics
which still needs to be specified.

\begin{figure}
\psscalebox{1.0 1.0} 
{
\begin{pspicture}(0,27.89293)(8.934142,30.527071)
\definecolor{colour1}{rgb}{0.7019608,0.7019608,0.7019608}
\psline[linecolor=colour1, linewidth=0.04](4.9170704,28.51707)(4.9170704,30.51707)
\psline[linecolor=colour1, linewidth=0.04](4.71707,29.31707)(4.9170704,29.71707)(5.11707,29.31707)
\psline[linecolor=colour1, linewidth=0.04](5.5170703,28.51707)(5.5170703,30.51707)
\psline[linecolor=colour1, linewidth=0.04](5.71707,29.71707)(5.5170703,29.31707)(5.3170705,29.71707)
\psline[linecolor=colour1, linewidth=0.04](0.3170703,28.51707)(0.3170703,30.51707)
\psline[linecolor=colour1, linewidth=0.04](0.11707031,29.31707)(0.3170703,29.71707)(0.5170703,29.31707)
\psline[linecolor=colour1, linewidth=0.04](0.9170703,28.51707)(0.9170703,30.51707)
\psline[linecolor=colour1, linewidth=0.04](1.1170703,29.71707)(0.9170703,29.31707)(0.71707034,29.71707)
\psline[linecolor=colour1, linewidth=0.04](1.9170703,28.51707)(1.9170703,30.51707)
\psline[linecolor=colour1, linewidth=0.04](1.7170703,29.31707)(1.9170703,29.71707)(2.1170702,29.31707)
\psline[linecolor=colour1, linewidth=0.04](2.5170703,28.51707)(2.5170703,30.51707)
\psline[linecolor=colour1, linewidth=0.04](2.7170703,29.71707)(2.5170703,29.31707)(2.3170702,29.71707)
\psbezier[linecolor=black, linewidth=0.04](0.9170703,30.50707)(0.9114975,30.289856)(0.7570703,30.23707)(0.6070703,30.2370703125)(0.45707032,30.23707)(0.33996353,30.289963)(0.3170703,30.50707)
\psbezier[linecolor=black, linewidth=0.04](2.5170703,30.50707)(2.5114975,30.289856)(2.3570702,30.23707)(2.2070704,30.2370703125)(2.0570703,30.23707)(1.9399636,30.289963)(1.9170703,30.50707)
\psbezier[linecolor=black, linewidth=0.04](0.3170703,28.519293)(0.32264313,28.736506)(0.4770703,28.789293)(0.6270703,28.789292534722225)(0.7770703,28.789293)(0.8941771,28.736399)(0.9170703,28.519293)
\psbezier[linecolor=black, linewidth=0.04](1.9170703,28.514849)(1.9226431,28.732061)(2.0770702,28.784847)(2.2270703,28.784848090277784)(2.3770704,28.784847)(2.494177,28.731955)(2.5170703,28.514849)
\psline[linecolor=colour1, linewidth=0.04](3.5170703,28.51707)(3.5170703,30.51707)
\psline[linecolor=colour1, linewidth=0.04](3.3170702,29.31707)(3.5170703,29.71707)(3.7170703,29.31707)
\psline[linecolor=colour1, linewidth=0.04](4.11707,28.51707)(4.11707,30.51707)
\psline[linecolor=colour1, linewidth=0.04](4.3170705,29.71707)(4.11707,29.31707)(3.9170704,29.71707)
\psbezier[linecolor=black, linewidth=0.04](4.11707,30.50707)(4.1114974,30.289856)(3.9570704,30.23707)(3.8070703,30.2370703125)(3.6570704,30.23707)(3.5399635,30.289963)(3.5170703,30.50707)
\psbezier[linecolor=black, linewidth=0.04](5.5170703,30.50707)(5.5114975,30.289856)(5.3570704,30.23707)(5.2070704,30.2370703125)(5.0570703,30.23707)(4.9399633,30.289963)(4.9170704,30.50707)
\psbezier[linecolor=black, linewidth=0.04](5.525959,28.51707)(5.5046425,28.823298)(5.172626,29.037071)(4.5126257,29.037070312499974)(3.8526258,29.037071)(3.5163715,28.861794)(3.5215147,28.520403)
\psbezier[linecolor=black, linewidth=0.04](4.1115146,28.514849)(4.0859656,28.658363)(4.2472577,28.782063)(4.515405,28.771514756944445)(4.7835526,28.760967)(4.9119477,28.741507)(4.9170704,28.517193)
\psline[linecolor=colour1, linewidth=0.04](7.9170704,28.51707)(7.9170704,30.51707)
\psline[linecolor=colour1, linewidth=0.04](7.71707,29.31707)(7.9170704,29.71707)(8.11707,29.31707)
\psline[linecolor=colour1, linewidth=0.04](8.517071,28.51707)(8.517071,30.51707)
\psline[linecolor=colour1, linewidth=0.04](8.717071,29.71707)(8.517071,29.31707)(8.31707,29.71707)
\psline[linecolor=colour1, linewidth=0.04](6.5170703,28.51707)(6.5170703,30.51707)
\psline[linecolor=colour1, linewidth=0.04](6.3170705,29.31707)(6.5170703,29.71707)(6.71707,29.31707)
\psline[linecolor=colour1, linewidth=0.04](7.11707,28.51707)(7.11707,30.51707)
\psline[linecolor=colour1, linewidth=0.04](7.3170705,29.71707)(7.11707,29.31707)(6.9170704,29.71707)
\psbezier[linecolor=black, linewidth=0.04](7.11707,30.50707)(7.1114974,30.289856)(6.9570704,30.23707)(6.8070703,30.2370703125)(6.65707,30.23707)(6.5399637,30.289963)(6.5170703,30.50707)
\psbezier[linecolor=black, linewidth=0.04](8.517071,30.50707)(8.5114975,30.289856)(8.35707,30.23707)(8.20707,30.2370703125)(8.057071,30.23707)(7.9399633,30.289963)(7.9170704,30.50707)
\psbezier[linecolor=black, linewidth=0.04](8.525959,28.51707)(8.5046425,28.823298)(8.172626,29.037071)(7.5126257,29.037070312499974)(6.852626,29.037071)(6.5163713,28.861794)(6.521515,28.520403)
\psbezier[linecolor=black, linewidth=0.04](7.1115146,28.514849)(7.0859656,28.658363)(7.2472577,28.782063)(7.515405,28.771514756944445)(7.7835526,28.760967)(7.9119477,28.741507)(7.9170704,28.517193)
\psline[linecolor=black, linewidth=0.02](0.017070312,28.31707)(0.21707031,28.117071)(1.2170703,28.117071)(1.4170703,27.91707)(1.6170703,28.117071)(2.6170702,28.117071)(2.8170702,28.31707)
\psline[linecolor=black, linewidth=0.02](3.3170702,28.31707)(3.5170703,28.117071)(5.9170704,28.117071)(6.11707,27.91707)(6.3170705,28.117071)(8.717071,28.117071)(8.91707,28.31707)
\rput[bl](1.2,29.4){$\cdots$}
\rput[bl](5.8,29.4){$\cdots$}
\rput[bl](9,29.4){$\cdots$}
\rput[bl](-0.4,27.5){chains of length $\ell=1$}
\rput[bl](4.3,27.5){chains of length $\ell=2$}
\end{pspicture}
}
\caption{The contractions in $[\ell_1, \ldots, \ell_K]$.}%
\label{figU2}
\end{figure}
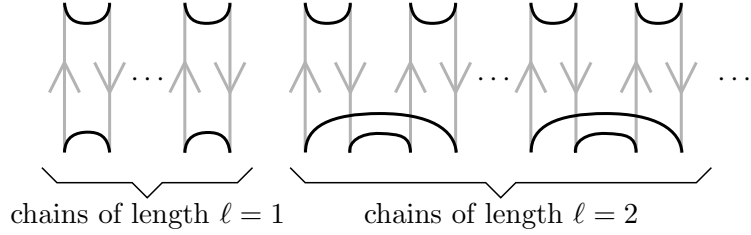

Connecting the arrows in this way, we obtain closed lines connecting~$4 \ell$ indices.
We refer to such a closed line as an $\ell$-chain
(on the right of Figure~\ref{figU1} a three-chain is depicted).
Using the freedom to permute the~$p$ pairs of arrows
(these permutations must be taken into account in our
end formula by a simultaneous symmetrization in the indices~$i_1, \ldots, i_p$ and~$j_1, \ldots, j_p$),
the resulting configuration of contractions is determined uniquely by chains of length~$1 \leq \ell_1 \leq \ell_2
\leq \cdots \leq \ell_K$, where~$K$ denotes the number of chains. Clearly,
\[ \ell_1 + \cdots + \ell_K = p \:. \]
For a convenient notation, we form the tuple
\[ (\ell_k) = (\ell_1, \ldots, \ell_K) \qquad \text{and set} \qquad
|(\ell_k)| :=\ell_1 + \cdots + \ell_K \:. \]
Then the group integral~\eqref{groupint} can be written as
\beq \label{gint}
\sum_{\text{$(\ell_k)$ with $|(\ell_k)|=p$}} c_{(\ell_k)}(N)\: [\ell_1, \ldots \ell_K] \:,
\eeq
where~$c_{(\ell_k)}(N)$ are combinatorial prefactors (which, due to the signs from the
Levi-Civita symbol, could be positive or negative), and the 
tuple with square brackets stands for the contribution to the group integral
\begin{align*}
[\ell_1, \ldots \ell_K] &=
\frac{1}{p!} \sum_{\sigma, \sigma' \in S(p)}
\delta^{i_{\sigma(1)}}_{l_{\sigma'(1)}}\: \delta^{k_{\sigma'(1)}}_{j_{\sigma(1)}} \;
\delta^{i_{\sigma(2)}}_{l_{\sigma'(2)}}\: \delta^{k_{\sigma'(2)}}_{j_{\sigma(2)}}
\cdots \delta^{i_{\sigma(n_1)}}_{l_{\sigma'(n_1)}}\: \delta^{k_{\sigma'(n_1)}}_{j_{\sigma(n_1)}} \\[-0.5em]
&\quad\,\times\; \delta^{i_{\sigma(n_1+1)}}_{l_{\sigma'(n_1+1)}}\: \delta^{k_{\sigma'(n_1+1)}}_{j_{\sigma(n_1+2)}}
\delta^{i_{\sigma(n_1+2)}}_{l_{\sigma'(n_1+2)}}\: \delta^{k_{\sigma'(n_1+2)}}_{j_{\sigma(n_1+1)}} \\
&\qquad\quad \times \cdots 
\delta^{i_{\sigma(n_1+2 n_21)}}_{l_{\sigma'(n_1+2 n_2+1)}}\: \delta^{k_{\sigma'(n_1+2 n_2+1)}}_{j_{\sigma(n_1+2 n_2+2)}}
\delta^{i_{\sigma(n_1+2 n_2+2)}}_{l_{\sigma'(n_1+2 n_2+2)}}\: \delta^{k_{\sigma'(n_1+2 n_2+2)}}_{j_{\sigma(n_1+2 n_2+1)}} \cdots \:,
\end{align*}
where~$n_\ell$ denotes the number chains of length~$\ell$.
The contractions are shown graphically
in Figure~\ref{figU2}.

The remaining task is to estimate how the combinatorial factors~$c_{(\ell_k)}$ scale in~$N$.
To this end, it is useful to denote the longest chain of a configuration~$(\ell_1, \ldots, \ell_K)$ by
\[ L(\ell_1, \ldots, \ell_K) := \ell_K \:. \]
Next, we consider the combinatorial factors for all configurations for which this longest chain
is longer than~$L$ and take their maximum,
\[ C_L := \max \big\{ |c_{(\ell_1, \ldots, \ell_K)}|  \;\big|\;  L(\ell_1, \ldots, \ell_K) \geq L \big\} \:. \]
Clearly, these constants are decreasing in~$L$,
\[ C_1 \geq C_2 \geq \cdots \geq C_p \geq C_{p+1} = 0 \]
(the last equality holds because all chains have length at most~$p$).
In the next lemma we estimate the constants~$C_\ell$. Since we are concerned
only with the $N$-dependence, we want to disregards constants depending only on~$p$, $K$
and the chains~$(\ell_1, \ldots, \ell_K)$. For a short notation, we write~$\lesssim$ for
smaller or equal up to a such constant, uniformly in~$N$. Conversely, we write~$\gg$ if this uniform inequality does not hold, no matter how large the constant is chosen.
\begin{Lemma} For any~$p \in \N$ and for all~$L \in \{2, \ldots, p\}$,
\beq \label{Ces}
C_L \lesssim \frac{C_1}{N} \:.
\eeq
\end{Lemma}
\Proof We proceed by finite induction in~$L$, beginning at~$L=p+1$ and decreasing~$L$ step by step.
In the case~$L=p+1$, the inequality~\eqref{Ces} holds trivially because~$C_{p+1}=0$.
Assume that the inequality holds for~$L+1$, i.e.\
\beq \label{ihyp}
C_{L+1}, \ldots, C_{p+1} \lesssim \frac{C_1}{N} \:.
\eeq
Our task is to show that the inequality also holds for~$L$, \eqref{Ces}.

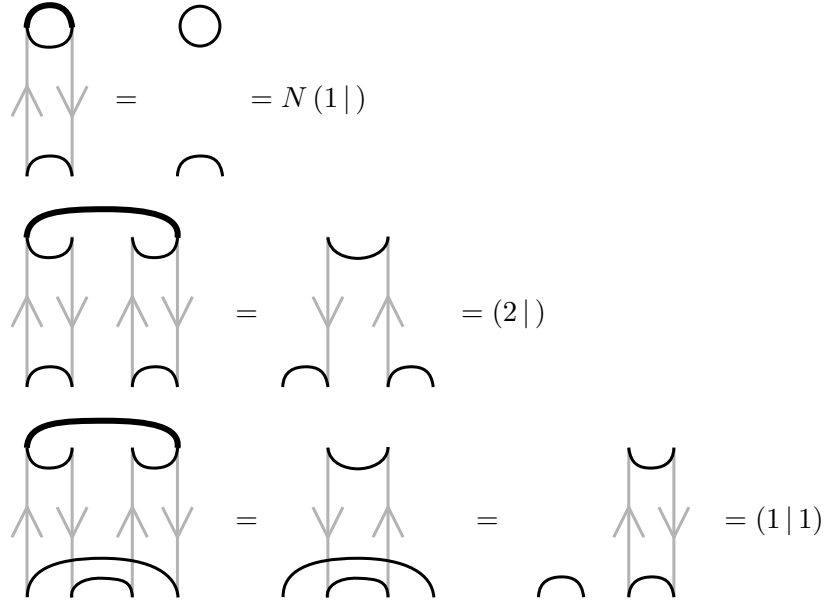
\begin{figure}
\psscalebox{1.0 1.0} 
{
\begin{pspicture}(0,25.55158)(9.055777,33.4862)
\definecolor{colour0}{rgb}{0.7019608,0.7019608,0.7019608}
\psline[linecolor=colour0, linewidth=0.04](0.22788818,31.172308)(0.22788818,33.17231)
\psline[linecolor=colour0, linewidth=0.04](0.027888184,31.97231)(0.22788818,32.372307)(0.42788818,31.97231)
\psline[linecolor=colour0, linewidth=0.04](0.8278882,31.172308)(0.8278882,33.17231)
\psline[linecolor=colour0, linewidth=0.04](1.0278882,32.372307)(0.8278882,31.97231)(0.6278882,32.372307)
\psline[linecolor=colour0, linewidth=0.04](0.22788818,28.372309)(0.22788818,30.372309)
\psline[linecolor=colour0, linewidth=0.04](0.027888184,29.172308)(0.22788818,29.572308)(0.42788818,29.172308)
\psline[linecolor=colour0, linewidth=0.04](0.8278882,28.372309)(0.8278882,30.372309)
\psline[linecolor=colour0, linewidth=0.04](1.0278882,29.572308)(0.8278882,29.172308)(0.6278882,29.572308)
\psbezier[linecolor=black, linewidth=0.04](0.8278882,33.162308)(0.8223154,32.945095)(0.66788816,32.892307)(0.5178882,32.89230834960937)(0.36788818,32.892307)(0.25078142,32.9452)(0.22788818,33.162308)
\psbezier[linecolor=black, linewidth=0.04](0.8278882,30.362309)(0.8223154,30.145094)(0.66788816,30.092308)(0.5178882,30.092308349609375)(0.36788818,30.092308)(0.25078142,30.1452)(0.22788818,30.362309)
\psbezier[linecolor=black, linewidth=0.04](0.22788818,31.17453)(0.233461,31.391745)(0.3878882,31.44453)(0.53788817,31.444530571831603)(0.6878882,31.44453)(0.80499494,31.391638)(0.8278882,31.17453)
\psbezier[linecolor=black, linewidth=0.04](0.22788818,28.370087)(0.233461,28.5873)(0.3878882,28.640085)(0.53788817,28.640086127387157)(0.6878882,28.640085)(0.80499494,28.587193)(0.8278882,28.370087)
\psline[linecolor=colour0, linewidth=0.04](1.6278882,28.372309)(1.6278882,30.372309)
\psline[linecolor=colour0, linewidth=0.04](1.4278882,29.172308)(1.6278882,29.572308)(1.8278881,29.172308)
\psline[linecolor=colour0, linewidth=0.04](2.227888,28.372309)(2.227888,30.372309)
\psline[linecolor=colour0, linewidth=0.04](2.4278882,29.572308)(2.227888,29.172308)(2.0278883,29.572308)
\psbezier[linecolor=black, linewidth=0.04](2.227888,30.362309)(2.2223153,30.145094)(2.0678883,30.092308)(1.9178882,30.092308349609375)(1.7678882,30.092308)(1.6507814,30.1452)(1.6278882,30.362309)
\psbezier[linecolor=black, linewidth=0.04](1.6278882,28.370087)(1.633461,28.5873)(1.7878882,28.640085)(1.9378881,28.640086127387157)(2.0878882,28.640085)(2.204995,28.587193)(2.227888,28.370087)
\psbezier[linecolor=black, linewidth=0.08](0.22344375,33.150085)(0.22901654,33.3673)(0.38344374,33.4512)(0.53344375,33.451197238498274)(0.6834437,33.4512)(0.80499494,33.37164)(0.8278882,33.15453)
\psbezier[linecolor=black, linewidth=0.04](2.227888,31.17453)(2.233461,31.391745)(2.3878882,31.44453)(2.5378883,31.444530571831603)(2.6878881,31.44453)(2.804995,31.391638)(2.8278883,31.17453)
\psbezier[linecolor=black, linewidth=0.08](0.22233263,30.358974)(0.23790544,30.706188)(0.7023326,30.736753)(1.2823327,30.736752794053817)(1.8623326,30.736753)(2.2763648,30.638144)(2.227888,30.356752)
\psline[linecolor=colour0, linewidth=0.04](4.227888,28.372309)(4.227888,30.372309)
\psline[linecolor=colour0, linewidth=0.04](4.4278884,29.572308)(4.227888,29.172308)(4.0278883,29.572308)
\psbezier[linecolor=black, linewidth=0.04](3.6278882,28.370087)(3.633461,28.5873)(3.7878883,28.640085)(3.9378881,28.640086127387157)(4.0878882,28.640085)(4.204995,28.587193)(4.227888,28.370087)
\psline[linecolor=colour0, linewidth=0.04](5.0278883,28.372309)(5.0278883,30.372309)
\psline[linecolor=colour0, linewidth=0.04](4.827888,29.172308)(5.0278883,29.572308)(5.227888,29.172308)
\psbezier[linecolor=black, linewidth=0.04](5.0278883,28.370087)(5.033461,28.5873)(5.187888,28.640085)(5.3378882,28.640086127387157)(5.4878883,28.640085)(5.604995,28.587193)(5.627888,28.370087)
\psline[linecolor=colour0, linewidth=0.04](0.22788818,25.572308)(0.22788818,27.572308)
\psline[linecolor=colour0, linewidth=0.04](0.027888184,26.372309)(0.22788818,26.772308)(0.42788818,26.372309)
\psline[linecolor=colour0, linewidth=0.04](0.8278882,25.572308)(0.8278882,27.572308)
\psline[linecolor=colour0, linewidth=0.04](1.0278882,26.772308)(0.8278882,26.372309)(0.6278882,26.772308)
\psbezier[linecolor=black, linewidth=0.04](0.8278882,27.56231)(0.8223154,27.345095)(0.66788816,27.292309)(0.5178882,27.292308349609375)(0.36788818,27.292309)(0.25078142,27.345201)(0.22788818,27.56231)
\psline[linecolor=colour0, linewidth=0.04](1.6278882,25.572308)(1.6278882,27.572308)
\psline[linecolor=colour0, linewidth=0.04](1.4278882,26.372309)(1.6278882,26.772308)(1.8278881,26.372309)
\psline[linecolor=colour0, linewidth=0.04](2.227888,25.572308)(2.227888,27.572308)
\psline[linecolor=colour0, linewidth=0.04](2.4278882,26.772308)(2.227888,26.372309)(2.0278883,26.772308)
\psbezier[linecolor=black, linewidth=0.04](2.227888,27.56231)(2.2223153,27.345095)(2.0678883,27.292309)(1.9178882,27.292308349609375)(1.7678882,27.292309)(1.6507814,27.345201)(1.6278882,27.56231)
\psbezier[linecolor=black, linewidth=0.08](0.22677708,27.56231)(0.24234988,27.909521)(0.6978882,27.922308)(1.2778882,27.922308349609374)(1.8578882,27.922308)(2.2763648,27.841476)(2.227888,27.560085)
\psline[linecolor=colour0, linewidth=0.04](4.227888,25.572308)(4.227888,27.572308)
\psline[linecolor=colour0, linewidth=0.04](4.4278884,26.772308)(4.227888,26.372309)(4.0278883,26.772308)
\psline[linecolor=colour0, linewidth=0.04](5.0278883,25.572308)(5.0278883,27.572308)
\psline[linecolor=colour0, linewidth=0.04](4.827888,26.372309)(5.0278883,26.772308)(5.227888,26.372309)
\psbezier[linecolor=black, linewidth=0.04](0.8223326,25.570086)(0.7967836,25.7136)(0.9580754,25.837301)(1.2262231,25.826752794053817)(1.4943707,25.816204)(1.6227657,25.796745)(1.6278882,25.57243)
\psbezier[linecolor=black, linewidth=0.04](2.236777,25.572308)(2.2154603,25.878536)(1.8834437,26.092308)(1.2234437,26.092308349609347)(0.5634437,26.092308)(0.22718927,25.917032)(0.23233263,25.575642)
\psbezier[linecolor=black, linewidth=0.04](5.0278883,30.372309)(5.0223155,30.155094)(4.7778883,30.082308)(4.627888,30.082308349609374)(4.477888,30.082308)(4.2507815,30.1452)(4.227888,30.362309)
\psbezier[linecolor=black, linewidth=0.04](5.636777,25.572308)(5.6154604,25.878536)(5.283444,26.092308)(4.6234436,26.092308349609347)(3.9634438,26.092308)(3.6271892,25.917032)(3.6323326,25.575642)
\psbezier[linecolor=black, linewidth=0.04](5.0278883,27.572308)(5.0223155,27.355095)(4.7778883,27.282309)(4.627888,27.282308349609377)(4.477888,27.282309)(4.2507815,27.345201)(4.227888,27.56231)
\psbezier[linecolor=black, linewidth=0.04](4.2223325,25.570086)(4.1967835,25.7136)(4.3580756,25.837301)(4.626223,25.826752794053817)(4.8943706,25.816204)(5.0227656,25.796745)(5.0278883,25.57243)
\psbezier[linecolor=black, linewidth=0.04](7.0278883,25.57453)(7.033461,25.791744)(7.187888,25.84453)(7.3378882,25.8445305718316)(7.4878883,25.84453)(7.604995,25.791637)(7.627888,25.57453)
\psline[linecolor=colour0, linewidth=0.04](8.227888,25.572308)(8.227888,27.572308)
\psline[linecolor=colour0, linewidth=0.04](8.027888,26.372309)(8.227888,26.772308)(8.427888,26.372309)
\psline[linecolor=colour0, linewidth=0.04](8.8278885,25.572308)(8.8278885,27.572308)
\psline[linecolor=colour0, linewidth=0.04](9.027888,26.772308)(8.8278885,26.372309)(8.627888,26.772308)
\psbezier[linecolor=black, linewidth=0.04](8.227888,25.57453)(8.233461,25.791744)(8.387888,25.84453)(8.537889,25.8445305718316)(8.687888,25.84453)(8.804995,25.791637)(8.8278885,25.57453)
\psbezier[linecolor=black, linewidth=0.04](8.8278885,27.56231)(8.822315,27.345095)(8.667889,27.292309)(8.517888,27.292308349609375)(8.367888,27.292309)(8.250781,27.345201)(8.227888,27.56231)
\pscircle[linecolor=black, linewidth=0.04, dimen=outer](2.5278883,33.17231){0.3}
\rput[bl](1.4,32.1){$=$}
\rput[bl](3.2,32.0){$= N\, (1 \,|\, )$}
\rput[bl](3,29.25){$=$}
\rput[bl](6,29.15){$= (2 \,|\, )$}
\rput[bl](3,26.5){$=$}
\rput[bl](6.2,26.5){$=$}
\rput[bl](9.5,26.4){$= (1 \,|\, 1)$}
\end{pspicture}
}
\caption{Contracting the indices~$i_1$ and~$k_1$.}%
\label{figU3}
\end{figure}

Following the procedure explained in the final example in~\cite{creutz},
we contract the indices~$i_1$ and~$l_1$. Using that the matrices are unitary,
we get a factor~$\delta^{k_1}_{j_1}$. The remainder is the group integral in the case~$p-1$.
Denoting the contraction by an arrow, we write the resulting formula symbolically as
\beq \label{directcontract}
\sum_{\text{$(\ell_k)$ with $|(\ell_k)|=p$}} \!\!\!\!\!\!\!\!\!\!c_{(\ell_k)}(N)\: [\ell_1, \ldots \ell_K]
\quad\longrightarrow\quad
\delta^{k_1}_{j_1} \!\!\!\!\!\!\!\!\!\!\!\!\!\!\!\sum_{\text{$(\ell_k)$ with $|(\ell_k)|={p-1}$}}
\!\!\!\!\!\!\!\!\!\!\!\!\!\!\!c_{(\ell_k)}(N)\: [\ell_1, \ldots \ell_K] \:.
\eeq
On the other hand, the contraction of the indices~$i_1$ and~$l_1$
can be performed for each contribution~$[\ell_1, \ldots, \ell_K]$.
This means symbolically that two arrows are connected at the top by an additional line.
The lower points of these arrows become the ``free'' indices~$k_1$ and~$j_1$.
We denote the resulting configuration by~$(q \:|\: \ell_1, \ldots, \ell_k)$, where~$q$ is the length
of the line connecting the two free indices, whereas~$\ell_1, \ldots, \ell_k$ are the lengths of
the closed chains. This construction is illustrated in Figure~\ref{figU3}.

More concretely, we let~$(\ell_1, \ldots, \ell_K)$ be a configuration with~$L(\ell_1, \ldots, \ell_K)=L \geq 2$.
We assume that this configuration violates~\eqref{Ces}, i.e.
\beq \label{contradict}
\big| c_{(\ell_1, \ldots, \ell_K)} \big| \gg C_1/N \:.
\eeq
Our special attention are the factors~$N$ which arise when contracting~$i_1$ and~$l_1$.
Such a factor arises from the formula
\[ \delta^i_j \, \delta^j_i = N \]
when a chain of length one is formed, as is depicted at the top of Figure~\ref{figU3} by a circle.
In fact, a factor~$N$ arises only if such a circle is formed. Moreover, contracting~$i_1$ and~$l_1$
gives rise to at most one factor~$N$.
Next, a circle is formed only if two adjacent arrows are contracted (as shown in the example
at the top of Figure~\ref{figU3}). In this case, we get an open line whose length is the same as that
of the corresponding chain. These findings are summarized by the formula
\beq \label{Lcontract}
[\ell_1, \ldots, \ell_K] \;\longrightarrow\; n_{\ell_K} \,\ell_K\, N \: [\ell_K \,|\, \ell_1, \ldots, \ell_{K-1}]
+ \sum_{q < \ell_K} \alpha_{q, \tilde{\ell}_k}\: [q \,|\, \tilde{\ell}_1, \ldots, \tilde{\ell}_{\tilde{K}}]
\eeq
with suitable combinatorial factors~$\alpha_{q, \tilde{\ell}_k}$ (whose form will be irrelevant).
For clarity, we note that the factor~$n_{\ell_K} \ell_K$ arises when counting the number of possible
contractions giving rise to a circle. Being independent of~$N$, this combinatorial factor is irrelevant
for our argument; we only need that this combinatorial factor is non-zero.

Comparing with~\eqref{directcontract}, where the open line has length one,
one concludes that the summand involving~$N$ in~\eqref{Lcontract} must be compensated
by other configurations. The configurations~$(\tilde{\ell}_1, \ldots, \tilde{\ell}_{\tilde{K}})$
with~$L(\tilde{\ell}_1, \ldots, \tilde{\ell}_{\tilde{K}})>L$ are not relevant in this respect,
because contracting them gives at most one factor~$N$ while by the induction hypothesis~\eqref{ihyp}
all combinatorial factors involve a factor~$1/N$. Therefore,
all the contributions obtained by contracting configurations with~$L(\tilde{\ell}_1, \ldots, \tilde{\ell}_{\tilde{K}})>L$
are of the order~$C_1 \,\O(N^0)$ and thus cannot compensate the summand involving~$N$ in~\eqref{Lcontract}.

Next, configurations~$(\tilde{\ell}_1, \ldots, \tilde{\ell}_{\tilde{K}})$
with~$L(\tilde{\ell}_1, \ldots, \tilde{\ell}_{\tilde{K}})=L$ cannot compensate the summand involving~$N$ in~\eqref{Lcontract}, simply because the condition~$[\ell_K \,|\, \ell_1, \ldots, \ell_{K-1}]
\simeq [\tilde{\ell}_{\tilde{K}} \,|\, \tilde{\ell}_1, \ldots, \tilde{\ell}_{K-1}]$ implies that the two configurations must be the same.

It remains to consider~$(\tilde{\ell}_1, \ldots, \tilde{\ell}_{\tilde{K}})$
with~$L(\tilde{\ell}_1, \ldots, \tilde{\ell}_{\tilde{K}})<L$. In this case, the contractions
do give rise to the configuration~$(\ell_K \,|\, \ell_1, \ldots, \ell_{K-1})$, but only if the
indices~$i_1$ and~$l_1$ lie in different chains. This does not give rise to a factor~$N$.
We thus obtain the estimate
\[ N\, \big| c_{(\ell_1, \ldots, \ell_K)} \big| \lesssim C_1 \:, \]
in contradiction to our assumption~\eqref{contradict}. This concludes the proof.
\QED

Using the estimate of this lemma in~\eqref{gint},
the group integral can be expanded as
\begin{align}
&\fint_{\SU(N)} \scrU^{i_1}_{j_1} \cdots \scrU^{i_p}_{j_p} \: \big( \scrU^{-1} \big)^{k_1}_{l_1} \cdots
\big( \scrU^{-1} \big)^{k_p}_{l_p} \:d\mu_\G(\scrU) 
= c_{(1,\ldots, 1)} [1, \ldots 1] \;\Big( 1+ O \Big( \frac{1}{N} \Big) \Big) \notag \\
&= c_{(1^p)} \: \frac{1}{p!} \sum_{\sigma, \sigma' \in S(p)}
\delta^{i_{\sigma(1)}}_{l_{\sigma'(1)}}\: \delta^{k_{\sigma'(1)}}_{j_{\sigma(1)}} \;
\cdots \delta^{i_{\sigma(p)}}_{l_{\sigma'(p)}}\: \delta^{k_{\sigma'(p)}}_{j_{\sigma(p)}}
\;\Big( 1+ O \Big( \frac{1}{N} \Big) \Big) \notag \\
&= c_{(1^p)} \: \sum_{\sigma \in S(p)}
\delta^{i_1}_{l_{\sigma(1)}}\: \delta^{k_{\sigma(1)}}_{j_1} \;
\cdots \delta^{i_p}_{l_{\sigma(p)}}\: \delta^{k_{\sigma(p)}}_{j_p}
\;\Big( 1+ O \Big( \frac{1}{N} \Big) \Big) \:, \label{preres}
\end{align}
where~$(1^p)$ is a short notation for the configuration with~$p$ chains of length one.
Contracting the indices~$i_1$ and~$l_1$ gives
\begin{align*}
& \fint_{\SU(N)} \scrU^{i_2}_{j_2} \cdots \scrU^{i_p}_{j_p} \: \big( \scrU^{-1} \big)^{k_2}_{l_2} \cdots
\big( \scrU^{-1} \big)^{k_p}_{l_p} \:d\mu_\G(\scrU) \\
&= N\: c_{(1^p)} \: \delta^{k_1}_{j_1} \sum_{\sigma \in S(p-1)}
\delta^{i_2}_{l_{\sigma(1)+1}}\: \delta^{k_{\sigma(1)+1}}_{j_1} \;
\cdots \delta^{i_p}_{l_{\sigma(p)+1}}\: \delta^{k_{\sigma(p)+1}}_{j_p}
\;\Big( 1+ O \Big( \frac{1}{N} \Big) \Big) \:.
\end{align*}
We thus obtain the inductive relation~$c_{(1^{p-1})} = N\: c_{(1^p)}$, which 
in view of~$c_{()}=1$ can be solved to obtain
\[ c_{(1^p)} = \frac{1}{N^p} \:. \]
Using this formula in~\eqref{preres} gives~\eqref{groupint}.
This concludes the proof of Lemma~\ref{lemmagroup}.

\Thanks{{{\em{Acknowledgments:}}
We would like to thank Gernot Akemann,
Yan Fyodorov, Mario Kieburg and Tilo Wettig for helpful comments on the integration over the
unitary group.
We are grateful to the referees for valuable suggestions.
We would like to thank the ``Universit\"atsstiftung Hans Vielberth'' for generous support.
N.K.'s research was also supported by the NSERC grant RGPIN~105490-2018.

\providecommand{\bysame}{\leavevmode\hbox to3em{\hrulefill}\thinspace}
\providecommand{\MR}{\relax\ifhmode\unskip\space\fi MR }
\providecommand{\MRhref}[2]{%
  \href{http://www.ams.org/mathscinet-getitem?mr=#1}{#2}
}
\providecommand{\href}[2]{#2}

\end{document}